\documentclass[preprint,12pt]{elsarticle}

\usepackage[english]{babel}
\usepackage{hyperref}
\usepackage{amsfonts} 
\usepackage{amsmath}
\usepackage{amssymb}
\usepackage{graphicx}
\usepackage{hyperref}
\usepackage{color}
\usepackage{natbib}  
\usepackage{bm} 
\usepackage{isomath} 
\usepackage{braket}
\usepackage{geometry}
\usepackage{fleqn}
\usepackage{txfonts}

\renewcommand{\H}{{\cal H}}
\newcommand{\vect}[1]{\vectorsym{#1}} 

\definecolor{dblue}{RGB}{0,0,0.8}

\begin{document}

\begin{frontmatter}



\title{Statistical physics of flux-carrying Brownian particles}


\author{Antonio A. Valido}
\ead{a.valido@iff.csic.es}
\address{Instituto de F\'isica Fundamental IFF-CSIC, Calle Serrano 113b, 28006 Madrid, Spain}

\date{\today}
\begin{abstract}
Chern-Simons gauge field theory has provided a natural framework to gain deep insight about many novel phenomena in two-dimensional condensed matter. We investigate the nonequilibrium thermodynamics properties of a (two-dimensional) dissipative harmonic particle when the Abelian topological gauge action and the (linear) Brownian motion dynamics are treated on an equal footing. We find out that the particle exhibits remarkable magneticlike features in the quantum domain that are beyond the celebrated Landau diamagnetism: this could be viewed as the non-relativistic Brownian counterpart of the composite excitation of a charge and magneticlike flux. Interestingly, it is shown that the properties of such flux-carrying Brownian particle are in good agreement with the classical statistical mechanics at sufficient high temperatures, as well as are widely consistent with the Third Law of thermodynamics in the studied dissipative scenarios. Our findings also suggest that its ground state may be far from trivial, i.e. it fakes a seemingly degenerate state.
\end{abstract}


\end{frontmatter}

\section{Introduction}\label{intro}
During the last decades there has been a renowned interest in two dimensional systems which may exhibit emergent striking phenomena, such as topological order \cite{wen20041}, ranging from condensed matter \cite{atland20101} to cold atoms \cite{goldman20161} and quantum information theory \cite{kitaev20051}. A prime example of this is the fractional quantum Hall effect \cite{stormer19991}, which is one of the most studied phenomena in condensed matter and whose understanding has promoted the development of many ground-breaking concepts and field theoretical approaches \cite{hansson20171}, among which we highlight the microscopic descriptions built upon the topological quantum field theory, or more specifically, based on the Chern-Simons gauge field theory. Besides the quantum Hall effect \cite{heinonen19911}, this theory has proved useful in the understanding of other phenomena connected to condensed matter systems, for instance, the origin of the anyonic statistics \cite{dunne19991} or high-temperature superconductivity \cite{chen19891}, as well as some purely theoretical applications \cite{witten19891}. Simultaneously, these works further motivated the study of introducing the Abelian Chern-Simons action (which just exists in two-spatial dimensions) in the traditional Maxwell's equations, leading to an unconventional (two-dimensional) electrodynamics \cite{moura20011}, known as topologically massive gauge field or Maxwell-Chern-Simons theory ($\text{CS-QED}_{2+1}$) \cite{deser19821}, that has been successfully applied to study new forms of gauge field mass generation \cite{dunne19991}, the dynamical Lorentz symmetry breaking \cite{hosotani19931,itoh19981,dillenschneider20061}, or the statistics transmutation \cite{matsuyama19901} (e.g. see \cite{ferreiros20181}) which have recently shown appealing applications in quantum computation theory \cite{pachos20121}.

Most recent discoveries like the topological insulators (TI) \cite{qi20111} have fostered the incipient question about how \textit{exotic symmetry structures} are shaped by the thermal noise and dissipative effects \cite{callan19921,cobanera20161,viyuela20151,sieberer20161}, which are ubiqutious in real life experiments and applications. For instance, it was shown that the topological order (mainly characterized by the quantized Hall conductance) of two-dimensional TI are generally degraded by the interaction with a bosonic thermal bath \cite{viyuela20141,zhang20171,shen20151}, though smarter characterizations of this new symmetry have recently indicated that the topological features can be resilient to moderate temperature effects \cite{viyuela20181}. Closely related works have also searched for observables signatures of spatial noncommutative effects in the conventional Brownian motion \cite{cobanera20161,santos20171,misaki20181}. 

Since the Chern-Simons theory lies at the heart of our nowadays understanding of two-dimensional topological matter \cite{hansson20171,qi20111,cho20111}, it seems natural to address the Chern-Simons gauge field in the study of the interplay between dissipation and new underlying symmetry structures. This constitutes a new microscopic approach in the realm of open quantum system theory \cite{weiss20121,caldeira20141,breuer20021}, which is complementary to a recent extended-environment treatment \cite{yao20171} exploiting the geometric magnetism instead \cite{campisi20121}. This approach permited in Ref.\cite{valido20191} to obtain from firts principles a single and unified effective microscopic description that captures novel aspects of the quantum Brownian motion within $\text{CS-QED}_{2+1}$, besides former well-known results \cite{grabert19841,haake19851,hanggi20051}. This description significantly disitingishes from previous studies, based on the traditional independent-oscillator model (e.g. Caldeira-Legget model) \cite{feynman19631,ford19881,valido20131,caldeira19831,grabert19881,devega20171,valido20132}, mainly owning to the fact that the Chern-Simons action ties charge particles to magnetic fluxes (which is nothing else but the manifestation of both parity and time invariance breaking) \cite{dunne19991,ferreiros20181}. This property translates into the system particles are subject to an (two-dimensional) environmental fluctuating force, say $\hat{\vect \xi}_{MCS}$, that follows an equal-time noncommutative relation
\begin{equation}
\left[  \hat \xi^{\alpha}_{MCS}, \hat \xi^{\beta}_{MCS}\right]\propto -i \kappa \epsilon_{\alpha\beta},\label{MCSEF}
\end{equation}
where $\kappa \in \mathbb{R}$ is the so-called Chern-Simons constant and $\epsilon_{\alpha\beta}$ stands for the two-dimensional Levi-Civita symbol. According to the linear response theory \cite{weiss20121,atland20101} this feature has an immediate consequence in the quantum open-system dynamics: the dissipative paticles exhibit a transverse reaction to the environmental forces likewise an ordinary Hall response \cite{callan19921,roy20081}, which ultimately  makes  them  undergo  a  vortexlike  Brownian dynamics \cite{valido20191}.

Based on this dissipative microscopic description, in the present paper we extensively examine the role played by the Abelian topological gauge action in the nonequilibrium thermodynamics and the pseudomagnetic properties of a single (two-dimensional) harmonic oscillator \cite{valido20191}. We find out that the dissipative particle is endowed with notable pseudomagnetic  properties  by virtue of the mentioned environmental Hall response. Unlike the classical Landau diamagnetism \cite{stohr20061} under dissipation \cite{jayannavar19811,dattagupta19971,kumar20091,bandyopadhyay20061,bandyopadhyay20091,bandyopadhyay20062,hong19911,li19901,kumar20141}, the dissipative harmonic oscillator is shown to develop a non-vanishing orbital magneticlike moment without the need of an external magnetic field or independent torque \cite{wan20061}, instead this (non-spontaneous) magnetization occurs as a consequence of certain work delivered by the Maxwell-Chern-Simons field environment. Conversely, the new dissipative microscopic description also predicts a vanishing average magneticlike moment in the high temperature domain, which is in accordance with the classical statistical mechanics, or more specifically, with the well-known Bohr-van Leeuwen (BvL) theorem \cite{savoie20151,pradhan20102,kaplan20091}. This is common to the (two-dimensional) damped harmonic oscillator moving in constrained spaces \cite{pradhan20102,kaplan20091} or in presence of an external time-independent magnetic field \cite{jayannavar19811,dattagupta19971}. However, we shall show that the nonequilibrium thermodynamics properties of our dissipative particle substantially differ from those of a charged magneto-oscillator following the conventional Brownian motion \cite{li19901,hong19911,kumar20091,bandyopadhyay20061,bandyopadhyay20091,bandyopadhyay20062,kumar20141}. In essence, our results stress out that in the quantum domain the environmental Chern-Simons action  "dresses" the dissipative particle with an intrinsic magneticlike flux, so it is effectively promoted from the ordinary damped harmonic oscillator to a flux-carrying Brownian particle.

The present paper is organized as follows. In Sec.\ref{DMCSM} the dissipative microscopic description governing the open-system dynamics is presented. From this, in Sec.\ref{SRPF} we compute a general closed-form expression for the harmonic oscillator partition function following the imaginary-time path integral approach, and widely discussed its properties. Then, the nonequilibrium thermodynamics quantities are studied analytically and numerically for two different dissipative scenarios: the strict Markovian dynamics in Sec.\ref{OHSDS}, as well as an instance of non-Markovian dynamics in \ref{LTSDS}, whilst the Sec.\ref{EIMP} addresses the average magneticlike properties. Finally, we summarize and draw the main conclusions in Sec.\ref{OCR}.

\section{Dissipative Maxwell-Chern-Simons model}\label{DMCSM}
Let us consider the dissipative dynamics of a non-relativistic system, which is composed of $N$ harmonic oscillators 
constrained to move in the $x-y$ plane and endowed with identical mass $m$ for seek of simplicity, in contact with a thermal field environment at time $t_{0}$. Their position and momentum operators shall be denoted by $(\hat{\vect q}_{i},\hat{\vect p}_{i})$ where $i\in \left\lbrace 1,N \right\rbrace $. Let us use the Greek letters and Einstein notation for the two spatial dimensions, and $\hat a_{\vect k}^{\dagger}$ ($\hat a_{\vect k}$) denotes the creation (annihilation) operator of the environmental $\vect k$-mode with excitation frequency (or dispersion relation) 
\begin{equation}
\omega_{\vect k}^{2}=c^2|\vect k|^2+\kappa^2, \label{ESPG}
\end{equation}
where $\vect k \in 2\pi/L\ \mathbb{Z}^2$ ($L$ is a characteristic length of the environment), and $c$ is the speed of light (or the sound velocity of the environment, in a more general dissipative scenario \cite{valido20132}). In its simplest version, it was shown in the Ref.\cite{valido20191} that the dissipative dynamics consistent with the non-relativistic $\text{CS-QED}_{2+1}$ is captured in the long-wavelength and low-energy regime (i.e. small displacement approximation) by the following low-lying microscopic Hamiltonian,
\begin{equation}
\hat \H =\sum_{i=1}^{N} \frac{1}{2m}\left(\hat{\vect p}_{i}-e\hat{\vect A}_{MCS}(\bar{\vect q}_{i})\right)^{2} +\hat V_{R}(\hat{\vect q}_{1},\cdots,\hat{\vect q}_{N})   +\sum_{\vect k}\hbar\omega_{\vect k}\hat a^{\dagger}_{\vect k}\hat a_{\vect k},
\label{HMCS}
\end{equation}
where $0<e$ mainly determines the coupling strength to the Maxwell-Chern-Simons field environment (e.g. the elementary positive charge in the point-particle case), and $\hat V_{R}$ is an effective renormalized potential, i.e.,
\begin{equation}
\hat V_{R}(\hat{\vect q}_{1},\cdots,\hat{\vect q}_{N}) =\hat V(\hat{\vect q}_{1},\cdots,\hat{\vect q}_{N}) -\frac{m}{2}\sum_{i,j=1}^{N}\phi_{\alpha\beta}(\Delta\bar{\vect q}_{ij})\hat q^{\alpha}_{i}\hat q^{\beta}_{j},\label{HCPR}
\end{equation}
with $\hat V(\hat{\vect q}_{1},\cdots,\hat{\vect q}_{N}) $ denoting the bare parabolic potential describing the confining and coupling interaction between system particles, and $\bar{\vect q}_{i}$ denoting the central position of the $i$th harmonic oscillator. Here $\phi_{\alpha\beta}$ stands for an environment-induced renormalization term
\begin{equation}
\phi_{\alpha\beta}(\Delta\bar{\vect q}_{ij})= \frac{\kappa^2}{m}\sum_{\vect k}\frac{\epsilon_{\alpha\lambda}\epsilon_{\beta\lambda'}}{m_{\vect k}}l_{\lambda}(\vect k)l_{\lambda'}(\vect k)\cos(\vect k \cdot \Delta\bar{\vect q}_{ij}),  \nonumber
\end{equation}
where $\Delta\bar{\vect q}_{ij}=\bar{\vect q}_{i}-\bar{\vect q}_{j} $ is the mean distance between the system oscillators, which determines a characteristic length for the interaction between the system particles mediated by the Maxwell-Chern-Simons field environment \cite{valido20132}, and $l_{\alpha}(\vect k)$ denotes the system-environment coupling coefficient given by
\begin{equation}
l_{\alpha}(\vect k)=-\frac{e \sqrt{m_{\vect k}\hbar  }\ \epsilon_{\alpha\beta} k_{\beta} \  }{2\pi L|\vect k|\omega_{\vect k}}f(\vect k),
\label{EIT} 
\end{equation}  
with $f(\vect k)$ being the system particle form factor (that is, the spatial Fourier transform of the particle charge distribution). Since the particle charge distribution is spatially symmetric in most interesting cases \cite{ford19881,valido20131}, in the following we shall assume an isotropic form factor $f(\vect k)=f(|\vect k|)$ in the Eqs. (\ref{HCPR}) and (\ref{EIT}). Furthermore, as similarly occurs in the conventional Brownian motion \cite{ford19881}, $f(|\vect k|)$ must be an (exponential or algebraic) decaying function in order to avoid divergences which may lead to the ultraviolet catastrophe \cite{weiss20121}.

Interestingly, $\hat{\vect A}_{MCS}(\bar{\vect q}_{i})$ (for $i\in\left[ 1,N\right] $) is the vector potential or dynamical gauge field given by \cite{valido20191}
\begin{equation}
\hat{\vect A}_{MCS}(\bar{\vect q}_{i})=\sum_{\vect k}f(\vect k)e^{i\vect k\cdot \bar{\vect q}_{i}}\check{\vect A}_{free}(\vect k)-\frac{\kappa}{2\pi}\sum_{\vect k}\frac{f(\vect k)e^{i\vect k\cdot \bar{\vect q}_{i}}}{\omega_{\vect k}}\big(i\vect k\times \check{\vect A}_{free}(\vect k)\big)\check{G}_{c}(\vect k) \ \vect k,  \label{GFMCS}  
\end{equation}
where $\check{G}_{c}(\vect k)$ and $\check{\vect A}_{free}(\vect k)$ stand, respectively, for the spatial Fourier transform of the two-dimensional Coulomb Green's function and the amplitude of the polarized plane wave of the free Maxwell-Chern-Simons gauge field in the propagation direction $\vect k$, i.e.
\begin{equation}
\check{\vect A}_{free}(\vect k)=\frac{1}{2\pi L\sqrt{2\omega_{\vect k}}}\big(\vect \varepsilon_{\vect k}\hat a_{\vect k}+\vect \varepsilon^{\dagger}_{-\vect k}\hat a_{-\vect k}^{\dagger}\big),
\nonumber
\end{equation}
whilst $\vect \varepsilon_{\vect k}$ is equivalent to the spatial Fourier coefficient of the usual polarization vector \cite{valido20191}. While the first term in the right-hand side of (\ref{GFMCS}) identically coincides with the vector potential obtained from the ordinary dipole approximation (i.e. $e^{i\vect k\cdot\hat{\vect q}_{i}}\approx 1$ for $i\in\{1,N\}$), the second term has no counterpart in previous dissipative microscopic descriptions \cite{kohler20131,yao20171} (e.g. in the independent-oscillator model \cite{feynman19631,ford19881,valido20131,caldeira19831,grabert19881,caldeira20141,devega20171}), and completely stems from the Chern-Simons action \cite{valido20191}. Importantly, the Hamiltonian model (\ref{HMCS}) is obtained after neglecting a back reaction upon the environment \cite{valido20191}, whose effects on the open-system dynamics eventually vanish in the asymptotic time limit by demanding sensible properties to the environment, i.e. a substantially broad spectrum and a continuous and finite system environment coupling strength \cite{valido20191}.

Besides the vector potential in dipole approximation, it is important to realize from the Eq.(\ref{GFMCS}) that the additional contribution in $\hat{\vect A}_{MCS}$ due to the Chern-Simon action explicitly contains the \textit{divergenless} component of the free Maxwell-Chern-Simons gauge field which encapsulates a pseudomagnetic field \cite{dunne19991,dillenschneider20061} (i.e. $\check{B}(\vect k)=i\vect k\times\check{\vect  A}_{free}(\vect k)$). The latter is rooted in a non-vanishing magneticlike flux that it is effectively attached to the system particles by the Chern-Simons action \cite{valido20191} (for instance, in $\text{CS-QED}_{2+1}$ static charges are able to produce simultaneously electric and magneticlike fields \cite{moura20011}). In this way, one may expect that the electric field $\hat{\vect E}_{MCS}$ retrieved by the vector potential (\ref{GFMCS}) (i.e. $\hat{\vect E}_{MCS}(\bar{\vect q}_{i},t)=-\partial_{t}\hat{\vect A}_{MCS}(\bar{\vect q}_{i},t)$ after replacing $\hat a_{\vect k}\rightarrow \hat a_{\vect k}e^{-i\omega_{\vect k}(t-t_{0})}$ in (\ref{GFMCS}) \cite{valido20191}) can split into a Maxwell and Chern-Simons part, i.e.
\begin{equation}
\hat{\vect E}_{MCS}(\bar{\vect q}_{i},t)=\hat{\vect E}_{dip}(\bar{\vect q}_{i},t)+\hat{\vect E}_{CS}(\bar{\vect q}_{i},t), \label{MCSEFI}
\end{equation}
for $i\in\left[ 1,N\right] $, where $\hat{\vect E}_{dip}$ represents the customary dipolar electric field of the system particles, whilst $\hat{\vect E}_{CS}$ can be though of as the non-conservative electric field which is self-consistently induced by a changing magneticlike flux according to the induction Faraday’s law \cite{valido20191}. The latter is the so-called Chern-Simons electric field, which is axially symmetric along the z-direction and yields the non-commutative property (\ref{MCSEF}) (we note that $\hat{\vect E}_{MCS}$ will play the role of the environmental fluctuating force, i.e. $\hat{\vect \xi}(t)=e\hat{\vect E}_{MCS}(t)$) as well as the ordinary Hall response mentioned in the introduction. Intuitively, the novel crucial effect of the dissipative Maxwell-Chern-Simons approach (\ref{HMCS}) could be summarized as follows: each system particle formally consists of a magneticlike flux along the z-axis that in turn gives rise to a non-trivial current of flux tubes across a  given  closed  surface,  and  which  ultimately  produces an intricate electromotive force (emf) responsible for $\hat{\vect E}_{CS}$ \cite{landau19711}. Compared to the ordinary Maxwell electrodynamics, the Chern-Simons action introduces a parity and time-reversal asymmetry in the Maxwell's equations (recall the Chern-Simons action breaks those symmetries in the Lagrangian description) that translates into explicitly breaking the rotational symmetry of the corresponding electric field lines within a localized region of space and time in much the same fashion as a changing external magnetic field does \cite{wan20061}. 

According to standard electrodynamics, one could expect that the aforementioned emf exerts certain torque upon each system particle \cite{wan20061, kobe19831,saha20081}. As a matter of fact, in Ref.\cite{valido20191} it was shown for a single harmonic oscillator (i.e. $N=1$) that the Langevin Markovian equation governing the quantum open-system dynamics resemblances a Brownian motion with an additional rotational force 
\begin{equation}
\hat{\vect F}_{CS}=-m\Omega_{CS}^2 \ \hat{\vect q}\times \vect e_{z},
\label{RFF}
\end{equation}
with $\vect e_{z}$ being the unit vector in the z-direction and $\Omega_{CS}$ determining the coarse-grained flux strength, and where the fluctuating force $\hat{\vect \xi}_{MCS}(t)$ satisfies an extended fluctuation-dissipation relation, e.g. in the high-temperature limit it expresses as follows
\begin{equation}
\frac{1}{2}\left\langle \left\lbrace \hat \xi_{MCS}^{\alpha}(t),\hat \xi_{MCS}^{\beta}(t')\right\rbrace \right\rangle = m\beta^{-1}\Bigg(\delta_{\alpha\beta}2\Gamma_{0}\delta(t-t') +\epsilon_{\alpha\beta}\Omega_{CS}^{2}\text{sgn}(t-t')\Bigg),\label{FDRF}
\end{equation}
where  $\Gamma_{0}$ denotes the usual dissipative coefficient and $\text{sgn}(x)$ stands for the sign of the argument $x$. The rotational force (\ref{RFF}) and the second term in the right-hand side of the Eq.(\ref{FDRF}) are direct consequences of the environmental Hall response manifested by (\ref{MCSEF}). As a result, the force (\ref{RFF}) makes the system particles undergo a vortexlike Brownian dynamics in the Markovian regime, as anticipated in the introduction. In this sense, the average motion of our dissipative system closely resemblances an array of interacting vortex-quasiparticles \cite{ao19991}, quantum rings subject to a static magnetitc field \cite{gumber20181}, or plane rotors in presence of an external time-dependent magnetic flux \cite{kobe19831,kobe19821,asorey19831,caio20161}. Clearly, by disregarding the Chern-Simons effects (i.e. $\kappa\rightarrow 0$), the Eq.(\ref{FDRF}) returns the usual fluctuation-dissipation relation characteristics of the (two-dimensional) \textit{conventional Brownian motion} in the Makovian limit \cite{grabert19841,haake19851,hanggi20051}.

Additionally, the electric field $\hat{\vect E}_{CS}$ produced by the environmental emf can do certain mechanical work on the system particles (e.g. it could change the particle kinetic angular momentum) that somehow could make the Maxwell-Chern-Simons heat bath act moderately as a \textit{work source}  as well. One could expect that such work would manifest in the nonequilibrium Helmholtz free energy of the flux-carrying particle in agreement with the standard thermodynamics \cite{kardar20071,callen19871}. We shall show in the strict Markovian dynamics case that the free energy exhibits a positive linear dependence in temperature at low energies that is intimately related to the rotational force (\ref{RFF}). This result will be further discussed in Sec.\ref{OHSDS} (see Eq.(\ref{FEOLT}) and subsequent discussion). In addition, the aforementioned torque will be presumably responsible for an orbital magneticlike moment of the system particles \cite{wan20061,kobe19831}. As mentioned in the introduction, we find in the Markovian dynamics regime that a single harmonic oscillator (i.e. $N=1$) develops a non-trivial orbital magneticlike moment (see Eq.(\ref{QOMM}) in Sec.\ref{EIMP}) which is mainly determined by the strength of the rotational force (\ref{RFF}). We shall look into this feature more carefully in Sec.\ref{EIMP}.

For our future purpose, it is convenient to carry out a gauge (G\"oppert-Mayer) transformation \cite{kohler20131,ford19881}
\begin{equation}
\hat U_{GM}= \text{exp}\Bigg(-i\frac{e}{\hbar}  \sum_{i=1}^{N} \hat{\vect q}_{i}\cdot\hat{\vect A}_{MCS}(\bar{\vect q}_{i})\Bigg),
\nonumber
\end{equation}
and rewrite the above microscopic description (\ref{HMCS}) in terms of the environment harmonic $\vect k$-oscillator, described by the operators $(\hat x_{\vect k}, \hat p_{\vect k})$ and endowed with mass $m_{\vect k}$, so that we arrive to an equivalent microscopic Hamiltonian which is the basis of the subsequent analysis \cite{valido20191}
\begin{eqnarray}
\hat \H&=&\sum_{i=1}^{N} \frac{\hat{\vect p}_{i}^{2}}{2m}+\hat V_{R}(\hat{\vect q}_{1},\cdots,\hat{\vect q}_{N}) \label{HMCSF2} \\
&+&\sum_{\vect k}\frac{1}{2m_{\vect k}}\Bigg(\bigg( \hat p_{\vect k}+\sum_{i=1}^{N}\big(\omega_{\vect k}l_{\alpha}(\vect k)\sin(\vect k \cdot \bar{\vect q}_{i}) -\kappa\epsilon_{\alpha\beta}l_{\beta}(\vect k)\cos(\vect k \cdot \bar{\vect q}_{i})\big)\hat q^{\alpha}_{i}\bigg) ^2  \nonumber \\
&+&\bigg( m_{\vect k}\omega_{\vect k} \hat x_{\vect k}-\sum_{i=1}^{N}\big(\omega_{\vect k}l_{\alpha}(\vect k)\cos(\vect k \cdot \bar{\vect q}_{i}) +\kappa\epsilon_{\alpha\beta}l_{\beta}(\vect k)\sin(\vect k \cdot \bar{\vect q}_{i})\big)\hat q^{\alpha}_{i}\bigg)^2\Bigg).
\nonumber
\end{eqnarray}
This Hamiltonian is manifestly positive-definite as long as the renomalized potential is prevented from being inverted, i.e. 
\begin{equation}
0\leq \hat V_{R}(\hat{\vect q}_{1},\cdots,\hat{\vect q}_{N}),
\label{CRPS}
\end{equation}
for any $\hat{\vect q}_{i}$ with $i\in\left[1,N \right] $. The above inequality constitutes a necessary condition for the open-system dynamics modeled by (\ref{HMCSF2}) may reproduce the relaxation process towards a thermal equilibrium state regardless of the reduced system initial conditions \cite{haake19851,weiss20121,valido20132,valido20191}. From this point onward we work within the parameter domain where expression (\ref{CRPS}) holds.

Before proceeding with our analysis of the nonequilibrium thermodynamics properties, let us draw some attention to the main features of the Hamiltonian model (\ref{HMCSF2}). By construction, the proposed microscopic description (\ref{HMCS}) takes the form of a minimal-coupling Hamiltonian of the desired system with a dynamical Abelian gauge field $\hat{\vect A}_{MCS}$ acting as a thermal field environment. This feature is common to the standard independent-oscillator model (e.g. Caldeira-Legget model) \cite{haake19851,feynman19631,ford19881,valido20131,caldeira19831,grabert19881,caldeira20141,devega20171}, and indeed, the latter is fully contained in our description as a particular instance \cite{kohler20131,valido20132,ford19881}. We shall see that we recover the well-known results of the nonequilibrium thermodynamics characteristics of the conventional Brownian motion when the Chern-Simons action is turned off ($\kappa \rightarrow 0$). Although the Chern-Simons constant can take either positive or negative continuous values in principle \cite{dunne19991}, we shall also show that the statistical physics properties remains invariant to this choice.

A comment about the experimental verification of our results is in order. Besides the Chern-Simons gauge field displays a remarkable duality with background magnetic fields like in the quantum Hall effect \cite{dunne19991}, its charactertistic effects in the long wavelength limit can be experimentally realized in cold Rydberg atoms by an appropriate manipulation of a radial electric (i.e. $E^{\alpha}=-\sum_{i=1}^{N}\mathcal{E}(\bar{\vect q_{i}})\hat q^{\alpha}_{i}$) and a magnetic field (i.e. $\vect B(\bar{\vect q_{i}})=\mathcal{B}(\bar{\vect q_{i}})\vect z$) \cite{baxter19961,zhang19961}. More specifically, the so-called R\"ontgen interaction has been broadly proposed to experimentally reproduce spatial-spatial and momentum-momentum noncommutativity for the charge particle \cite{zhang20061,zhang20041}. This is the interaction between a magnetic field and the magnetic dipole moment carried by the atom, i.e. $\hat H_{Ront}(\bar{\vect q})\propto \dot{\hat{\vect q}}\cdot(\hat{\vect d}\times\vect B(\bar{\vect q}))$, where $\hat{\vect q}$ refers to the center-of-mass coordinate and $\hat{\vect d}$ is the electric dipole moment of the atom \cite{baxter19961,zhang19961}. Following a similar argument, the R\"ontgen mechanism could be used to simulate the undelying time-reverse asymmetry and parity violation giving rise to the noncommutative dissipative geometry manifested by the Eq.(\ref{MCSEF}). Upon an appropriate arrangement of the ordinary electric and magnetic fields, a large ensemble of Rydberg atoms in high quantum number states (displaying long coherence times) could regard an experimental candidate for the interesting system: a single Brownian particle would represent the center-mass motion of a single structureless Rydberg atom. Since the Chern-Simons constant $\kappa$ will be proportional to $\mathcal{B}(\bar{\vect q_{i}})$ \cite{zhang20061,zhang20041}, we will need a good spatial resolution of the magnetic field in order to tune the Chern-Simons effects between atoms at will. Recalling $\kappa$ is considered sufficiently weak compared to the coupling strength $e$ in the desired parameter regime, there is no need of a strong magnetic field. Although it may be seem challenging to meet these experimental conditions, in the last decades there has been a huge progress in the control and manipulation of Rydberg atoms \cite{saffman20101}, which makes our findings amenable to verification in a series of future experiments in the realm of Brownian motion. On the other side, we would like to emphasize that the Maxwell-Chern-Simons electrodynamics naturally emerges in a wide range of appealing situations in planar condensed matter physics, so that our present treatment is not restricted to the aforementioned Rydberg-atom platform. For instance, it was shown that a system composed of charged particles constrained to move on an infinite plane and subjected to the ordinary 3D electromagnetic interaction leads to an effective Maxwell-Chern-Simons electromagnetic field (after dimensional reduction) \cite{marino19931}, or alternatively, it can be shown that the low-lying description of the massive Thirring model maps onto the Maxwell-Chern-Simons gauge theory (via a bosonization procedure) \cite{fradldn19941}.

\section{Nonequilibrium thermodynamics}\label{SRPF}  
Now we turn the attention to the nonequilibrium thermodynamics properties of the flux-carrying particle. For a better exposition, we shall concentrate the attention on the single harmonic oscillator case (i.e. $N=1$) with frequency $\omega_{0}$ and placed at the origin of the coordinate system ($\bar{q}_{i}^{\alpha}=0$ with $\alpha=1,2$). The potential renormalization appearing in the Eq.(\ref{HCPR}) then takes a diagonal form, i.e. $\phi_{\alpha\beta}=\phi\delta_{\alpha\beta}$ with $\alpha,\beta=1,2$.

As it is commonly considered in the study of the open-system statistical physics or nonequilibrium thermodynamics properties of the conventional Brownian motion \cite{weiss20121,hanggi20081}, we assume that the global system, composed of the interesting system and the Maxwell-Chern-Simons field environment, is in the usual canonical ensemble of equilibrium statistical mechanics at a temperature characterized by $\beta=1/k_{B}T$ (i.e. $\hat \rho_{\beta}\propto e^{-\beta \hat \H}$). In this context, the nonequilibrium thermodynamics quantities (e.g. free energy, internal energy or entropy) related to the dissipative harmonic oscillator, namely flux-carrying particle, are determined from the partition function defined as follows \cite{kumar20141,weiss20121,hanggi20081,ingold20091,ingold20121},
\begin{equation}
Z_{flux}(\beta)=\frac{Z(\beta)}{Z_{MCS}(\beta)},
\label{PFEI}
\end{equation} 
where $Z(\beta)$ and $Z_{MCS}(\beta)$ represent, respectively, the partition functions of the global system and the Maxwell-Chen-Simons field environment in a canonical thermal equilibrium state at inverse temperature $\beta$. Starting from the independent-oscillator or anomalous dissipative model, the approach based on the Eq.(\ref{PFEI}) has been extensively used to analyze the nonequilibrium thermodynamic properties of a broad class of dissipative systems: the damped free particle \cite{hanggi20081,ingold20091,hanggi20061,ingold20121}, the damped harmonic oscillator \cite{ford20071,bandyopadhyay20101} or charged-oscillator particle in presence of an external uniform magnetic field \cite{kumar20141,bandyopadhyay20102,bandyopadhyay20101,bandyopadhyay20061}, as well as the deviation from the standard thermodynamics in the strong coupling regime within the framework known as the Hamiltonian of mean force \cite{hilt20111}. Following the standard approach of statistical mechanics, $Z_{flux}(\beta)$ may be computed by means of the imaginary-time path integral formalism \cite{ingold20021, weiss20121,kumar20141,kumar20091,bandyopadhyay20061,zinn20101}: this is obtained by integrating out the environmental degrees of freedom in the Euclidean action associated to the microscopic Hamiltonian (\ref{HMCSF2}). Specifically, the global partition function for the single-oscillator case reads \cite{weiss20121}
\begin{equation}
Z(\beta)=\oint \mathcal{D}\vect q(\cdot)\oint \mathcal{D}\underline{\vect x}(\cdot)\ \text{exp}\left\lbrace -\mathcal{S}^{(E)}[\vect q(\cdot),\underline{\vect x}(\cdot)]/\hbar\right\rbrace ,
\label{PFEII}
\end{equation}
with $\underline{\vect x}(\tau)$ denoting $\prod_{\vect k} \vect x_{\vect k}(\tau)$ and $\mathcal{S}^{(E)}$ being the Euclidean action of the whole system
\begin{equation}
\mathcal{S}^{(E)}=\mathcal{S}_{S}^{(E)}+\mathcal{S}_{MCS,I}^{(E)}=\int_{0}^{\hbar\beta}d\tau \ \mathcal{L}^{(E)} ,
\label{EAS}
\end{equation}
where $\mathcal{S}_{S}^{(E)}$ is the usual Euclidean action of a two-dimensional harmonic oscillator with bare frequency $\omega_{0}$ \cite{weiss20121}, and we have introduced the Euclidean Lagrangian of the single particle case retrieved by the microscopic model (\ref{HMCSF2}), i.e.,
\begin{equation}
\mathcal{L}^{(E)}=\mathcal{L}_{S}^{(E)}+\mathcal{L}_{MCS,I}^{(E)},
\label{EAE}
\end{equation}
with
\begin{eqnarray}
\mathcal{L}_{S}^{(E)}&=&\frac{1}{2}m\dot{\hat{\vect q}}^2+\frac{1}{2}m\omega_{0}^{2}\hat{\vect q}^{2},
\nonumber \\
\mathcal{L}_{MCS,I}^{(E)}&=&\sum_{\vect k}\frac{m_{\vect k}}{2}\Bigg(\Bigg(\dot{\hat{x}}_{\vect k}^{2}+i\kappa\epsilon_{\alpha\beta} \frac{l_{\beta}(\vect k)}{m_{\vect k}}\hat{q}_{\alpha} \Bigg)^2+\omega^2_{\vect k}\left( \hat{x}_{\vect k}-\frac{l_{\alpha}(\vect k) }{m_{\vect k}}\hat{q}_{\alpha}\right)^2\Bigg),
\nonumber
\end{eqnarray}
where the overdot denotes differentiation with respect to the imaginary time $\tau$ \cite{zinn20101}. Notice that the renomalization term in (\ref{HCPR}) has been absorbed by the Euclidean Lagrangian $\mathcal{L}_{MCS,I}^{(E)}$ in order to compose the square of the first binomial in the right-hand side.

Thanks to the proposed dissipative microscopic model (\ref{HMCSF2}) is quadratic in the canonical variables of both the interested system and environment, the whole path integral (\ref{PFEII}) is Gaussian, and thus, the partition function of the flux-carrying particle at a finite temperature can be readily worked out by using well-known identities from the Gaussian (imaginary-time) path integrals \cite{weiss20121,zinn20101}. We write down directly the result and refer the interested reader to the \ref{app1} for further details. Concretely, we find that the reduced partition function, from which all nonequilibrium thermodynamics quantities will be computed, can be expressed in terms of the bosonic Matsubara frequencies $\nu_{n}=2\pi n/\hbar \beta$ (with $n\in \mathbb{N}^{+}$) as follows
\begin{equation}
Z_{flux}(\beta)=Z_{0}^{2}(\beta)\left(1+ \frac{\tilde{\Lambda}_{\parallel}(0)}{\omega_{0}^2}\right)^{-1} \prod_{n=1}^{\infty}\Bigg[\frac{(\nu_{n}^2+\omega_{0}^2+\tilde{\Delta}(\nu_{n}))^{2}}{(\nu_{n}^2+\omega_{0}^2+\tilde{\Delta}(\nu_{n})+\tilde{\Lambda}_{||}(\nu_{n}))^{2}-|\tilde{\Lambda}_{\perp}(\nu_{n})|^2}\Bigg],\label{PFSII}
\end{equation}
where $Z_{0}(\beta)$ is identical to the partition function of the conventional damped one-dimensional harmonic oscillator with frequency $\omega_{0}$ \cite{weiss20121}, i.e., 
\begin{equation}
Z_{0}(\beta)=\frac{1}{\hbar\beta\omega_{0}}\prod_{n=1}^{\infty}\Bigg[\frac{\nu_{n}^2}{\nu_{n}^2+\omega_{0}^2+\tilde{\Delta}(\nu_{n}) }\Bigg],
\end{equation}
and $\tilde{\Delta}(\nu_{n})$ completely identifies with the Laplace transform of the environment dynamical susceptibility owning to the conventional Brownian motion, which can be given in terms of the Laplace transform of the usual friction kernel $\tilde{\gamma}(s)$ \cite{weiss20121,caldeira20141,grabert19881}, i.e.
\begin{equation}
\tilde{\Delta}(\nu_{n})=|\nu_{n}|\tilde{\gamma}(|\nu_{n}|) =\frac{2}{m\pi}\int_{0}^{\infty}d\omega\frac{J(\omega)}{\omega}\frac{\nu_{n}^2}{\nu_{n}^2+\omega^2},\label{SDCI}
\end{equation}
whereas the kernels $\tilde{\Lambda}_{||}(\nu_{n})$ and $\tilde{\Lambda}_{\perp}(\nu_{n})$ represents respectively the longitudinal and transverse dynamical susceptibilities arising from the environmental Chern-Simons action, which in the single-oscillator case take the form
\begin{eqnarray}
\tilde{\Lambda}_{||}(\nu_{n})&=&\kappa^2\frac{\tilde{\Delta}(\nu_{n})}{|\nu_{n}|^2}, \label{SDCII} \\
\tilde{\Lambda}_{\perp}(\nu_{n})&=&\kappa\frac{\tilde{\Delta}(\nu_{n})}{2\nu_{n}}, \label{SDCIII}
\label{SDCIII}
\end{eqnarray}
and $J(\omega)$ stands for the standard definition of the spectral density,
\begin{equation}
J(\omega)=\frac{e^2\hbar}{16\pi L^2}\sum_{\vect k}\omega_{\vect k}f^2(|\vect k|)\delta(\omega-\omega_{\vect k}), 
\label{SPDI}
\end{equation}
where $\omega_{\vect k}$ is given by the dispersion relation (\ref{ESPG}). Hence, all the effects due to the environmental emf discussed in Sec.\ref{DMCSM} must be fully characterized by (\ref{SDCII}) and (\ref{SDCIII}). Concretely, the Chern-Simons dynamical susceptibility $\tilde{\Lambda}_{\perp}(\nu_{n})$ quantifies the transverse response to the fluctuating force, so it completely encodes the environmental Hall response discussed in Secs. \ref{intro} and \ref{DMCSM}.

Interestingly, the real-time Fourier transforms of the expressions (\ref{SDCII}) and (\ref{SDCIII}) (obtained via analytic continuation $s\rightarrow -i\omega+0^{+}$) reveal that the power spectral density of the Chern-Simons dynamical susceptibilities is $1/f$-type (i.e. $\check{\Delta}
_{\parallel}(\omega)\propto 1/\omega^{2}$ and $\check{\Delta}_{\perp}(\omega)\propto 1/\omega$), which is common to many sources of noise in condensed matter physics and quantum information \cite{paladino20141}, for instance magnetic flux noise in SQUIDs. We shall see that this feature has profound consequences in the nonequilibrium thermodynamics properties: it implies that the influence of the environmental Chern-Simons action in the statistical mechanics may be substantially diminished in the high-temperature regime (recall $\nu_{n}\propto \beta^{-1}$) as compared to the conventional dissipative mechanism described by $\tilde{\Delta}(\nu_{n})$. In other words, the Chern-Simons effects may have no significant influence on the nonequilibrium thermodynamics of the dissipative harmonic oscillator in the classical limit. This will be shown to occur in the two dissipative scenarios studied in Secs. \ref{OHSDS} and \ref{LTSDS}. 

In addition, the $1/f$-type feature gives rise to a time-local contribution in the effective Euclidean action (see the Eq.(\ref{SEASI}) in appendix \ref{app1}) that can make the longitudinal Chern-Simons dynamical susceptibility manifest in the first term of the right-hand side of the Eq.(\ref{PFSII}). Indeed, the infrared contribution of the longitudinal Chern-Simons dynamical susceptibility is intimately related to the aforementioned potential renomalization, that is,
\begin{equation}
\phi=\lim_{s\rightarrow 0}\tilde{\Lambda}_{||}(s)=\frac{2\kappa^2}{m\pi}\int_{0}^{\infty}\frac{J(\omega)}{\omega^3}d\omega.
\label{RPCS}
\end{equation}
Note that this renormalization term will represent a minor correction to the nonequilibrium thermodynamics quantities since $\phi<\omega_{0}$ in order to the subsidiary condition (\ref{CRPS}) be hold. Although this could be explicitly absorbed in the definition of the partition function $Z_{\text{flux}}(\beta)$, we will keep the exact dependence thorough the following discussion to emphasize its role in the statistical physics. Furthermore, it is worthwhile to notice that $\phi$ seems to blows up at the origin $\omega=0$. However, there is no such infrared divergence in an appropriate description since the environmental spectrum is gaped by $\kappa$ as is illustrated by (\ref{ESPG}), and thus, the spectral density formally vanishes for $0\leq \omega<\kappa$.

From the Eqs. (\ref{PFSII}), (\ref{SDCI}), (\ref{SDCII}), and (\ref{SDCIII}) directly follow that the nonequilibrium thermodynamics properties of the reduced system, and the dissipative dynamics as well, are ultimately determined by the choice of the spectral density (\ref{SPDI}), as similarly occurs in the conventional Brownian motion. Since we are interested in a dissipative scenario where the reduced density matrix of the flux-carrying particle may be well characterized by a thermal equilibrium state, we must require further constraints to the spectral density to prevent spurious situations (e.g. a non-positive density of states for the flux-carrying particle \cite{hanggi20081}). For instance, in the Langevin-equation approach we must demand that the real-time Fourier transform of the retarded Green's function, denoted by $\check{\vect G}_{R}(\omega)$, dictating the quantum open-system dynamics has no isolated pole lying outside of the environmental dense spectrum in order to guarantee the contour-ordered Green's function of the interested system satisfies the Kubo-Martin-Schwinger boundary condition \cite{valido20191}. In the present framework, we likewise demand the covariance matrix of the Gaussian path integral, which retrieves the reduced partition function $Z_{flux}(\beta)$, is positive-definite, or equivalently,
\begin{equation}
|\tilde{\Lambda}_{\perp}(s)|<s^2+\omega_{0}^2+\tilde{\Delta}(s)+\tilde{\Lambda}_{||}(s) \ \ \text{with} \ \ s\in \mathbb{R}^{+},
\label{SCDI}
\end{equation} 
which ensures, together with the condition (\ref{CRPS}), that the expression (\ref{PFSII}) will be a well-defined partition function \cite{hanggi20051} (see the Eqs. (\ref{RPFA}) and (\ref{ZSCM}) in the \ref{app1}). Intuitively, the subsidiary condition (\ref{SCDI}) basically manifests that, in order to the flux-carrying particle be in a thermal equilibrium state from standard thermodynamics \cite{callen19871}, the dissipation mechanism must dominate the open-system dynamics against the enviromental Hall response. This will be more clear in the future discussions. Finally, we could also require that the memory kernels decay in the ultraviolet limit \cite{weiss20121,kumar20141}, e.g. $\tilde{\Delta}(s),\tilde{\Lambda}_{||}(s),\tilde{\Lambda}_{\perp}(s)\sim s^{-l}$ for $s\rightarrow \infty$ with $2<l$, to certify the infinite product in (\ref{PFSII}) will converge. 

At first sight, the Eq.(\ref{PFSII}) (together with (\ref{SDCII}) and (\ref{SDCIII})) reveals that the Chern-Simons effects will apparently manifest at second-order in the Chern-Simons coupling strength compared with the harmonic oscillator bare frequency, so that the environmental Hall response may substantially influence the open-system statistical physics of the dissipative particle, at least in the low-temperature regime. In the following sections we address the analysis of the deviation of the nonequilibrium thermodynamics from the damped two-dimensional harmonic oscillator for two distinct dissipative scenarios. More concretely, we extensively study both the strict Markovian dissipative evolution characterized by the Ohmic spectral density, and the non-Markovian dynamics characteristic of the thermal harmonic noise, that is the Lorentzian environmental power spectrum.

\subsection{Strict Ohmic spectral density}\label{OHSDS}
Let us begin our analysis with the extensively studied strict Ohmic dissipative scenario. Thus, the spectral density takes the well-known form $J_{ohm}(\omega)=2m \gamma_{0} \omega$  where $\gamma_{0}$ denotes the classical friction coefficient which essentially quantifies the system-environment coupling strength \cite{hanggi20051}. Substituting this in the Eqs. (\ref{SDCI}), (\ref{SDCII}) and (\ref{SDCIII}); we directly arrive at
\begin{eqnarray}
\tilde{\Delta}(s)&=& 2 \gamma_{0} s, \label{SDCO} \\
\tilde{\Lambda}_{||}(s)&=&  \frac{2 \gamma_{0}\kappa^2}{s} , \label{SDCOD} \\
\tilde{\Lambda}_{\perp}(s)&=& \kappa \gamma_{0} \label{SDCODD}.
\end{eqnarray}
At this stage, we must stress out that to be our treatment physically consistent with a dissipative Markovian picture, it must be satisfied the following conditions
\begin{equation}
\frac{\kappa}{\omega_{0}},\frac{\gamma_{0}}{\omega_{0}}\ll 1,
\label{SSCI}
\end{equation}
which reflects nothing but the system-environment interaction and environment spectrum gap must be sufficiently small compared to the the bare frequency of the interesting oscillator. As a counterpart, from the Eq.(\ref{SDCOD}) follows that the longitudinal Chern-Simons dynamical susceptibility diverges in the infrared limit (i.e. $\beta\rightarrow \infty$), and byproduct, the nonequilibrium thermodynamics quantities will diverge as well. Although the latter manifests that the strict Ohmic spectral choice may fail to describe the properties of the flux-carrying particle at absolute zero temperature \cite{valido20131}, it permits to derive closed-form expressions for the nonequilibrium thermodynamics quantities that provide useful intuition for a broad set of values of the problem parameters. For instance, the strict Ohmic spectral choice has been extensively used to study the conventional Brownian motion in both absence and presence of an external uniform magnetic field \cite{li19901,bandyopadhyay20091,dattagupta19971,bandyopadhyay20061}, despite the free and internal energies are logaritmic divergent \cite{li19901} or the heat capacity may take on negative values \cite{ingold20121}. 

Owing to the dissipative kernels (\ref{SDCO}), (\ref{SDCOD}) and (\ref{SDCODD}) have an algebraic form, one may readily see that the reduced partition function (\ref{PFSII}) turns into an infinitive product of rational expressions. This infinite product may be conveniently manipulated to be expressed in terms of the Euler representation of the Gamma function \cite{weiss20121,kumar20141}, denoted by $\Gamma(x)$, via making use of its algebraic properties \cite{abramowitz19641}. Concretely, the reduced partition function can be cast in the following form (see \ref{app2} for further details) 
\begin{equation}
Z_{flux}(\beta)=Z_{0}^{2}(\beta)\left(1+ \frac{\phi}{\omega_{0}^2}\right)^{-1} \frac{\prod_{i=1}^{3}\Gamma\left(1+\frac{ R_{i}\hbar \beta}{2\pi} \right) \Gamma\left(1+\frac{ R_{i}'\hbar \beta}{2\pi} \right)}{\Big[\Gamma\left(1+\frac{ r_{1}\hbar \beta}{2\pi}  \right)\Gamma\left(1+\frac{ r_{2}\hbar \beta}{2\pi}  \right)\Big]^2},\label{PFSO} 
\end{equation}
with $r_{i}$ given by, 
\begin{equation}
r_{1,2}=-\gamma_{0}\pm \sqrt{\gamma_{0}^2-\omega_{0}^2},     \nonumber 
\end{equation}
and $R_{i}$ and $R'_{i}$ (for $i\in\left\lbrace 1,2,3 \right\rbrace $) being the roots of the polynomial
\begin{equation}
Q(R)=R^2\Bigg(R^2+\omega_{0}^2+2\gamma_{0}R+\frac{2\gamma_{0}\kappa^2}{R}\Bigg)^{2}-\kappa^2\gamma_{0}^2R^2. \label{POSD}
\end{equation}
It can be seen from (\ref{POSD}) that both $R_{1,2}$ and $R'_{1,2}$ boil down into $r_{1,2}$ in the limit of vanishing Chern-Simons action (i.e. $\kappa\rightarrow 0$) whereas $R_{3}=R_{3}'=0$, so that $Z_{flux}(\beta)$ converges to the partition function $Z_{0}^{2}(\beta)$ of the damped two-dimensional harmonic oscillator (deduced from the standard independent-oscillator model in the Markovian regime \cite{kumar20091}) when the Chern-Simons effects are disregarded. 

To get a better understanding about the influence of the environmental Chern-Simons action, it is convenient to carry out a perturbative analysis in the small parameter $\kappa/\omega_{0}\ll 1$, which is in full agreement with a Markovian treatment (recall that the latter holds when the subsidiary condition (\ref{SSCI}) is satisfied). More concretely, it is important to realize that (\ref{POSD}) is proportional to the determinant of the Laplace transform of the inverse of the retarded Green's function (see the Eq.(\ref{ZSCM}) in \ref{app1}), i.e. 
\begin{equation}
Q(s)=s^2\det\Big(\tilde{\vect G}_{R}^{-1}(s)\Big).
\label{POSDI}
\end{equation}
Since we are interested in the strict Makovian dynamics regime, we may require $\tilde{\vect G}_{R}(s)$ to take a Breit-Wigner resonance shape around certain renormalized frequency $\Omega_{0}$ without loss of generality \cite{valido20191,alamoudi19991,alamoudi19981,anisimov20091}. The latter can be accomplished by doing a first-order Taylor expansion of the longitudinal part of the dynamical susceptibility appearing in (\ref{POSD}), i.e.,
\begin{equation}
2\gamma_{0}s+\frac{2\gamma_{0}\kappa^2}{s}\approx \frac{2\gamma_{0} \kappa^2}{\Omega_{0}}+2\gamma_{0}\Omega_{0}+2\Gamma_{0}(s-\Omega_{0}),
\nonumber 
\end{equation}
where
\begin{equation}
\Gamma_{0}=\gamma_{0}-\frac{\gamma_{0}\kappa^2}{\Omega_{0}^{2}}, 
 \label{GOCS0}
\end{equation}
and the renormalized frequency is the real root of the following algebraic equation
\begin{eqnarray}
\Omega_{0}^3-\omega_{0}^2\Omega_{0}-4\gamma_{0}\kappa^2=0.
\label{OWF0}
\end{eqnarray}
We note that the rotational force (\ref{RFF}) is exclusively consequence of the second term in the right-hand side of (\ref{POSD}), so we may rephrase the effect of the environmental Hall response in terms of the rotational force strength,
\begin{equation}
\Omega_{CS}=\pm\sqrt{\gamma_{0}\kappa}.
 \label{GOCS0RF}
\end{equation} 
Following this approach, the roots of the polynomial (\ref{POSD}) finally read
\begin{eqnarray}
R_{1/2}&\simeq&-\Gamma_{0}\pm \sqrt{\Gamma_{0}^2-\Omega_{CS}^{2}-\Omega_{0}^2}, \label{POSDR} \\
R_{1/2}'&\simeq& -\Gamma_{0}\pm \sqrt{\Gamma_{0}^2+\Omega_{CS}^{2}-\Omega_{0}^2}, \nonumber \\
R_{3}&=&R_{3}'=0.   \nonumber
\end{eqnarray}
Taking $\kappa$ identically to zero, it is clear from (\ref{GOCS0}), (\ref{OWF0}), and (\ref{GOCS0RF}) that we obtain $\Omega_{0}=\omega_{0}$, $\Gamma_{0}=\gamma_{0}$, and $\Omega_{CS}=0$, as expected. In this way, all the influence of the longitudinal Chern-Simons dynamical susceptibility is completely encoded by the parameters $\Gamma_{0}$ and $\Omega_{0}$, whereas the environmental Hall effects are solely encapsulated by the parameter $\Omega_{CS}$. This Breit-Wigner approach will be further exploited in Sec.\ref{SRPF} to study the average magneticlike properties of the flux-carrying particle. A continuation, we examine the impact of both longitudinal and transverse Chern-Simons effects upon the nonequilibrium thermodynamics quantities of the flux-carrying Brownian particle provided by the reduced partition function (\ref{PFSO}) endowed with the approximated roots (\ref{POSDR}). 

\subsubsection{Free energy}\label{Sfree}
By replacing the Eq.(\ref{PFSO}) in the formal definition of the Helmholtz free energy in terms of the partition function \cite{kumar20141,weiss20121,hanggi20081}, i.e. $F_{flux}(\beta)=-\beta^{-1}\ln Z_{flux}(\beta)$, one may readily obtain the deviation of the free energy
\begin{eqnarray}
\Delta F(\beta)&=& F_{flux}(\beta)-2F_{0}(\beta) \label{FEO} \\
&=& \beta^{-1}\log\left(1+ \frac{\phi}{\omega_{0}^2}\right) -\beta^{-1}\sum_{i=1,2}\Bigg(\log\Gamma\left(1+\frac{ R_{i}\hbar \beta}{2\pi} \right) \nonumber \\
&+&\log\Gamma\left(1+\frac{ R_{i}'\hbar \beta}{2\pi} \right)\Bigg) +2\beta^{-1}\sum_{i=1,2}\log\Gamma\left(1+\frac{r_{i}\hbar \beta}{2\pi} \right), \nonumber
\end{eqnarray}
where $F_{0}(\beta)$ coincides identically with the free energy of the damped one-dimensional harmonic oscillator in the Markovian limit \cite{weiss20121,hanggi20081,bandyopadhyay20061,kumar20091,bandyopadhyay20102,bandyopadhyay20091}. In words, the novel effects in the free energy due to the environmental Chern-Simons action are completely contained by the terms on the right-hand side of Eq.(\ref{FEO}). Let us focus the attention on the high- and low- temperature regime. In the former limit (i.e., $ \hbar \Gamma_{0}\beta\ll 1$), the gamma function in (\ref{FEO}) may be approximated by $\ln\Gamma(z)\simeq-\ln z-\gamma_{E}z +\pi^2z^2/12$ with $\gamma_{E}$ being the Euler-Mascheroni constant \cite{abramowitz19641}. Thus, one can show that the expression (\ref{FEO}) for the free energy simplifies to
\begin{equation}
\Delta F(\beta)=\beta^{-1}\log\left(1+ \frac{\phi}{\omega_{0}^2}\right)+\frac{2\hbar \gamma_{E}}{\pi}(\gamma_{0}-\Gamma_{0}) +\frac{(2\gamma_{0}^{2}-2\Gamma_{0}^{2}-\omega_{0}^2+\Omega_{0}^2)\hbar^2\beta}{12}+\mathcal{O}\left( \beta^{2}\right), \label{FEOHT}
\end{equation}
where the first term in the right-hand side clearly identifies with the contribution owning to the aforementioned potential renormalization. By ignoring this, the above equation (\ref{FEOHT}) unveils that the longitudinal Chern-Simons effects (encoded in $\Gamma_{0}$ and $\Omega_{0}$) lead the free energy of the flux-carrying particle to asymptotically reach (slightly) larger values in comparison to the damped two-dimensional harmonic oscillator in the classical limit. Nevertheless, the third term in the right-hand side of (\ref{FEOHT}), which depends linearly on the inverse temperature, reveals that the influence of the Chern-Simons action substantially diminishes in the classical limit in agreement with the previous discussion in Sec.\ref{SRPF}. 

Although it is not shown here, the contribution of the rotational strength represent a fourth-order term $\Omega_{CS}^4$ that is rapidly decreasing with $\sim \beta^{-3}$. From this is clear that the dissipative mechanism contained by the longitudinal dynamical susceptibility widely exceed the Hall response. The latter can be understood by going back to the general expression of the partition function (\ref{PFSII}) once we have done the Breit-Wigner approximation: the influence of the transverse dynamical susceptibility $\tilde{\Lambda}_{\perp}(s)$ in the strict Markovian scenario remains constant whilst the longitudinal counterpart grows linearly with temperature (recall $\nu_{n}\propto \beta^{-1}$). We shall see in Sec.\ref{EIMP} that this feature is consistent with the results obtained alternatively for the average magneticlike moment of the flux-carrying particle. Furthermore, this observation is in agreement with the results obtained in Ref.\cite{valido20191}, where it was shown that the environmental Hall effect in the high-temperature regime represents a fourth-order correction to the Markovian dissipative dynamics of the conventional damped harmonic oscillator. 

Similarly, the Eq.(\ref{FEO}) in the low temperature regime, i.e. $ \hbar \Gamma_{0}\beta\gg 1$, can be approximated with the help of \cite{abramowitz19641} 
\begin{equation}
\ln\Gamma(z)\simeq z\log z-\frac{1}{2}\log z-z+\frac{1}{2}\log(2\pi)+\frac{1}{12 z}-\frac{1}{360 z^3},
\nonumber
\end{equation} 
which after some straightforward manipulation retrieves the leading low-temperature behavior
\begin{eqnarray}
\Delta F(\beta)&=&\beta^{-1}\log\left(1+ \frac{\phi}{\omega_{0}^2}\right)+\frac{2\hbar(\gamma_{0}-\Gamma_{0})}{\pi} \label{FEOLT} \\
&-&\frac{\hbar}{2\pi}\sum_{i=1,2}\Bigg(R_{i}\log\left(  \frac{\hbar R_{i}\beta}{2\pi}\right) +R'_{i}\log\left(  \frac{\hbar R'_{i}\beta}{2\pi}\right) -2r_{i}\log\left(  \frac{\hbar r_{i}\beta}{2\pi}\right) \Bigg)\nonumber\\ 
&-&\frac{\beta^{-1}}{2}\log\left(\frac{\Omega_{0}^{4}}{\omega_{0}^{4}}-\frac{\Omega_{CS}^{4}}{\omega_{0}^4} \right) +\frac{2\pi (\Gamma_{0}\omega_{0}^2\Omega_{0}^2-\gamma_{0}(\Omega_{0}^4-\Omega_{CS}^4))\beta^{-2}}{3\hbar\omega_{0}^{2}(\Omega_{0}^{4}-\Omega_{CS}^{4})}+\mathcal{O}\left( \beta^{-4}\right). \nonumber
\end{eqnarray}
A quick glance reveals that the third term in the right-hand side of the Eq.(\ref{FEOLT}) introduces a logarithmic divergence at absolute zero temperature ($\beta\rightarrow\infty$) in the strict Markovian dissipative scenario. As was previously explained, this anomalous behavior is intimately related to the fact that the longitudinal Chern-Simons dynamical susceptibility presents an infrared divergence.

Unlike the charged magneto-oscillator following the conventional Brownian motion
\cite{kumar20091,bandyopadhyay20061,bandyopadhyay20091,bandyopadhyay20062,hong19911,li19901,kumar20141}, we may also observe that the free energy of the flux-carrying particle exhibits a linear dependence with temperature (see the fourth term in the right-hand side of Eq.(\ref{FEOLT})). The role of the latter is made clear by doing a perturbative analysis in the small parameter $\kappa/\omega_{0}\ll 1$ once substituted (\ref{GOCS0}), (\ref{GOCS0RF}) and $\Omega_{0}\approx \omega_{0}+2\gamma_{0}\kappa^2/\omega_{0}^{2}$, i.e.,
\begin{eqnarray}
\log\left(\frac{\Omega_{0}^{4}}{\omega_{0}^{4}}-\frac{\Omega_{CS}^{4}}{\omega_{0}^4} \right) &\simeq& 4\log\left(\frac{\Omega_{0}}{\omega_{0}}\right) -   \left( \frac{\Omega_{CS}}{\Omega_{0}}\right) ^{4} \label{AFRR}\\
&=&\frac{\gamma_{0}\kappa^2(8\omega_{0}-\gamma_{0})}{\omega_{0}^4} +\frac{8\gamma_{0}^2\kappa^{4}(\gamma_{0}-\omega_{0})}{\omega_{0}^{7}} +\mathcal{O}\left( \left( \frac{\kappa}{\omega_{0}}\right)^6\right)  \nonumber ,
\end{eqnarray}
which unveils that such linear term constitutes a negative contribution to the free energy up to next leading order in the Chern-Simons coupling strength (see the subsidiary condition (\ref{SSCI})). This feature has a deep consequence on the thermodynamic entropy $S_{flux}(\beta)$ of the flux-carrying particle: it make the entropy to take a  (positive) non-vanishing constant value in the low temperature regime as similarly occurs in degenerate systems \cite{kardar20071}. Using the Maxwell's thermodynamics relations
\begin{equation}
S_{flux}(\beta)=-k_{B}\frac{\partial F_{flux}(\beta)}{\partial \beta^{-1}}, \label{MTR}
\end{equation}
we may see in fact that the deviation of the particle entropy $\Delta S$ from the conventional Brownian motion does not cancel at low temperatures (if we ignore the aforementioned infrared divergence). This behavior of the entropy shall be further discussed in Sec.\ref{SEntropy}.

Interestingly, one may also observe from (\ref{AFRR}) that the second term in the right-hand side is negative and is purely due to the environmental Hall response. In turn, this means that the Hall effects contributes positively to the free energy. Recalling the discussion in Sec.\ref{DMCSM}, this novel feature could be attributed to the fact that the rotational force (\ref{RFF}) stemming from the Hall response is enable to deliver certain mechanical work that may significantly contribute to the particle free energy \cite{landau19711,roy20081}. To further understand the operational meaning of this we may devise an experiment in which we are able to manipulate at will the Chern-Simons constant whilst the dissipative microscopic description (\ref{HMCSF2}) remains valid (e.g. flux-quench experiments are excluded \cite{nakagawa20161}), so that we may think of the \textit{reversible isothermal} process, after switching on quasistatically the Chern-Simons action, by which the dissipative harmonic oscillator is taken from an initial to a final thermal equilibrium state with the same temperature $\beta^{-1}$. Physically, the Maxwell-Chern-Simons field environment is assumed to relax very fast such that all process of interest within it are essentially quasistatic (which is consistent with a dissipative Markovian dynamics). According to standard thermodynamics (e.g., see the Clausius inequality of classical thermodynamics) \cite{callen19871,jarzynski20111}, a positive variation of the free energy in such scenario must represent the minimum work added to the system by the environmental Hall effects. The Eq.(\ref{FEOLT}) certainly manifests that the Maxwell-Chern-Simons environment may supply the interesting system with some power at low energies. We shall show in Sec.\ref{LTSDS} that this observation also coincides with the results in the low-temperature limit obtained for the dissipative scenario characterized by the Lorentzian-type spectral density (see Fig.\ref{Fig1}). Furthermore, following the same line of thinking, the asymptotic value of $\Delta F(\beta\rightarrow \infty)$ represents the minimum work needed for coupling the dissipative harmonic oscillator to the Maxwell-Chern-Simons field environment \cite{ford20061}, as well as the minimum energy necessary to set up the aforementioned emf responsible for the rotational force. Hence, the expression (\ref{AFRR}) alternatively tells us that the the formation of the flux-carrying particle is not an spontaneous process from the thermodynamics point of view (which is in agreement with the Second Law of thermodynamics).

\subsubsection{Internal energy}
Now we study the difference of the internal energy between the flux-carrying particle with respect to the damped two-dimensional harmonic oscillator, i.e. $\Delta U(\beta)=U_{flux}(\beta)-2 U_{0}(\beta)$ (with $U_{0}(\beta)$ corresponding to the internal energy of the damped one-dimensional harmonic oscillator in the Markovian regime). Specifically, we find 
\begin{eqnarray}
\Delta U(\beta)&=&-\frac{\partial}{\partial\beta}\ln \frac{Z_{flux}(\beta)}{Z_{0}^{2}(\beta)} \label{IEO} \\
&=&-\frac{\hbar}{2\pi}\sum_{i=1,2}\Bigg(R_{i}\psi\left(1+\frac{ R_{i}\hbar \beta}{2\pi}\right) +R_{i}'\psi \left(1+\frac{ R_{i}'\hbar \beta}{2\pi}\right)\Bigg)  +\frac{\hbar}{\pi}\sum_{i=1,2}r_{i}\psi\left(1+\frac{ r_{i}\hbar \beta}{2\pi}\right) ,\nonumber
\end{eqnarray}
where $\psi(z)$ denotes the digamma function with argument $z$ \cite{gradshteyn20141}. Let us analyze the asymptotic behavior of Eq.(\ref{IEO}) in the high- and low- temperature limit. In the first case we have $z\ll 1$, and the digamma function may be approximated by $\psi(z)\simeq -1/z-\gamma_{E}+\pi^2z/6$ \cite{abramowitz19641,hilt20111}, so that we obtain the internal energy with the leading Chern-Simons correction reading
\begin{equation}
\Delta U(\beta)=\frac{2\hbar \gamma_{E}}{\pi}(\gamma_{0}-\Gamma_{0}) +\frac{(2\gamma_{0}^{2}-2\Gamma_{0}^{2}-\omega_{0}^2+\Omega_{0}^2)\hbar^2\beta}{6}+\mathcal{O}\left( \beta^{2}\right).\label{IEOHT}
\end{equation}
Clearly, the Eq.(\ref{IEOHT}) shows that the internal energy of the flux-carrying particle may be slightly higher than the internal energy of the damped two-dimensional harmonic oscillator. As similarly occurs to the free energy in the high-temperature limit, there is no trace of the rotational strength parameter up to next leading order in the inverse temperature, which indicates that the Chern-Simons effects in the internal energy is mainly due to the dissipative mechanism contained in the longitudinal dynamical susceptibility (e.g. see the second term in the right-hand side of (\ref{IEOHT})). Furthermore, the influence of the latter substantially decay for increasing temperatures, and eventually, saturates to an asymptotic constant value.

Once again the situation drastically changes in the low-temperature regime. In this limit, we may approach $\psi(z)\simeq \log(z)-1/2z-1/12 z^2+1/120 z^4$ \cite{abramowitz19641,hilt20111}, which inserted in the expression (\ref{IEO}) yields
\begin{eqnarray}
\Delta U(\beta)&=&-\frac{\hbar}{2\pi}\sum_{i=1,2}\Bigg(R_{i}\log\left( \frac{\hbar R_{i}\beta}{2\pi}\right) +R'_{i}\log\left( \frac{\hbar R'_{i}\beta}{2\pi}\right) -2r_{i}\log\left( \frac{\hbar r_{i}\beta}{2\pi}\right) \Bigg) \nonumber\\
&-&\frac{2\pi(\Gamma_{0}\omega_{0}^2\Omega_{0}^2-\gamma_{0}(\Omega_{0}^4-\Omega_{CS}^4))\beta^{-2}}{3\hbar \omega_{0}^2(\Omega_{0}^4-\Omega_{CS}^4)} +\mathcal{O}\left( \beta^{-4}\right), \label{IEOLT}
\end{eqnarray}
where the first term represents a logarithmic divergence at absolute zero temperature, as previously anticipated. On the other hand, a perturbative analysis of the second term of the right-hand side of (\ref{IEOLT}) in the small parameter $\kappa/\omega_{0}\ll 1$ unveils that this essentially constitutes a positive contribution, i.e.,
\begin{eqnarray}
\frac{\Gamma_{0}\omega_{0}^2\Omega_{0}^2-\gamma_{0}(\Omega_{0}^4-\Omega_{CS}^4)}{\hbar \omega_{0}^2(\Omega_{0}^4-\Omega_{CS}^4)}&\simeq& \frac{\Gamma_{0}\omega_{0}^2-\gamma_{0}\Omega_{0}^2}{\hbar\omega_{0}^2\Omega_{0}^2} +\frac{\Gamma_{0}}{\hbar \Omega_{0}^2}\left(\frac{\Omega_{CS}}{\Omega_{0}} \right)^{4} \label{IELTAP} \\
&\approx& -\frac{\gamma_{0}\kappa^2}{\hbar\omega_{0}^4},
\nonumber
\end{eqnarray}
which implies that the internal energy of the flux-carrying particle will be larger than the damped two-dimensional harmonic oscillators for increasing low temperatures. This indicates that the Maxwell-Chern-Simons field environment behaves as a heat source \cite{callan19921} (i.e. a heat bath): the environmental Chern-Simons effects will also deliver heat upon the system particle that transforms into kinetic energy, which is characteristics of a dissipative Brownian motion \cite{hanggi20051}. As similarly occurs for the free energy, we shall also see in Sec.\ref{LTSDS} that the internal energy deviation shows an initial growth with temperature for a non-Markovian dissipative dynamics characterized by the Lorentz-type spectral density. Finally, observe that the logarithmic divergences of both internal and free energies are identical, i.e. compared the first-term in the right-hand side of (\ref{IEOLT}) with the third contribution of the free energy deviation (\ref{FEOLT}), and they will compensate in the entropy definition.

\subsubsection{Entropy} \label{SEntropy}
Let now address the change of the thermodynamic entropy caused by the Chern-Simons action. Having determined the deviation of the the free energy (\ref{FEO}) and internal energy (\ref{IEO}) of the flux-carrying particle, the entropy variation can be now defined as usually,
\begin{eqnarray}
\Delta S(\beta)&=&S_{flux}(\beta)-2S_{0}(\beta) \label{TEE} \\
&=&k_{B}\beta\left(\Delta U(\beta)-\Delta F(\beta) \right), \nonumber
\end{eqnarray}
where $S_{flux}(\beta)$ and $S_{0}(\beta)$ denote, respectively, the thermodynamic entropy of the flux-carrying particle and the entropy of the damped (one-dimensional) harmonic oscillator. For future discussions it is also convenient to introduce the (reversible) heat delivered by the flux-carrying particle at a constant inverse temperature $\beta$ after switching on quasistatically the Chern-Simons action \cite{hilt20111,kardar20071},
\begin{equation}
\Delta Q(\beta)=\Delta F(\beta)-\Delta U(\beta).
\label{HLT}
\end{equation}
It is important to notice that $\Delta Q(\beta)$ in a reversible isothermal process represents the minimum heat exchanged between the system and the Maxwell-Chern-Simons environment \cite{callen19871}.

Plugging the equations (\ref{FEOHT}) and (\ref{IEOHT}) in (\ref{TEE}), we obtain a closed-form expression for the entropy change in the high-temperature limit,
\begin{equation}
\Delta S(\beta)=-k_{B}\log\left(1+ \frac{\phi}{\omega_{0}^2}\right)  +\frac{(2\gamma_{0}^{2}-2\Gamma_{0}^{2}-\omega_{0}^2+\Omega_{0}^2)\hbar^2k_{B}\beta^{2}}{12} +\mathcal{O}\left(\beta^{3} \right) ,\label{ESHT}
\end{equation}
where the influence of the first term in the right-hand side owing to the potential renormalization is relatively small according to the subsidiary condition (\ref{CRPS}). Interestingly, by ignoring the latter, the deviation of the entropy from the conventional Brownian motion rapidly vanishes for growing temperatures (i.e. $\beta \rightarrow 0$). Gathering up this result with the previous observation from the free and internal energies, our analysis apparently indicates that in the strict Markovian dissipative scenario and high-temperature limit the flux-carrying particle largely shares the nonequilibrium thermodynamics properties of the conventional damped harmonic oscillator, so that we essentially recover the results from the classical statistical mechanics.

Once again, by substituting Eqs. (\ref{FEOLT}) and (\ref{IEOLT}) in (\ref{TEE}), we compute the entropy deviation in the low temperature limit and for weak Chern-Simons effects
\begin{eqnarray}
\Delta S(\beta)&= &-k_{B}\log\left(1+ \frac{\phi}{\omega_{0}^2}\right) -\frac{2\hbar k_{B}(\gamma_{0}-\Gamma_{0})\beta}{\pi} +\frac{k_{B}}{2}\log\left(\frac{\Omega_{0}^{4}}{\omega_{0}^{4}}-\frac{\Omega_{CS}^{4}}{\omega_{0}^4} \right) \nonumber \\
&-&\frac{4\pi k_{B} (\Gamma_{0}\omega_{0}^2\Omega_{0}^2-\gamma_{0}(\Omega_{0}^4-\Omega_{CS}^4))\beta^{-1}}{3\hbar\omega_{0}^{2}(\Omega_{0}^{4}-\Omega_{CS}^{4})} +\mathcal{O}\left( \beta^{-3}\right),  \label{ESLT} 
\end{eqnarray}
which is manifestly divergent at absolute zero temperature as similarly occurs for the free and internal energies in the strict Markovian dynamics. Now one may clearly see from (\ref{ESLT}) that the longitudinal Chern-Simons dynamical susceptibility is the main responsible for this divergence by noting that the latter disappears owing to $\gamma_{0}=\Gamma_{0}$ when $\tilde{\Lambda}_{\parallel}(s)$ is disregarded (see the Eq.(\ref{GOCS0})).

In spite of such anomalous feature, the behavior of the entropy is in conformity with the Second law of thermodynamics. From the Eqs. (\ref{FEOLT}), (\ref{IEOLT}) and (\ref{HLT}) follow that the heat exchanged with the Maxwell-Chern-Simons field environment at absolute zero temperature is negative, i.e. $\Delta Q(\beta\rightarrow \infty)\rightarrow -4\hbar(\gamma_{0}-\Gamma_{0})/\pi$ (see the Eq.(\ref{GOCS0})). This result is in complete agreement with the second law of thermodynamics formulated in terms of Clausius's inequality (i.e. $\Delta Q\leq T\Delta S$): no heat can be taken from the bath at absolute zero temperature \cite{nieuwenhuizen20021}. Furthermore, it is important to realize that from the Eqs. (\ref{AFRR}) and (\ref{HLT}) one may see that the heat intake of the dissipative particle from the Maxwell-Chern-Simons field environment is comparatively larger than the delivered environmental work outlined in Sec.\ref{Sfree} (i.e. $0\leq -\Delta Q$ in the low temperature regime). This implies that the entropy increases with temperature which is characteristic of a conventional heat bath in standard thermodynamics \cite{callen19871}. On the other side, it shall be shown in the next section (see the Eq.(\ref{HCLTE})) that the heat capacity vanishes as temperature goes to zero in complete agreement with the Third Law of thermodynamics \cite{kumar20091,bandyopadhyay20102,bandyopadhyay20091,hanggi20061} (though it diverges in the absolute-zero temperature limit). This can be seen from (\ref{ESLT}) by using the Maxwell's thermodynamic relation,
\begin{equation}
C_{flux}(\beta)=\beta^{-1}\frac{\partial S_{flux}(\beta)}{\partial \beta^{-1}}. \nonumber
\end{equation}
Since $S_{flux}(\beta)$ decays linearly with temperature as follows from (\ref{ESLT}), the contribution of the Chern-Simons action to the heat capacity will vanish linearly as well.

Intriguingly, by disregarding the discussed infrared divergence and the potential renormalization $\phi$, the equation (\ref{ESLT}) reveals that the entropy of the flux-carrying particle is dominated for a non-vanishing constant value at low temperatures (see the third term in the right-hand side). The latter represents a positive contribution to the entropy up to next leading order of the small parameter $\kappa/\omega_{0}\ll 1$ as follows from the expression (\ref{AFRR}), and completely coincides with the observation in the dissipative non-Markovian scenario as well (see the discussion around the figure \ref{Fig2} in Sec.\ref{LTSDS}). This feature recalls the result obtained from degenerate systems \cite{kardar20071}, and it suggests that the flux-carrying particle may have a non-trivial ground state different from the damped harmonic oscillator, which must be somehow related to the fact that the coupling of the dissipative oscillator with the Chern-Simons action may effectively alter its statistics \cite{matsuyama19901}. This point will deserve further attention in a treatment of the statistical mechanics in a dissipative scenario free of the mentioned infrared singularity.  

\subsubsection{Heat capacity}
Finally, we present the result related to the heat capacity of the flux-carrying particle. Given the internal energy (\ref{IEO}), one may readily obtain this by using the standard thermodynamics relations \cite{kumar20141,weiss20121,hanggi20081}, i.e.,
\begin{eqnarray}
C_{flux}(\beta)&=&k_{B}\frac{\partial U_{flux}(\beta)}{\partial\beta^{-1}} \nonumber \\
&=& 2C_{0}(\beta) +\frac{\hbar^2 k_{B}\beta^2}{(2\pi)^2}\Bigg(\sum_{i=1,2}R_{i}^2\psi'\left(1+\frac{ R_{i}\hbar \beta}{2\pi}\right) +R_{i}'^2 \psi' \left(1+\frac{ R_{i}'\hbar \beta}{2\pi}\right)\nonumber \\
&-&2\sum_{i=1,2}r_{i}^2\psi'\left(1+\frac{ r_{i}\hbar \beta}{2\pi}\right) \Bigg), \label{HCE}
\end{eqnarray}
where $C_{0}(\beta)$ stands for the heat capacity of the usual damped (one-dimensional) harmonic oscillator in the Markovian dynamics regime, and we have defined $\psi'(z)=\frac{d\psi(z)}{dz}$. We next study the behavior of (\ref{HCE}) in the interesting asymptotic limits. Making use of the expression $\psi'(z)\simeq 1/z+1/2z^2+1/6z^3 -1/30z^5 $ valid for large arguments $z$, we find that the low-temperature heat capacity up to the next leading order in the inverse temperature expresses as follows
\begin{equation}
C_{flux}(\beta)=2C_{0}(\beta)+\frac{2\hbar k_{B}(\gamma_{0}-\Gamma_{0})\beta}{\pi}-\frac{4\pi k_{B} (\Gamma_{0}\omega_{0}^2\Omega_{0}^2-\gamma_{0}(\Omega_{0}^4-\Omega_{CS}^4))\beta^{-1}}{3\hbar\omega_{0}^{2}(\Omega_{0}^{4}-\Omega_{CS}^{4})} +\mathcal{O}\left( \beta^{-3}\right),\label{HCLTE}
\end{equation}
where $C_{0}(\beta)$ is expected to decay linearly with temperature \cite{bandyopadhyay20091,bandyopadhyay20102,kumar20141,hanggi20061}. As similarly discussed for the entropy, the absolute-zero temperature divergence of the heat capacity can be essentially attributed to the infrared singularity exhibited by the longitudinal Chern-Simons dynamical susceptibility in the strict Markovian dynamics. 

Interestingly, the above expression (\ref{HCLTE}) shows that the heat capacity of the flux-carrying Brownian particle vanishes as the temperature approaches its absolute zero value if we disregard the aforementioned anomalous behavior. In other words, at least for weak Chern-Simons constant values (i.e. $\kappa/\omega_{0}\ll 1$), the Eqs. (\ref{ESLT}) and (\ref{HCLTE}) suggest that the proposed microscopic description (\ref{HMCSF2}) is able to reproduce the Third Law of thermodynamics up to a renormalization factor.

In the opposite limit of high-temperature, the first derivative of the digamma function in (\ref{HCE}) approaches to $\psi'(z)\simeq 1/z^2+\pi^2/6-z+\pi^4 z^2/30$, and thus, the heat capacity takes the form
\begin{equation}
\Delta C(\beta)=-\frac{(2\gamma_{0}^{2}-2\Gamma_{0}^{2}-\omega_{0}^2+\Omega_{0}^2)\hbar^2k_{B}\beta^{2}}{6} +\mathcal{O}\left( \beta^{3}\right).
\label{HCHTE}
\end{equation}
Recalling the results from the previous sections, the Eq.(\ref{HCHTE}) manifests once again that the nonequilibrium thermodynamics properties of the flux-carrying particle largely overlap with those of the damped two-dimensional harmonic oscillator at sufficient high temperatures.

To recap, the deviation of the flux-carrying particle thermodynamic properties from the conventional Brownian motion significantly change from the high- to the low- temperature regime. It turns out that in the former limit the dominant behavior of the deviation of the free energy, internal energy, entropy and heat capacity is to decay as an inverse power of the temperature. This eventually makes the Chern-Simons influence be negligible in the classical limit, and as a consequence, the thermodynamics properties of the flux-carrying particle closely approach to those from the damped two-dimensional harmonic oscillator. This is consistent with the observation that the Chern-Simons effects are significantly damped in the high-temperature regime owning to the fact that the Chern-Simons dynamical susceptibilities display $1/f$-type power spectral density (see the Eqs. (\ref{SDCII}) and (\ref{SDCIII})). By contrast, the flux-carrying particle in the quantum domain exhibits a quite different behavior to the conventional Brownian motion in presence of an external magnetic field \cite{ kumar20091,bandyopadhyay20061,bandyopadhyay20091,bandyopadhyay20062,hong19911,li19901}: the free energy of the flux-carrying particle has an additional linear dependence with temperature that leads the entropy to take a finite constant value in the low-temperature regime. The entropy also presents an anomalous divergence which is rooted in an infrared singularity characteristic of the longitudinal Chern-Simons dynamics susceptibility in the strict Markovian scenario. Disregarding this infrared divergence in both the entropy and the heat capacity, the nonequilibrium thermodynamics quantities of the flux-carrying particle seem to reproduce the results expected from the Third law of thermodynamics. In parallel, it was discussed that the influence of the potential renormalization in (\ref{HCPR}) just regards a rescaling of the flux-carrying particle partition function because the subsidiary condition (\ref{CRPS}).

\subsection{Lorentzian-type spectral density}\label{LTSDS}
The foregoing analysis holds when the system-environment coupling strength is small compared to the strength of the harmonic confining potential, and thus, the environmental Chern-Simons effects are sufficiently weak (according to the subsidiary conditions (\ref{SSCI})). Let us turn the attention to a more general dissipative scenario which presumably brings about non-Markovian dissipative effects, so that the system-environment coupling is expected to substantially influence the nonequilibrium thermodynamic properties. Basically, we find that the nonequilibrium thermodynamic quantities of the flux-carrying particle in the studied non-Markovian scenario widely agree with the discussion in Secs. \ref{SRPF} and \ref{OHSDS}, and importantly, the entropy and heat capacity seem to reproduce the results in conformity with the Third Law of thermodynamics for a strong system-environment coupling as well (up to a renormalization factor).

\begin{figure}[]
\begin{center}
\includegraphics[scale=0.29]{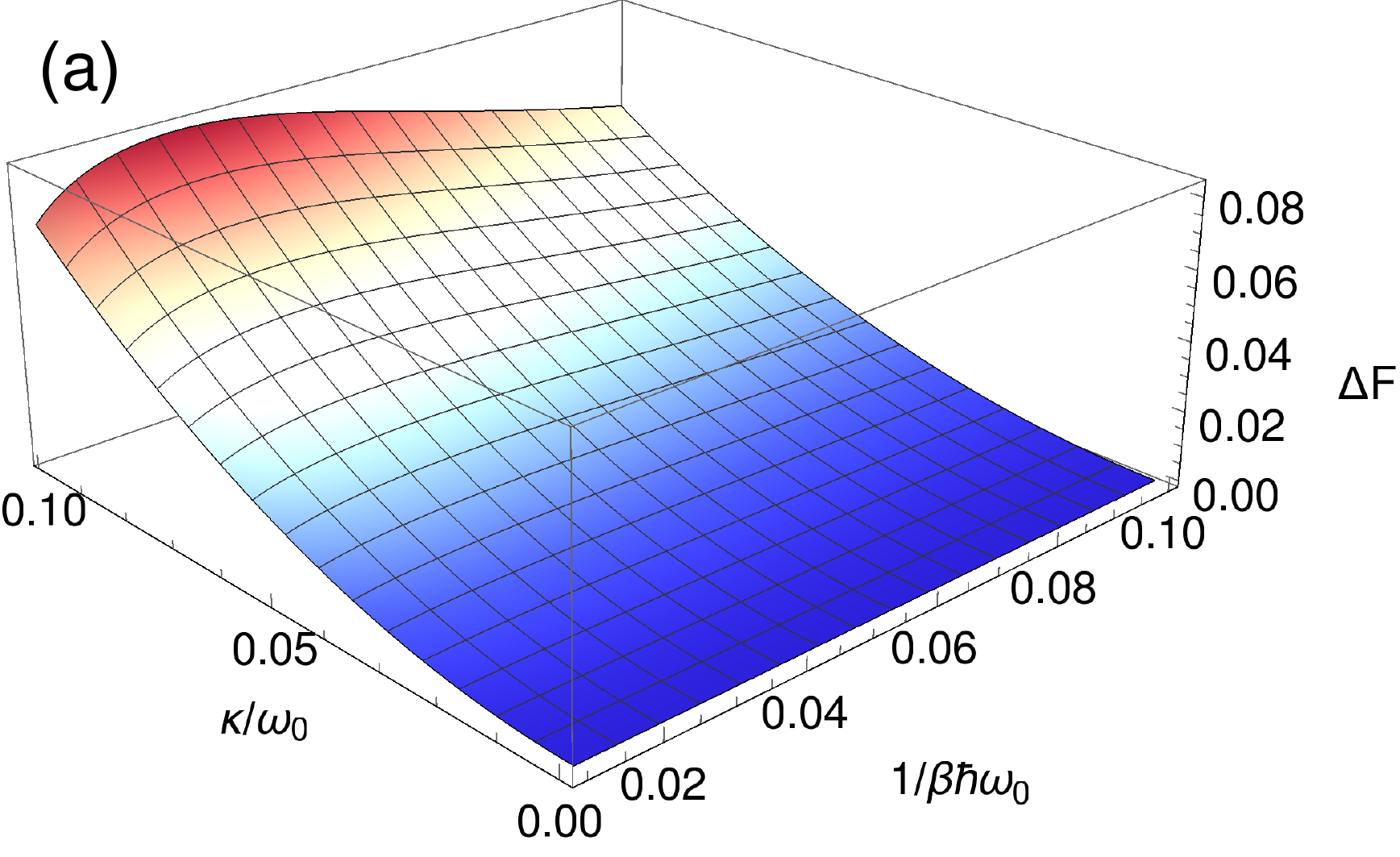}
\includegraphics[scale=0.29]{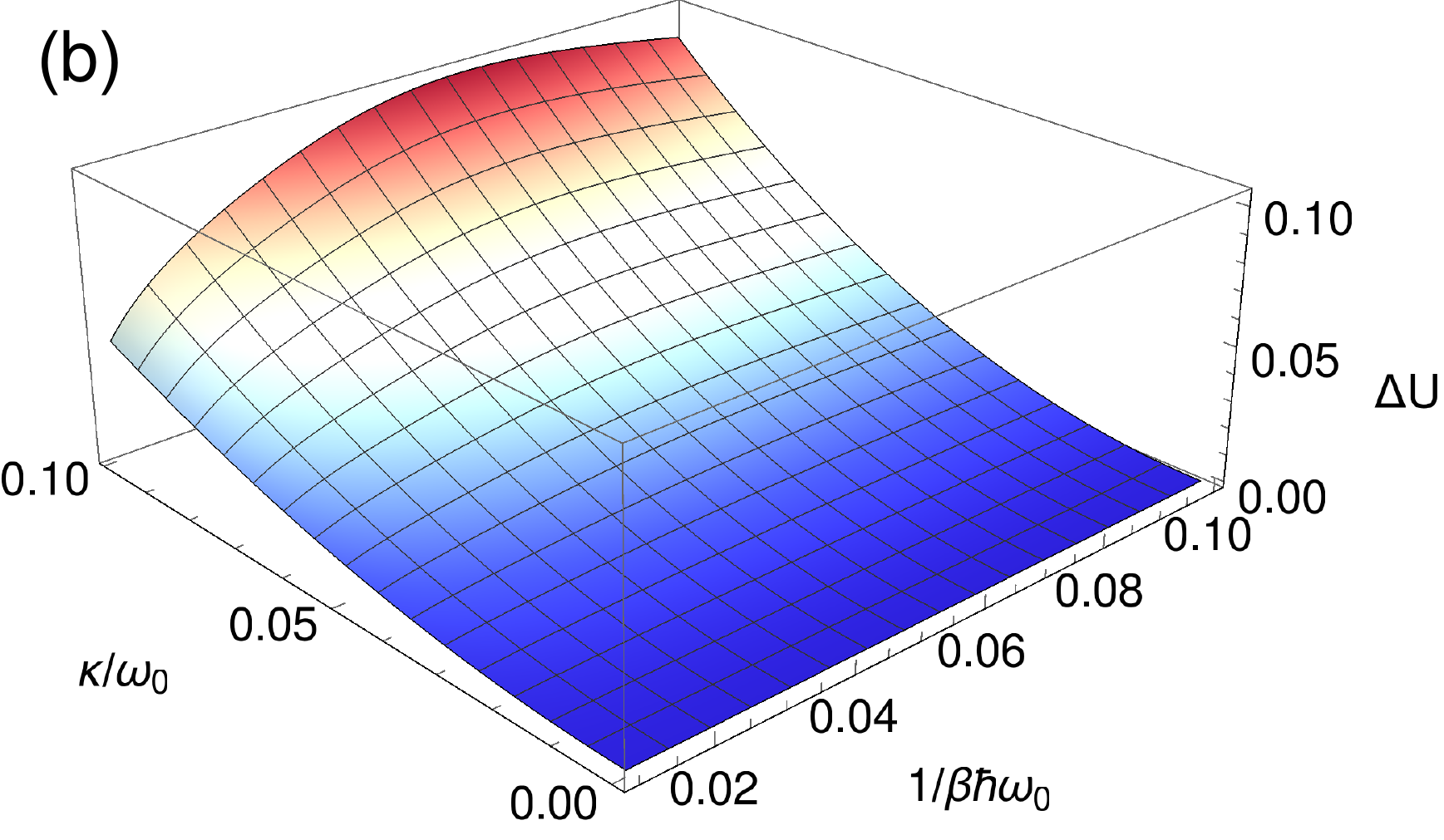}
\includegraphics[scale=0.29]{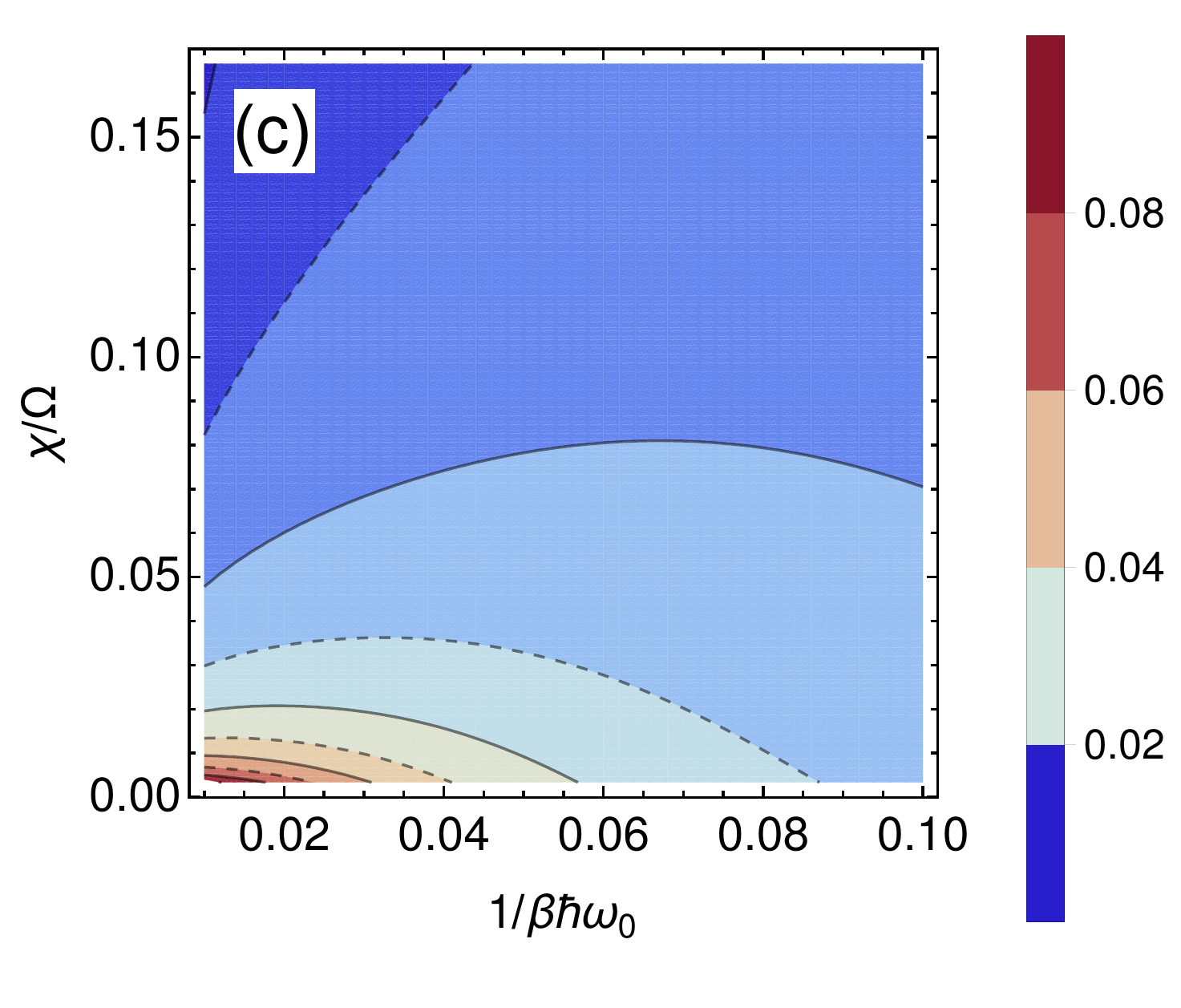}
\end{center}
\caption{(color online). (a and b) Three-dimensional plots of the deviation of the free and internal energies in the strong system-environment coupling regime as a function of the Chern-Simons constant $\kappa$ and the inverse temperature $\beta^{-1}$. We have fixed the problem parameters as $\omega_{0}=10m=10$ and $\chi/\Omega=1/30$ and $\gamma_{0}/\omega_{0}=1/2$, in the natural units $\hbar=k_{B}=1$. (c) Contour plot of the deviation of the free energy as a function of the spectral-density parameter $\chi$ and the inverse temperature $\beta^{-1}$. The parameters were fixed as to the upper and cental plots, and the Chern-Simons action strength was fixed at $\kappa/\omega_{0}=1/20$.\label{Fig1}} 
\end{figure}

Going beyond the weak system-environment coupling limit, we address the intricate dissipative dynamics dictated by the Lorentzian-type spectral density \cite{bandyopadhyay20102,kumar20141}
\begin{equation}
J_{lorentz}(\omega)= \frac{2m\gamma_{0}\Omega^{4}\omega}{\chi^2\omega^2+(\Omega^2-\omega^2)^2},
\label{SPDII}
\end{equation}
where $\gamma_{0}$ is the friction coefficient as before, $\Omega$ and $\chi$ are the frequency and damping parameters of a thermal harmonic noise. The spectral density (\ref{SPDII}) has been extensively used to describe systems driven by a colored noise having a Lorentzian power spectrum \cite{dykman19911}. In particular, it was employed to address the nonequilibrium thermodynamics of the damped two-dimensional harmonic oscillator in presence of an external uniform magnetic field \cite{kumar20141,bandyopadhyay20102}. Replacing this in Eq.(\ref{SDCI}) and using the standard tables of integration, one may see that the conventional dynamical susceptibility takes the form \cite{kumar20141}
\begin{equation}
\tilde{\Delta}(s)= \frac{2\gamma_{0}\Omega^2}{\chi}\frac{s(s+\chi)}{s^2+s\chi+\Omega^2}\label{DKLSD}, 
\end{equation}
whereas the Chern-Simons dynamical susceptibilities are now given by, 
\begin{eqnarray}
\tilde{\Lambda}_{||}(s)&=&\frac{2\gamma_{0}\Omega^2\kappa^2}{\chi} \frac{s+\chi}{s(s^2+s\chi+\Omega^2)}, \label{LCHLTSP} \\
\tilde{\Lambda}_{\perp}(s)&=&\frac{\gamma_{0}\Omega^2\kappa}{\chi} \frac{s+\chi}{s^2+s\chi+\Omega^2}. \label{LCHLTSPI}
\end{eqnarray}
Notably, one may readily see from (\ref{LCHLTSP}) and (\ref{LCHLTSPI}) that the environmental Chern-Simons effects in the low temperature regime ($s\ll 1$) are powered for the case of structured environments $\chi/\Omega\ll 1$ \cite{kumar20141}: the dominant behavior of the longitudinal Chern-Simons dynamical susceptibility goes like $\sim 1/s$.

\begin{figure*}[]
\begin{center}
\includegraphics[scale=0.28]{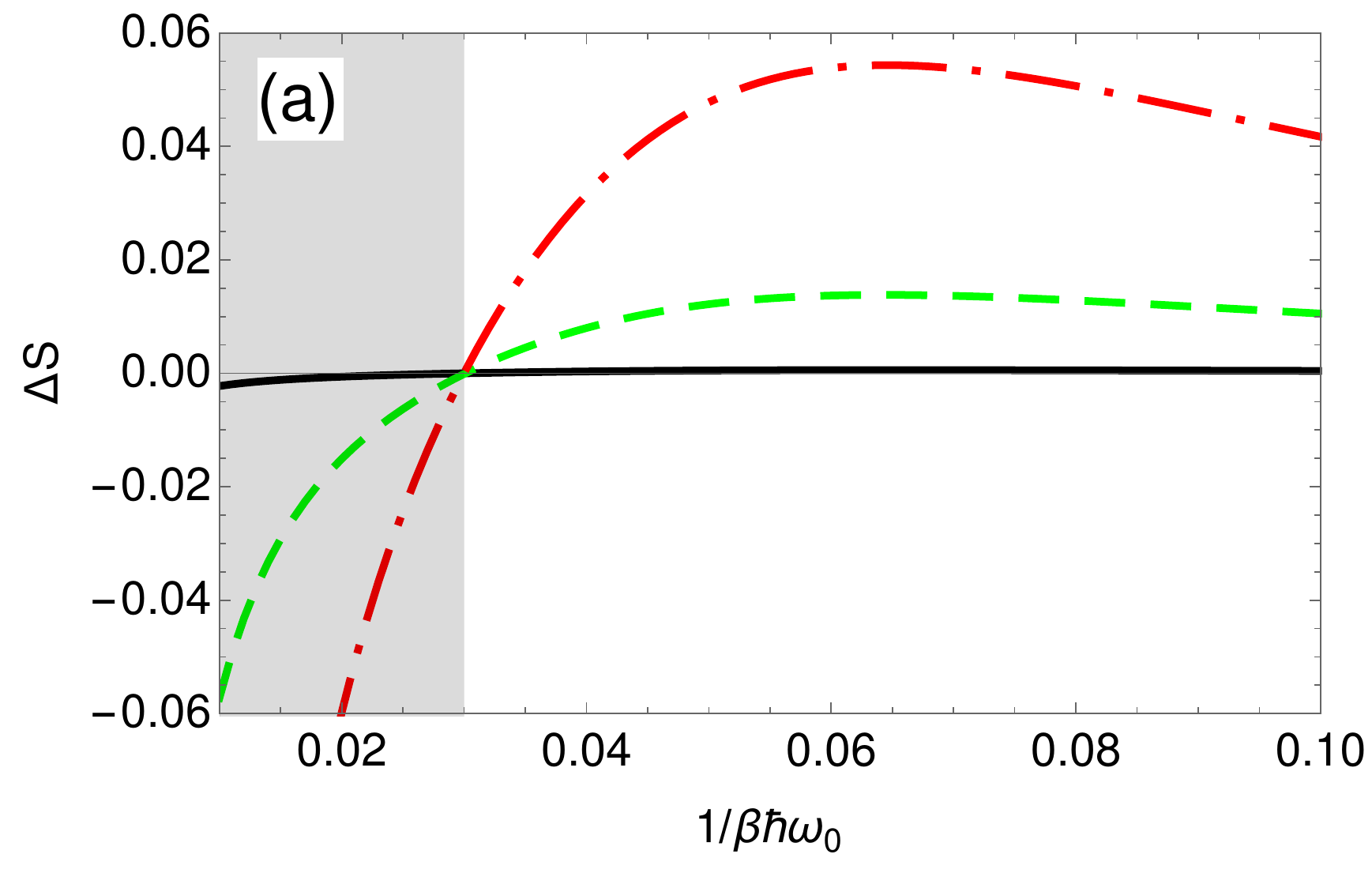}
\includegraphics[scale=0.28]{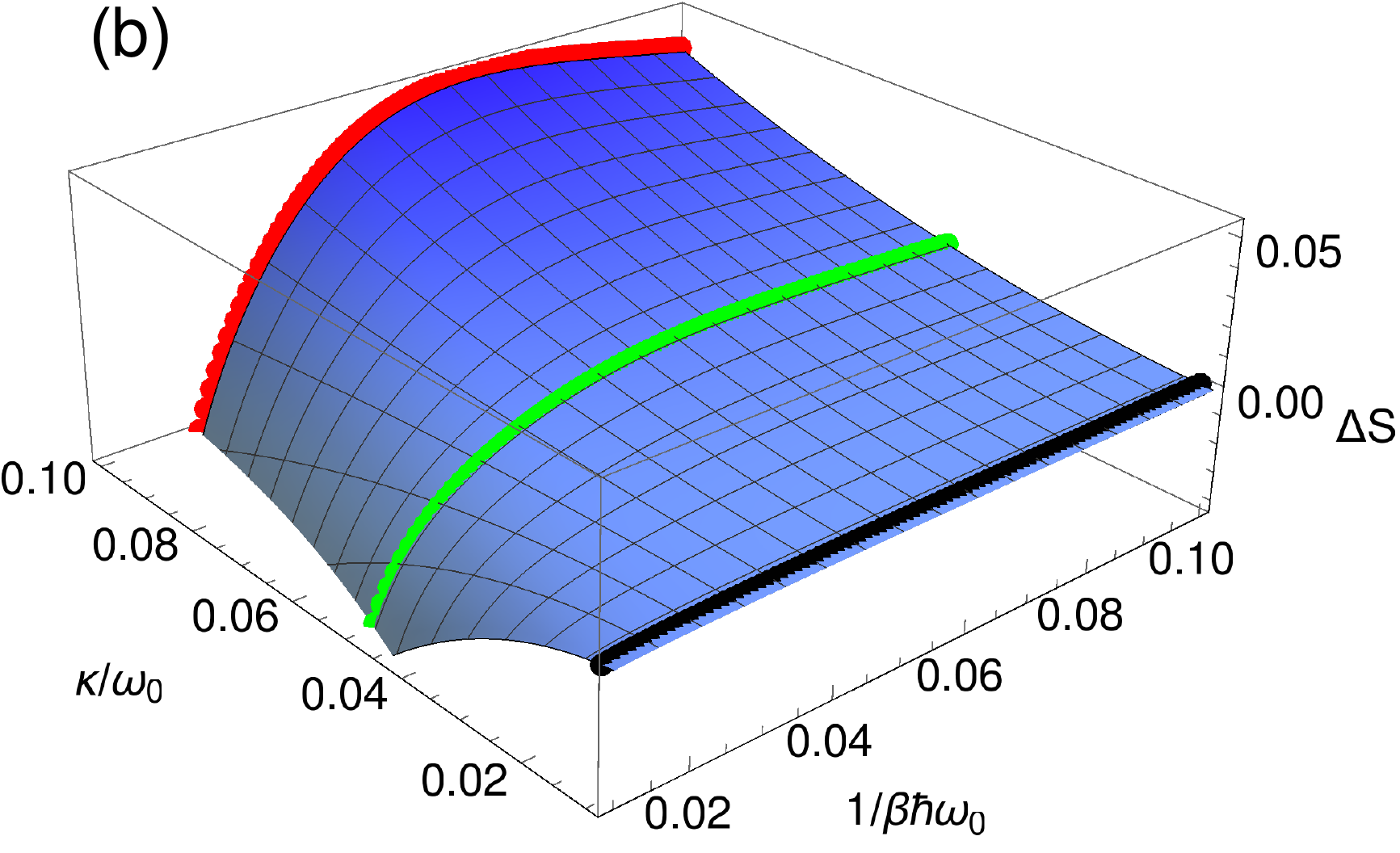}
\includegraphics[scale=0.28]{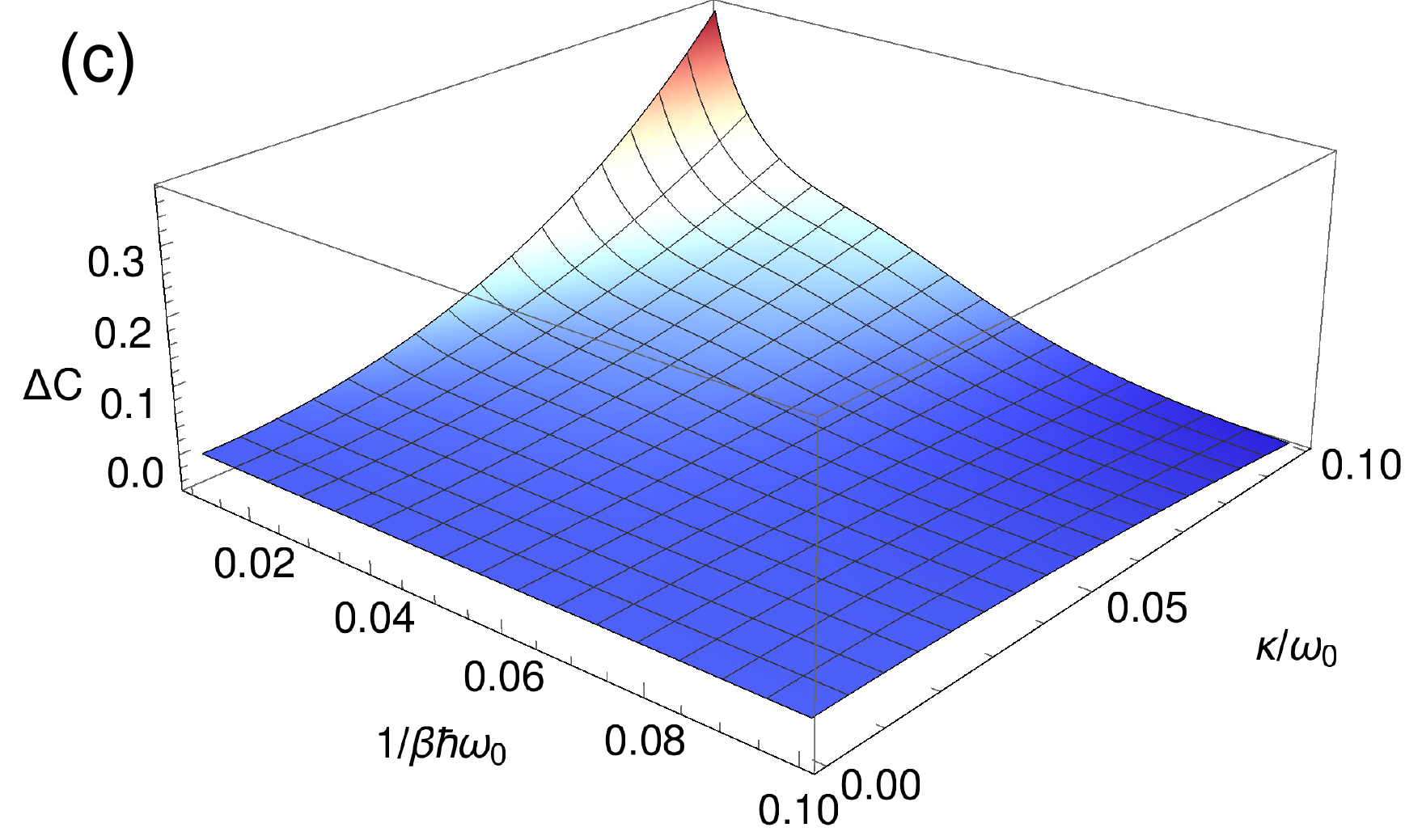}
\end{center}
\caption{(color online). (a) The entropy deviation as a function of the inverse temperature $\beta^{-1}$ for fixed values of the Chern-Simons constant. The black solid, green dashed, and red dashed-dotted lines represent the entropy deviation for $\kappa/\omega_{0}=0.01$, $\kappa/\omega_{0}=0.05$, $\kappa/\omega_{0}=0.1$, respectively. The shadow area corresponds to spurious results due to the infrared divergence. (b) Three-dimensional plot of the entropy deviation as a function of the Chern-Simons action strength $\kappa$ and the inverse temperature $\beta^{-1}$. The lines pictured in the left panel have been represented here as well. (c) Similarly, three-dimensional plot of the heat capacity deviation  as a function of the Chern-Simons constant $\kappa$ and the inverse temperature $\beta^{-1}$. In all the plots, we have fixed the rest of parameters as $\omega_{0}=10m=10$, and $\chi/\Omega=1/30$ and $\gamma_{0}/\omega_{0}=1/2$, in the natural units $\hbar=k_{B}=1$.\label{Fig2}} 
\end{figure*}

The price to pay for considering the spectral choice (\ref{SPDII}) is that the longitudinal Chern-Simons susceptibility (\ref{LCHLTSP}) presents an infrared singularity similarly as occurs for the Markovian case, and as a consequence, the nonequilibrium thermodynamic quantities of the flux-carrying particle will diverge at the absolute zero temperature as well. This can be further understood by noting that the denominator of the general algebraic expression of the partition function (\ref{PFSII}) becomes infinite since $\tilde{\Lambda}_{||}(s)$ diverges, whereas the numerator remains finite when we take the absolute zero temperature limit: that is, $Z_{flux}\rightarrow 0$ for $\beta\rightarrow \infty$. In a more profound sense, the strict Ohmic and Lorentzian-type dissipative spectral densities show this anomalous behavior mainly due to they do not formally take account the dimensionality ($d=2$) of the Maxwell-Chern-Simons environment, which roughly characterizes the power of the spectral density at low energies \cite{valido20132}, i.e. $J(\omega)\sim \omega^{d}$. Although these spectral densities may represent a broad class of quantum dissipative systems, they do not regard a fully physical description for two-dimensional systems in the low-temperature limit. Furthermore, it is important to notice that to be the choice (\ref{SPDII}) consistent with the relation dispersion (\ref{ESPG}) of the environmental excitations, we must demand the following subsidiary condition is fulfilled,
\begin{equation}
\kappa\ll \Omega,
\label{LTSDCI}
\end{equation}
which guarantees nothing else but the interesting system will mainly interact with the environmental modes that are well contained in the spectrum of the Maxwell-Chern-Simons environment. Note that this condition does not prevent us of taking large values for the parameter $\gamma_{0}/\omega_{0}\sim 1$ which roughly characterizes a strong system-environment interaction.

\begin{figure}[]
\begin{center}
\includegraphics[width=0.5\columnwidth]{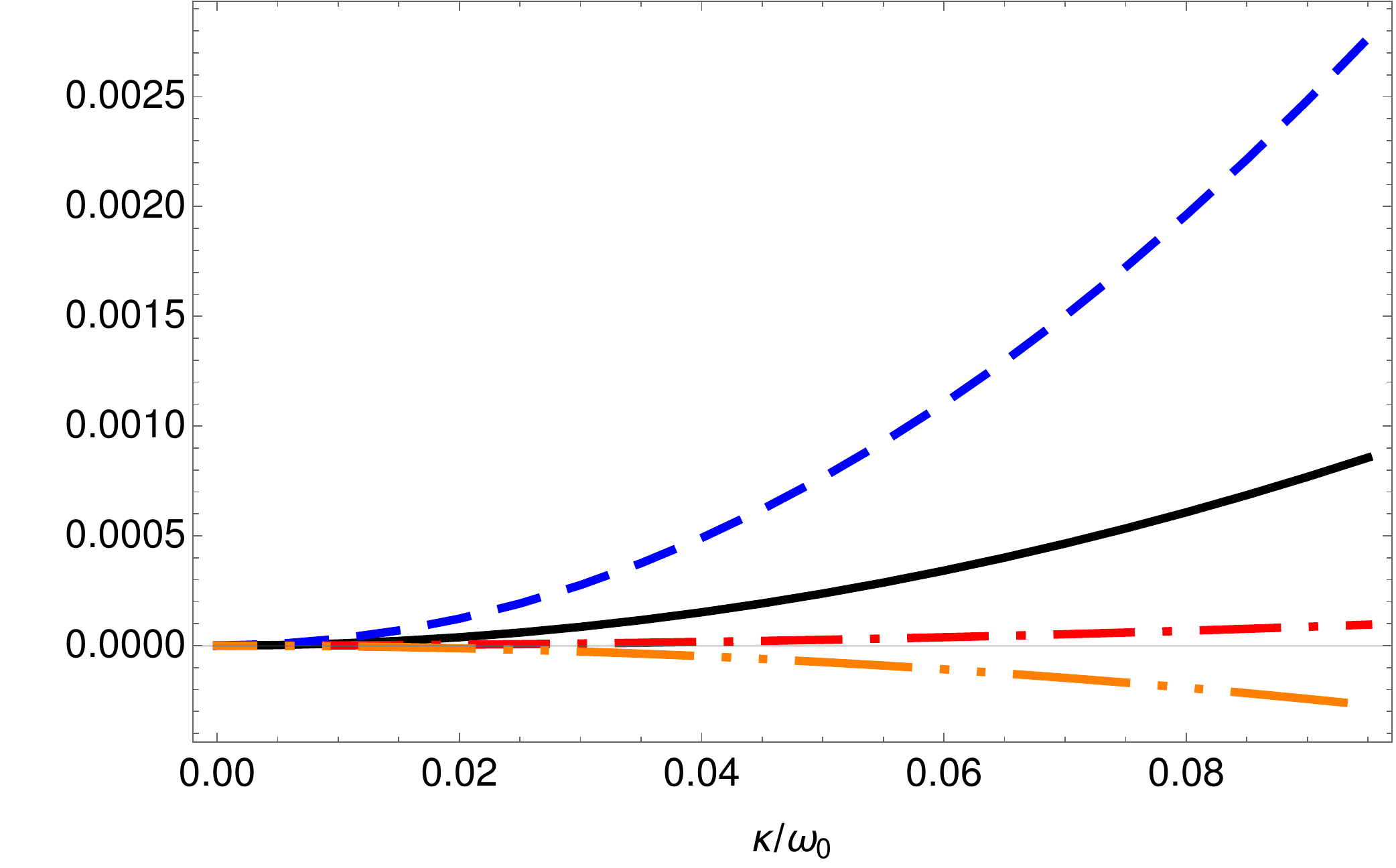}
\end{center}
\caption{(color online). Nonequilibrium thermodynamic quantities in the high temperature regime as a function of the Chern-Simons action strength $\kappa$ for a fixed temperature $\beta^{-1}=2\hbar\omega_{0}$. The solid black, dashed blue, dot-dashed red, and double dot-dashed orange depict the deviations $\Delta F$, $\Delta U$, $\Delta S$ and $\Delta C$, respectively. The rest of parameters were fixed as to the Fig.\ref{Fig1}. \label{Fig3}} 
\end{figure}

Similarly to the (strict) Ohmic spectral density, the reduced partition function (\ref{PFSII}) becomes a rational function once substituted the environmental kernels (\ref{DKLSD}), (\ref{LCHLTSPI}) and (\ref{LCHLTSPI}), so this can be calculated by following an identical procedure as to compute the expression (\ref{PFSO}) in the Sec.\ref{OHSDS} (see appendix \ref{app2} for further details). The partition function of the flux-carrying Brownian particle now reads,
\begin{equation}
Z_{flux}(\beta)=Z_{0}^{2}(\beta)  \left(1+ \frac{\phi}{\omega_{0}^2}\right)^{-1} \frac{\prod_{i=1}^{5}\Gamma\left(1+\frac{R_{i}\hbar \beta}{2\pi} \right) \Gamma\left(1+\frac{R_{i}'\hbar \beta}{2\pi} \right)}{\prod_{i=1}^{2}\Big[\Gamma\left(1+\frac{r_{i}\hbar \beta}{2\pi} \right) \Gamma\left(1+\frac{r_{i}'\hbar \beta}{2\pi} \right)  \Big]^2},\label{PFSL} 
\end{equation}
where $Z_{0}(\beta)$ is the partition function of the damped one-dimensional harmonic oscillator (a closed-form expression for $Z_{0}$ can be found in \cite{kumar20141}), and $\left\lbrace r_{i}, r'_{i}\right\rbrace $ as well as $\left\lbrace R_{i},R_{i}'\right\rbrace $ are the roots of the following polynomials, respectively,
\begin{eqnarray}
P(r)&=& (r^2+\omega_{0}^2)(r^2+\chi r+\Omega^2)+\frac{2\gamma_{0}\Omega^2}{\chi}r(r+\chi), \nonumber \\
Q(R)&=&\left( RP(R)+\frac{2\kappa^2\gamma_{0}\Omega^2}{\chi}(R+\chi)\right) ^{2}-\frac{\kappa^2 \gamma_{0}^2\Omega^4}{\chi^2}R^2(R+\chi)^2.\label{RII}
\end{eqnarray}
As expected the Eq.(\ref{PFSL}) returns the result of the independent-oscillator model for a vanishing Chern-Simons action (i.e. $Q(R)\rightarrow (r P(r))^2$ when $\kappa\rightarrow 0$). Owning to the partition function (\ref{PFSL}) closely resemblances the expression (\ref{PFSO}) for the Ohmic case, it is readily to see that the deviation of the nonequilibrium thermodynamics quantities $\Delta F(\beta)$, $\Delta U(\beta)$, $\Delta S(\beta)$, and $\Delta C(\beta)$ for the Lorentzian-type spectral scenario can be cast in an identical form as shown in Eqs. (\ref{FEO}), (\ref{IEO}), (\ref{TEE}) and (\ref{HCE}) respectively, but rather in terms of the roots of the polynomials (\ref{RII}). Instead of writing down the full expression, we illustrate in the figures (\ref{Fig1}), (\ref{Fig2}) and (\ref{Fig3}) the results obtained for their numerical computation. For seek of simplicity, the influence of the potential renormalization shall be neglected throughout the following discussion.

In figure (\ref{Fig1}) we may observe a similar behavior of $\Delta F(\beta)$ and $\Delta U(\beta)$ (see upper and central panel) as predicted by the Eqs. (\ref{FEOLT}) and (\ref{IEOLT}) in the Markovian case at low temperatures. First, it is clear that the free and internal energies (slightly) increase for rising Chern-Simons action strength which is in accordance with the fact that certain work must be done upon the dissipative harmonic oscillator in order to establish the environmental emf that eventually leads to the formation of the flux-carrying Brownian particle: there is not spontaneous creation of the flux-carrying particle understood in the thermodynamics sense (i.e. a non-spontaneous process implies an increase of the free energy). Furthermore, we find out that this effect intensifies when the environment becomes structured. In particular, the figure (\ref{Fig1}) illustrates a contourplot of the free energy deviation as a function of the temperature and the characteristic parameter $\chi/\Omega$. One may readily see from this plot that for a given constant temperature the $\Delta F$ may reach larger values as $\chi/\Omega$ becomes smaller, which is in complement agreement with our previous observation about the Eqs. (\ref{LCHLTSP}) and (\ref{LCHLTSPI}).

By fixing $\kappa$ to a constant value, we may also appreciate that the free energy follows a linear decayment for higher temperatures than $1/\beta\hbar\omega_{0}\approx0.04$, which is better illustrated below in the plot (\ref{Fig2}) for the entropy (see the green dashed and red dashed-dotted lines in the right and central panels). The latter figure reveals that the entropy roughly saturates to a constant value before diverging at absolute-zero temperature (see shadow area in the right panel of figure (\ref{Fig2})), which according to the Maxwell's thermodynamic relation (\ref{MTR}) means that the dominant behavior of the free energy of the flux-carrying particle is rougly linear at low temperatures. Interestingly, this linear feature displayed by the free energy has no counterpart in the charged magneto-oscillator coupled to a conventional heat bath \cite{hong19911,kumar20091,bandyopadhyay20061,bandyopadhyay20091,bandyopadhyay20062,li19901,dattagupta19971}, and is intimately related to the mechanism by which the environmental Hall response deliver some energy that transforms into mechanical work added to the flux-carrying particle, which subsequently supports our previous observation about that the Maxwell-Chern-Simons field environment may behave as a work source. At this point, it is important to realize that the dissipation mechanism dominates the open-system dynamics against the environmental Hall effects according to the subsidiary condition (\ref{SCDI}), and thus, the heat delivered from the environment to the dissipative particle will be larger in comparison to such work. Additionally, we may also see from the figure (\ref{Fig1}) that the free and internal energies at higher temperatures show an algebraic decayment instead, which is in agreement with the power dependence with the inverse temperature found previously in the strict Markovian case.

The figure (\ref{Fig2}) similarly depict the deviation of the entropy and the heat capacity for a strong system-environment coupling. A quick glance reveals that both thermodynamics quantities diverge for decreasing temperatures and higher Chern-Simons constant, e.g. the entropy abruptly drops below $1/\beta\hbar\omega_{0}\approx0.04$ (see the shadow area in the left panel). Although it is not appreciated from figure (\ref{Fig1}), this also occurs for the free and internal energies. Let us emphasize once again that such divergence is due to we are treating a dissipative non-Markovian model (i.e. Lorentzian spectral density) for which the longitudinal Chern-Simons dynamical susceptibility contains an infrared divergence (see the Eq.(\ref{LCHLTSP})). 

As mentioned above, from the right and central panels in figure (\ref{Fig2}), we may observe in the low energy regime that the entropy of the flux-carrying particle remains almost constant in temperature for different values of the Chern-Simons action strength (see the green dashed and red dashed-dotted lines) which is a hallmark of a linear behavior in the free energy in complete agreement with the previous discussion in Secs.\ref{Sfree} and \ref{SEntropy}. This is better illustrated in the central panel, where the entropy deviation is plot as a function of the temperature and the Chern-Simons constant. Discarding the aforementioned anomalous behavior, these figures additionally suggest that the entropy reproduces the Third Law of thermodynamics: it takes a finite constant value as we approach the zero temperature limit \cite{kardar20071}. On the other side,  the heat capacity, which is depicted in the left panel of figure (\ref{Fig2}), clearly displays a similar behavior to the entropy: it vanishes even for large values of the system-environment coupling (i.e. $\gamma_{0}/\omega_{0}\sim 1$) in the low-temperature regime, though diverges at absolute zero temperature (as previously anticipated). That is, this plot reveals that the heat capacity for the studied non-Makovian dissipative scenario cancels at low energy in conformity with the Third Law of Thermodynamics. Accordingly, this observation together with the behavior exhibited by the entropy seems to indicate that the dissipative Maxwell-Chern-Simons model (\ref{HMCSF2}) retrieves results which are widely in conformity with the Third Law of thermodynamics in the studied non-Markovian dissipative scenario as well.

Finally, the figure (\ref{Fig3}) illustrates the behavior of the nonequilibrium thermodynamics quantities as a function of the Chern-Simons constant in the high-temperature regime. Interestingly, we may observe for a fixed temperature ($\gamma_{0}\beta\ll 1$) that the free and internal energies, as well as the entropy, slightly grow with the Chern-Simons action strength in agreement with the results found for the strict Ohmic spectral density (see the Eqs. (\ref{FEOHT}), (\ref{IEOHT}), and (\ref{ESHT})), as opposite to the heat capacity which decreases (see the Eq.(\ref{HCHTE})). Although it is not shown here, we also observe that all the nonequilibrium thermodynamics quantities exhibit a similar dependence with temperature as previously shown for the Markovian case: $\Delta S$, $\Delta U$, $\Delta F$ and $\Delta C$ in the high-temperature limit go as a power of the inverse temperature. Hence, the deviation of the nonequilibrium quantities from the conventional Brownian motion ultimately saturate to a constant value (which just depends of the spectral density parameters) in the studied non-Markovian dissipative scenario as well.

\section{Environment-induced orbital magneticlike moment}\label{EIMP}

So far we have explored the statistical mechanics of the flux-carrying particle, which surprisingly reveals that in the classical limit it largely shares the nonequilibrium thermodynamics properties of the damped two-dimensional harmonic oscillator, so that we recover the results in agreement with the classical statistical mechanics. Now the stationary magneticlike features induced by the Maxwell-Chern-Simons field environment are addressed when the quantum dissipation is fully characterized by a strict Markovian dynamics \cite{hanggi20051}. As stated in the Sec.\ref{DMCSM} as well as extensively discussed in Ref.\cite{valido20191}, the crucial effect of the Chern-Simons action in the dissipative description (\ref{HMCSF2}) is to effectively attach a changing magneticlike flux to the system harmonic oscillators that gives rise to an ordinary Hall effect responsible for the environmental rotational force (\ref{RFF}), which in turn is behind of the previously discussed environmental work delivered upon the flux-carrying particle. Consequently, the harmonic oscillator undergoes a vortexlike motion that may induce a non-zero orbital magneticlike moment, as similarly occurs in the conventional Brownian particles in presence of external magnetic fields \cite{saha20081,kaplan20091,jayannavar19811}. Interestingly, we shall show in this section that the average orbital magneticlike moment of the flux-carrying particle vanishes in the high-temperature limit in conformity with the BvL theorem as well as the results of the previous sections.

First, we define the orbital magneticlike moment $\hat{\vect  M}$ (in SI units) proportional to the orbital angular momentum as usually \cite{roy20081,stohr20061,dattagupta19971}
\begin{equation}
\hat{\vect  M}=-\frac{e}{2m}\hat{\vect q}\times \hat{\vect p}.
\label{OMM}
\end{equation}
Our main interest is to compute the average value of (\ref{OMM}) when the global system is in the canonical equilibrium state $\hat \rho_{\beta}\propto e^{-\beta \hat \H}$, as previously assumed in Sec.\ref{SRPF}. It is convenient to express $\hat{\vect  M}$ in terms of the cross-correlation functions
\begin{equation}
\Delta_{\alpha\lambda} (t)=\frac{1}{2}\left\langle \left\lbrace\hat q_{\alpha}(t),\hat q_{\lambda}(0) \right\rbrace  \right\rangle_{\beta},
\label{STCCR}
\end{equation}
as follows
\begin{equation}
\left\langle \hat{ M}_{z}\right\rangle_{\beta} =\lim_{t\rightarrow 0}\frac{e}{2} (\dot{\Delta}_{12}(t)-\dot{\Delta}_{21}(t) )=e\frac{d\Delta_{12}(t)}{d t}\bigg|_{t=0},
\label{AVOMM}
\end{equation}
where we have made use of the property $\Delta_{12}(t)=\Delta_{21}(-t)$, and $\left\langle \cdot\right\rangle_{\beta} $ denotes the average on the global canonical equilibrium state $\hat \rho_{\beta}$. Fortunately, we can further employ the imaginary-time path integral framework presented in Sec.\ref{SRPF} to compute the expression for the equilibrium correlations functions (\ref{STCCR}). Starting from the Euclidean action (\ref{EAS}) once again, this tasks is accomplished by introducing an additional linear source term which may be interpreted as a fictitious external (two-dimensional) force $\vect F$ \cite{weiss20121,ingold20021,zinn20101}, that yields the generating functional
\begin{equation}
Z_{flux}(\beta,\vect F)=Z_{MCS}^{-1}(\beta)\oint \mathcal{D}\vect q(\tau)\oint \mathcal{D}\underline{\vect x}(\tau)\ e^{-\hbar^{-1}\big(S^{(E)}[\vect q(\tau),\underline{\vect x}(\tau)]+\int_{0}^{\hbar \beta} d\tau \ \vect q(\tau)\cdot\vect F(\tau)\big)}.
\nonumber
\end{equation}
In the imaginary time, the correlation function is then obtained from the following well-known identity \cite{ingold20021,atland20101,callan19921},
\begin{equation}
\left\langle \hat{q}_{\alpha}(\tau_{1})\hat{q}_{\lambda}(\tau_{2}) \right\rangle_{\beta} =\frac{1}{Z_{flux}(\beta)}\frac{\delta^2 Z_{flux}(\beta,\vect F)}{\delta F_{\alpha}(\tau_{1})\delta F_{\lambda}(\tau_{2})} \Bigg|_{\vect F=0} ,
\label{CCFZ}
\end{equation}
where $\tau_{1}>\tau_{2}$ in imaginary time, and $Z_{flux}(\beta)$ is the partition function given by (\ref{PFSII}).

Following a similar procedure as to compute the partition function in Sec.\ref{SRPF}, one can show that the Eq.(\ref{CCFZ}) yields the following fluctuation-dissipation relation (a detail derivation is sketched in the \ref{app3}),
\begin{equation}
\Delta_{\alpha\lambda} (t) =\frac{\hbar}{2\pi m}\int_{-\infty}^{\infty}d\omega  \ e^{-i\omega t} \coth\left( \frac{\omega\hbar\beta}{2}\right)\text{Im}\Big(\check{\vect G}_{R}\Big)_{\alpha\lambda}(\omega),
\label{FDRPCC}
\end{equation}
where  $\text{Im}\vect A$ denotes the imaginary-like part of the complex matrix $\vect A$, i.e. $\text{Im}\vect A=-i/2(\vect A-\vect A^{\dagger})$ (with $\dagger$ denoting the Hermitian conjugate transpose \cite{horn19901}), and $\check{\vect G}_{R}(\omega)$ represents the real-time Fourier transform of the retarded Green's function of the quantum open-system dynamics. Accordingly, the inverse of the latter is given by (see the Eq.(\ref{ZSCM}) in \ref{app1})
\begin{equation}
\Big(\check{\vect G}_{R}^{-1}\Big)_{\alpha\beta}(\omega)=\delta_{\alpha\beta}\big(-(\omega+i0^{+})^2+\omega_{0}^2\big)+\check{\Sigma}_{\alpha\beta}(\omega+i0^{+}) , \nonumber
\end{equation}
with $\check{\vect \Sigma}(\omega)$ standing for the real-time Fourier representation of the retarded self-energy, also known as the environmental dynamical susceptibility, i.e. 
\begin{equation}
\check{\Sigma}_{\alpha\beta}(\omega)=\delta_{\alpha\beta}\big(\check{\Delta}(\omega)+\check{\Lambda}_{\parallel}(\omega)\big)+\epsilon_{\alpha\beta}\check{\Lambda}_{\perp}(\omega).
\label{RSE}
\end{equation}
The above frequency-dependent kernels are obtained from the Laplace transform of the dynamical susceptibilities presented in Sec.\ref{SRPF} (see the Eqs. (\ref{SDCI}), (\ref{SDCII}) and (\ref{SDCIII})) via analytic continuation \cite{atland20101,weiss20121}, e.g.
\begin{equation}
\check{\Delta}(\omega)=  \lim_{\varepsilon\rightarrow 0^{+}} \tilde{\Delta}(s=-i\omega+\varepsilon).
\nonumber
\end{equation}
Notice that the expression (\ref{FDRPCC}) is the counterpart of the fluctuation-dissipation theorem in the conventional Brownian motion \cite{grabert19841,valido20132,hanggi20051}.

Here, we are mainly interested in the strict Markovian dynamics for which the real-time Fourier transform of the retarded Green's function must express as a rational function, such that it just displays simple complex poles. As was similarly done in Sec.\ref{OHSDS}, this is equivalent to approximate $\check{\vect G}_{R}(\omega)$  by a Breit-Wigner resonance shape \cite{valido20191,alamoudi19991,alamoudi19981}, i.e., $\check{\vect G}_{R}(\omega)\simeq \check{\vect G}_{BW}(\omega+i0^{+})$ with
\begin{equation}
\Big(\check{\vect G}_{BW}^{-1}\Big)_{\alpha\beta}(\omega)=\delta_{\alpha\beta}\big(-\omega^2+\Omega_{0}^2-i2\Gamma_{0}\omega\big)+\epsilon_{\alpha\beta}\Omega_{CS}^2 ,
\end{equation}
which has simple poles $\lambda_{\mp}=\Gamma_{0}\mp \eta$ with
\begin{equation}
\eta=\left( \Gamma_{0}^2-\Omega_{0}^2+i\Omega_{CS}^{2}\right)^{\frac{1}{2}},
\end{equation}
where $\Omega_{0}$, $\Gamma_{0}$, and $\Omega_{CS}$ are given in Sec.\ref{OHSDS}. Recall that we are dealing with the system-environment weak coupling limit in which the subsidiary condition (\ref{SSCI}) holds (which in turns implies $\Gamma_{0}/\Omega_{0}\ll 1$ and $\Omega_{CS}/\Omega_{0}\ll 1$). Accordingly, this permits to compute an exact solution for the thermal correlation functions (\ref{FDRPCC}) by using the
standard contour integration techniques \cite{spiegel19931}. Concretely, we find
\begin{equation}
\frac{d\Delta_{12}(t)}{d t}=\text{Re}\{\dot{S}_{0}(t)+\dot{S}_{CS}(t)\} -\frac{2\hbar\Omega_{CS}^2}{m\beta}\dot{S}_{CS}^{(q)}(t), \label{TDCR} 
\end{equation}
where $\text{Re}\{z\}$ denotes the real part of the complex number $z$, and we have defined the following auxiliary functions 
\begin{eqnarray}
S_{0}(t)&=&\frac{\hbar}{4m\eta}\bigg(\coth\left(\frac{i\beta\hbar\lambda_{-}}{2}\right)e^{-\lambda_{-}t} -\coth\left(\frac{i\beta\hbar\lambda_{+}}{2}\right)e^{-\lambda_{+}t} \bigg),  \label{C0RE} \\  
S_{CS}(t)&=&-\frac{\Omega_{CS}^2\hbar}{m\beta(\Omega_{0}^4+\Omega_{CS}^4)} -\frac{i\hbar}{2m\beta(\Omega_{0}^2-i \Omega_{CS}^2)\eta}\bigg(\lambda_{+}e^{-\lambda_{-}t}-\lambda_{-}e^{-\lambda_{+}t} \bigg), \label{CSRE}
\end{eqnarray}
and the quantum correction
\begin{equation}
S_{CS}^{(q)}(t)=\sum_{n=1}^{\infty}\frac{e^{-\nu_{n}t}\left(\left(\nu_{n}^2+\Omega_{0}^2\right)^2+\Omega_{CS}^4+4\Gamma_{0}^2\nu_{n}^2\right)}{(\nu_{n}^2-\lambda_{+}^2)(\nu_{n}^2-\lambda_{-}^2)(\nu_{n}^2-\lambda_{+}^{\dagger 2})(\nu_{n}^2-\lambda_{-}^{\dagger 2})}.  \nonumber \\ \label{QSRE} 
\end{equation}

\begin{figure}[]
\begin{center}
\includegraphics[width=0.60\columnwidth]{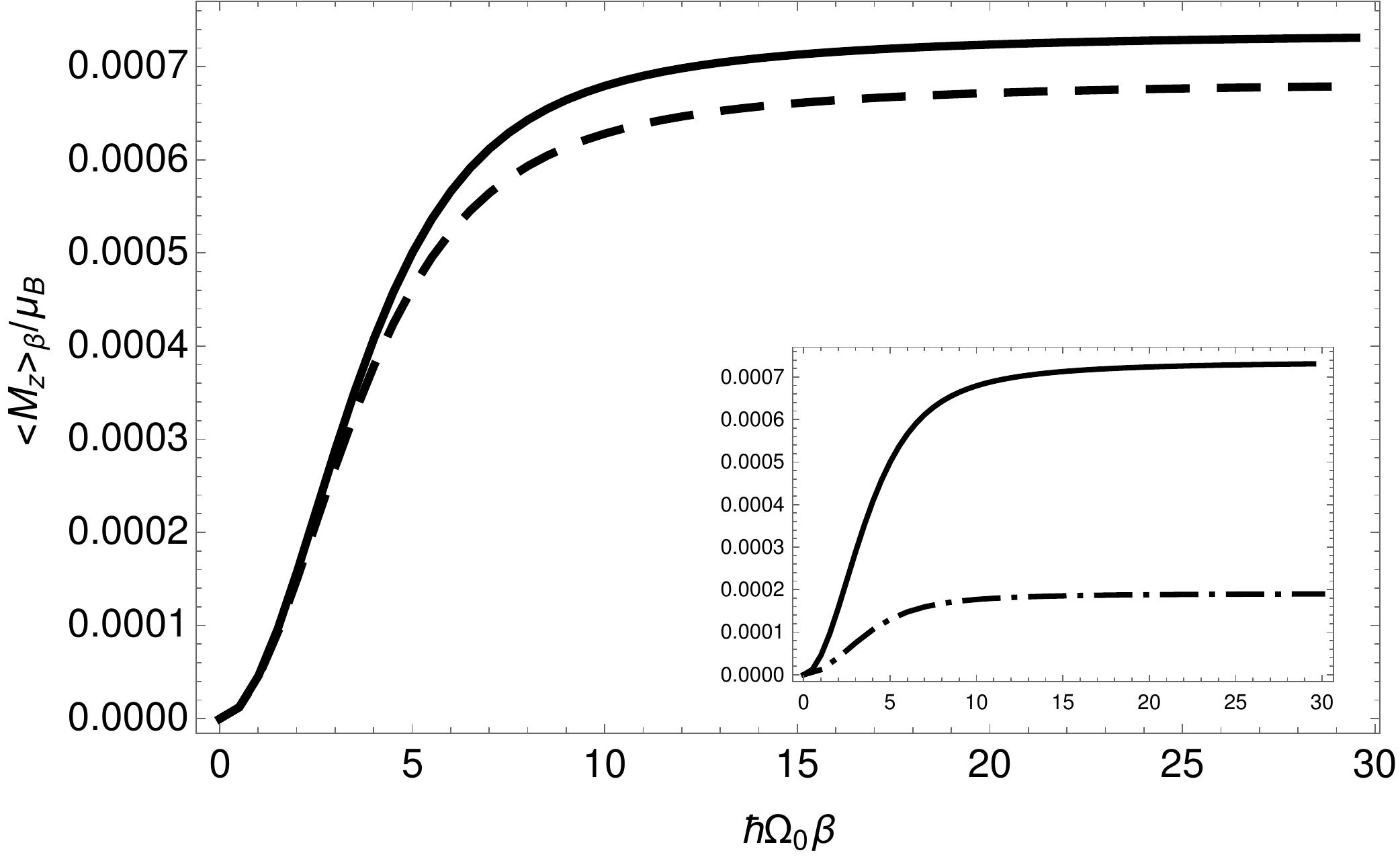}
\end{center}
\caption{The average value of the orbital magneticlike moment (in units of the Bohr magneton $\mu_{B}=e\hbar/2m$) as a function of the inverse temperature. The solid and dashed balck lines are, respectively, the results obtained for the dissipative coefficients $\Gamma_{0}=\Omega_{CS}$ and $\Gamma_{0}=2\Omega_{CS}$ at a fixed value of the renormalized oscillator frequency $\Omega=20\Omega_{CS}$. The dot-dashed black line in the inset corresponds to $\Omega_{0}=40\Omega_{CS}$, whereas the solid line is the same as in the main plot. The rest of parameters were chosen as $\Omega_{CS}=m/2=1/2$ in the natural units $\hbar=k_{B}=e=1$. \label{Fig4}} 
\end{figure}

Replacing the expressions (\ref{C0RE}), (\ref{CSRE}) and (\ref{QSRE}) in (\ref{AVOMM}) and approximating $\coth \left(i\beta\hbar\lambda_{\pm}/2\right)\simeq \mp 1$ as well as $\dot{S}_{CS}^{(q)}(t)$ up to next leading order in temperature, after some straightforward manipulation we obtain the average value of the orbital magneticlike moment (\ref{OMM}) in the low-temperature limit (i.e. $1\ll \Gamma_{0}\beta$), 
\begin{eqnarray} 
\left\langle  \hat{M}_{z} \right\rangle_{\beta} &=& e \hbar\left(\frac{\Omega_{CS}}{\Omega_{0}} \right)^2\Bigg(\frac{\Omega_{0}^{2}}{2\pi m(\Omega_{0}^2-\Gamma_{0}^{2})}   \nonumber\\
&+&\frac{\Omega_{0}^2\Gamma_{0}}{4\pi m (\Gamma_{0}^{2}-\Omega_{0}^2)^\frac{3}{2}} \Bigg(i\pi+\log\left( \frac{\Gamma_{0}+\sqrt{\Gamma_{0}^{2}-\Omega_{0}^2}}{\Gamma_{0}-\sqrt{\Gamma_{0}^{2}-\Omega_{0}^2}}\right)\Bigg)\Bigg) \nonumber \\
&+&\mathcal{O}\left( \left(\frac{\Omega_{CS}}{\Omega_{0}} \right)^4\right),
\label{QOMM}
\end{eqnarray}
which clearly pinpoints the environmental Hall response, manifested through the rotational strength parameter $\Omega_{CS}$, as responsible for a non-trivial intrinsic orbital magneticlike moment of the dissipative harmonic oscillator. Further, the positive value of the average magnetic moment can be understood by realizing that the rotational force giving rise to the orbital path is applied in the clockwise direction (see the Eq.(\ref{RFF})), which means that the induced magneticlike moment must be oriented in the counterclockwise direction for a particle with negative elemental charge. In view of this result and the discussion in Sec.\ref{Sfree}, we may conclude that the work done by the Maxwell-Chern-Simons field environment is devoted to generate the effective magnetization of the system particle.

The situation drastically change as one goes to the high-temperature regime (i.e. $\Gamma_{0}\beta\ll 1$): we find that the average magnetic moment cancels. This can be seen from the Eq.(\ref{TDCR}) by substituting $\coth \left(i\beta\hbar\lambda_{\pm}/2\right)\simeq -2i/\lambda_{\pm}\beta$ and by noting that the quantum corrections must become zero. Intriguingly, one could naively expect that the mean value of the magnetic moment does not vanish in the classical limit since the rotational force persists in the classical domain, however the results of Sec.\ref{OHSDS} indicates the other way around. It was shown that the effects in the nonequilibrium thermodynamics quantities owning to the environmental Hall response are effectively damped in the classical limit, since there is no significant contribution to the free and internal energies, which therein indicates that the mean value of the angular momentum of the flux-carrying particle eventually cancels, and thus, the average orbital magnetic moment as well by construction. As was explained in Sec.\ref{Sfree}, this could be mainly attributed to the fact that the environmental Hall response in the strict Markovian scenario remains constant whilst the dissipative effects contained in the longitudinal dynamical susceptibility grows linearly with temperature.

The behavior of the magneticlike moment of the flux-carrying particle is better illustrated by the figure (\ref{Fig4}), where its average value is plot as a function of the inverse temperature for distinct choices of the dissipative coefficient (see main plot) and renormalized harmonic frequency (see the inset). A quick glance reveals that the orbital magneticlike moment undergoes a smooth transition from a zero value in the high-temperature regime to a non-vanishing constant (determined by (\ref{QOMM})) in the quantum limit. Paying further attention, one may see for a fixed $\beta$ that the latter decreases for growing values of $\Omega_{0}$ as well as $\Gamma_{0}$, which is in agreement with the observation that the dominant effects in the strict Markovian dynamics are 
due to the dissipate mechanism and harmonic confinement, so that the relative influence of the rotational force in the open-system dynamics is effectively diminished.  

Finally let us emphasize that our results are in conformity with the so-called BvL theorem, albeit there is no need. Formally, this theorem states that the average magnetic moment of classical charged particles in the presence of an external \textit{time-independent} magnetic field is zero in thermal equilibrium \cite{savoie20151,pradhan20102,kaplan20091,caldeira20141}. Contrary to the latter, the root of the orbital magneticlike moment here is an environmental emf which is generated by a changing magneticlike flux \cite{wan20061}: the magneticlike moment arising from the dissipative microscopic description (\ref{HMCSF2}) is created by a \textit{dynamical} gauge field. Indeed, the rotational force (\ref{RFF}) closely resemblances the force term $\hat{\vect F}^{*}$ due to an azimuthal electric field produced by certain time-dependent magnetic field $\hat{\vect B}(t)$ in the ordinary Maxwell electrodynamics \cite{saha20081}, e.g.
\begin{equation}
\hat{\vect F}^{*}\propto\hat{\vect q}\times \frac{d  \hat{\vect B}(t)}{dt}.
\nonumber
\end{equation}
Hence, our flux-carrying Brownian particle has no direct comparison with the two-dimensional Brownian particle in presence of an external uniform magnetic field \cite{jayannavar19811}, so the dissipative microscopic model (\ref{HMCSF2}) are not constrained to the fundamental conclusion of the BvL theorem (despite the results found for the Markovian case are in agreement with this), and thus, it does not prevent us of finding out a non-vanishing magneticlike moment even at high temperature. For instance, it was shown for (closed-system) fermionic particles subject to the $\text{CS-QED}_{2+1}$ that the permanent magneticlike field induced by the Chern-Simons action may survive in the high-temperature regime \cite{itoh19981}, so one could expect here that the average orbital magneticlike moment remains non-trivial in the classical limit for a dissipative scenario where the transverse Chern-Simon effects were dominant, such as for a sufficiently strong coupling to the Chern-Simons action (e.g. $\kappa\gamma_{0}/\omega_{0}^{2}\gg 1$). Nonetheless, the latter could occur at expense of the dissipative dynamics of the flux-carrying particle departs from the low-lying effective description (\ref{HMCSF2}) \cite{valido20191} as well as deviates from the standard thermodynamics \cite{nieuwenhuizen20021} (recall the subsidiary condition (\ref{SCDI})).

\section{Concluding remarks}\label{OCR}

Following the imaginary-time path integral approach, we have extensively examined the statistical physics emerging from a novel microscopic description for two-dimensional dissipative harmonic systems which has been recently derived in the framework of the non-relativistic $\text{CS-QED}_{2+1}$ theory. This analysis has revealed that the properties of the single dissipative particle in the quantum domain substantially differ from those of the damped harmonic oscillator characteristic of the conventional Brownian motion. Essentially, the novel environmental Chern-Simons effects endow the dissipative particle with an orbital magneticlike moment, such that it could be thought of as a Brownian particle "dressing" a magneticlike flux which, however, has no counterpart in the traditional Landau diamagnetism. Unlike the quantum statistical mechanics of a charged magneto-oscillator coupled to a conventional heat bath, the dominant behavior of the free energy in the quantum limit is linear in temperature and the entropy may eventually saturate to a constant value at low energies, which indicates that the ground state of the flux-carrying particle bears certain resemblance to a degenerate state. Additionally, we discuss that by virtue of the environmental Hall response the Maxwell-Chern-Simons field environment play a role beyond the standard bosonic heat bath and may simultaneously act as a work source as well, which is behind of the environment-induced magnetization. Nonetheless, the heat intake of the dissipative particle from the environment is demanded to be larger than this work in order the Maxwell-Chern-Simons field environment behaves as a heat bath consistent with standard thermodynamics.

We have also found out that the environmental Chern-Simons effects have not significant impact in either the thermodynamics quantities or the orbital magneticlike moment in the high temperature limit for the studied dissipative scenarios: strict Markovian dynamics and a non-Markovian instance (i.e. we have explored both the weak and the strong system-environment coupling situations), instead they represent a small correction to the well-known properties of the damped two-dimensional harmonic oscillator. In this way, we recover the results in conformity with the BvL theorem (that is, a vanishing orbital magnetic moment in the classical limit). We argue that this is a consequence of the fact that the dynamical susceptibilities encoding the environmental Chern-Simons effects are energetically dampened in the high-temperature limit. Furthermore, we discuss that this ultimately occurs because such susceptibilities feature a power spectral density that is characteristic of a $1/f$-type noise. As a counterpart, the longitudinal Chern-Simons dynamical susceptibility presents an infrared divergence which leads the partition function of the flux-carrying particle to eventually vanish, and in turn, the nonequilibrium thermodynamics quantities diverge at absolute zero temperature in the studied dissipative scenarios. Up to a renormalization of this anomalous behavior, our results suggest that the present dissipative microscopic model widely reproduces the Third Law of thermodynamics, at least for a weak Chern-Simons action strength.

Recent advances in experimental techniques for tailoring non-classical behaviors of mesoscopic quantum objects coupled to quantum environments \cite{hanggi20091} have opened new avenues to experimentally study and test various fundamental concepts of the conventional Brownian motion, such as the energy equipartition theorem \cite{li20101}, the non-equilibrium thermodynamics \cite{millen20141} or the diffusion coefficients \cite{otsuka20091}. Thanks to this progress the results presented here could be amenable to experimental test in a new series of upcoming experiments in the realm of Brownian motion. In particular, it would be appealing to verify our predictions about the cancellation of the magneticlike moment in the high temperature regime, which is a question of major interest in the Brownian dynamics theory. Additionally, it would be interesting to further investigate the free energy and entropy changes owning to the Chern-Simons gauge field, in order to elucidate if this could eventually serve as an indirect assesment of some kind of "charge-flux" duality. On the other side, though a perfect realization of the Chern-Simons effects may be still challenging, it is well known that this theory describes the true electromagnetic interaction in several situations of planar condensed matter systems \cite{hansson20171,qi20111,marino19931,fradldn19941}. For instance, our approach is not excluded to capture fundamental features of the dissipative dynamics of charged particles subjected to magnetic induction fields in the limit of the magnetic dipole approximation.  

As the Chern-Simons theory resides on the heart of nowadays two-dimensional condensed matter theory, our approach regards a rather fundamental description of the statistical physics confined in the plane. For instance, the present description may be particularly relevant in understanding the quantum thermodynamics of systems composed of particles displaying certain charge-flux duality. On the other side, the proposed microscopic description has clear points of contact with the incipient artificial gauge field theory, and regards a natural candidate to study systems coupled to dynamically generated synthetic fields, or more challenging, it could be used as a paradigmatic model to address energy transport properties as well as electrical conductivity effects characteristic of charge-flux excitations coupled to thermal heat baths and magnetic flux noise. Yet, as similarly occurs in the conventional model, the basic concept of the flux-carrying Brownian particle could eventually find  applications in the understanding of the dissipative dynamics of collective degrees of freedom exhibiting anyon statistics. We hope that our present results may inspire future works along these lines.

\section*{acknowledgments}
The author is grateful to S. Kohler, A. G\'omez-Le\'on, T. Stauber and A. Cortijo for useful discussions. The author warmly thanks J. J. Garc\'ia-Ripoll for enlightening discussions and support through the development of the present work. This material is based upon work supported by the Air Force Office of Scientific Research under award number FA2386-18-1-4019.

\appendix

\section{Derivation of the reduced partition function}\label{app1}
In this appendix we illustrate the derivation of the reduced partition function given by Eq.(\ref{PFSII}), which appears in Sec.\ref{SRPF}. Following closely the program presented in \cite{weiss20121,zinn20101}, this is achieved by firstly introducing the imaginary-time Fourier transform of the system and environmental modes, denoted by $\tilde{\vect q}$ and $\tilde x_{\vect k}$ respectively,
\begin{eqnarray}
\hat x_{\vect k}(\tau)&=&\frac{1}{\hbar \beta}\sum_{n=-\infty}^{+\infty}\tilde x_{\vect k}(\nu_{n})e^{i\nu_{n}\tau},\label{FourRep} \\
\hat{\vect q}(\tau)&=&\frac{1}{\hbar \beta}\sum_{n=-\infty}^{+\infty}\tilde{\vect  q}(\nu_{n})e^{i\nu_{n}\tau},\nonumber
\end{eqnarray}
with the Matsubara frequencies $\nu_{n}=2\pi n/\hbar \beta$, and secondly, rewriting the spatial operator $\tilde{x}_{\vect k}(\nu_{n})$ of the environmental $\vect k$-mode in terms of the quantum fluctuations $\tilde{y}_{\vect k}(\nu_{n})$, i.e.,
\begin{equation}
\tilde{x}_{\vect k}(\nu_{n})=\bar{x}_{\vect k}(\nu_{n})+\tilde{y}_{\vect k}(\nu_{n}),
\label{FFEI}
\end{equation}
with $\bar{x}_{\vect k}(\nu_{n})$ corresponding to the stationary path of the Euclidean action of the environmental $\vect k$-mode (\ref{EAE}), and which obeys the Euclidean classical equations of motion
\begin{equation}
0=-m_{\vect k}\nu_{n}^2\bar x_{\vect k}(\nu_{n})-\nu_{n}\kappa\epsilon_{\alpha\beta}l_{\beta}(\vect k) \tilde{ q}_{\alpha}(\nu_{n})  -\omega^{2}_{\vect k}\left(m_{\vect k}\bar x_{\vect k}(\nu_{n})-l_{\alpha}(\vect k) \tilde q_{\alpha}(\nu_{n})\right). \label{ECE1}
\end{equation} 
By solving the Eq.(\ref{ECE1}) in terms of the spatial system operator $\tilde{q}_{\alpha}(\nu_{n})$, we obtain
\begin{eqnarray}
\tilde{x}_{\vect k}(\nu_{n})
&=&\frac{\omega_{\vect k}^2 l_{\alpha}(\vect k)-\nu_{n}\kappa \epsilon_{\alpha\beta}l_{\beta}(\vect k)}{m_{\vect k}(\nu_{n}^2+\omega_{\vect k}^2)}\tilde{q}_{\alpha}(\nu_{n})+\tilde{y}_{\vect k}(\nu_{n}), \nonumber \\
&=&\frac{C_{\alpha}(\vect k)}{m_{\vect k}(\nu_{n}^2+\omega_{\vect k}^2)}\tilde{q}_{\alpha}(\nu_{n})+\tilde{y}_{\vect k}(\nu_{n}), \nonumber
\end{eqnarray}
as well as the hermitian conjugate transpose (denoted by $\dagger$),
\begin{equation}
\tilde{x}_{\vect k}^{\dagger}(\nu_{n})=\frac{C_{\alpha}^{\dagger}(\vect k)}{m_{\vect k}(\nu_{n}^2+\omega_{\vect k}^2)}\tilde{q}_{\alpha}^{\dagger}(\nu_{n})+\tilde{y}_{\vect k}^{\dagger}(\nu_{n}).
\nonumber
\end{equation}

Plugging this result into the Euclidean action (\ref{EAS}), and appealing to the fact that the linear term in $\tilde{y}_{\vect k}(\nu_{n})$ must vanish since $\tilde{x}_{\vect k}(\nu_{n})$ is a stationary point \cite{weiss20121,zinn20101}, after some straightforward manipulation the system-environment action in the Fourier representation may be decomposed as follows 
\begin{eqnarray}
\mathcal{S}_{MCS,I}^{(E)}[\tilde{\vect q}, \underline{\bar{\vect x}}+ \underline{\tilde{\vect y}}]=\mathcal{S}_{MCS}^{(E)}[\underline{\tilde{\vect y}}]+\mathcal{S}_{I}^{(E)}[\tilde{\vect q}],
\label{EASI}
\end{eqnarray}
where the first term in the right-hand side identifies with the environmental partition function
\begin{equation}
Z_{MCS}(\beta)=\oint \mathcal{D}\underline{\vect y}(\cdot)\ \text{exp}\left\lbrace -\mathcal{S}_{MCS}^{(E)}[\underline{\vect y}(\cdot)]/\hbar\right\rbrace ,
\label{PFMCS}
\end{equation}
whilst the second term takes the following form,
\begin{eqnarray}
&&\mathcal{S}_{I}^{(E)}[\tilde{\vect q}]=  \frac{\hbar\beta}{2}\sum_{n=-\infty}^{\infty}\sum_{\vect k}\frac{\tilde{q}_{\alpha}^{\dagger}(\nu_{n})\tilde{q}_{\beta}(\nu_{n})}{m_ {\vect k}(\nu_{n}^2+\omega_{\vect k}^2)^2}\Bigg(\big(-i\nu_{n}C_{\alpha}^{\dagger}(\vect k)-i(\nu_{n}^2+\omega_{\vect k}^2)\kappa\epsilon_{\alpha\lambda}l_{\lambda}(\vect k)\big)\big(i\nu_{n}C_{\beta}(\vect k) \nonumber \\
&+&i(\nu_{n}^2+\omega_{\vect k}^2)\kappa\epsilon_{\beta\lambda'}l_{\lambda'}(\vect k)\big) +\omega_{\vect k}^2\left(C_{\alpha}^{\dagger}(\vect k)-(\nu_{n}^2+\omega_{\vect k}^2)l_{\alpha}^{\dagger}(\vect k)\right)
\left(C_{\beta}(\vect k)-(\nu_{n}^2+\omega_{\vect k}^2)l_{\beta}(\vect k) \right)  \Bigg) \nonumber \\
&=&\frac{\hbar\beta}{2}\sum_{n=-\infty}^{\infty}\sum_{\vect k}\frac{\tilde{q}_{\alpha}^{\dagger}(\nu_{n})\tilde{q}_{\beta}(\nu_{n})}{m_ {\vect k}(\nu_{n}^2+\omega_{\vect k}^2)^2}\Bigg((\nu_{n}^2+\omega_{\vect k}^2)C_{\alpha}^{\dagger}(\vect k)C_{\beta}(\vect k)+(\nu_{n}^2+\omega_{\vect k}^2)^2\big(\kappa^2\epsilon_{\alpha\lambda}l_{\lambda}(\vect k)\epsilon_{\beta\lambda'}l_{\lambda'}(\vect k) \nonumber \\
&+&\omega_{\vect k}^{2} l_{\alpha}^{\dagger}(\vect k)l_{\beta}(\vect k)\big) -(\nu_{n}^2+\omega_{\vect k}^2)\bigg[C_{\alpha}^{\dagger}(\vect k)\big(\omega_{\vect k}^2l_{\beta}(\vect k)-\kappa\nu_{n}\epsilon_{\beta\lambda'}l_{\lambda'}(\vect k)\big)
+C_{\beta}(\vect k)\left(\omega_{\vect k}^2l_{\alpha}^{\dagger}(\vect k)-\kappa\nu_{n}\epsilon_{\alpha\lambda}l_{\lambda}(\vect k)\right)\bigg] \Bigg) \nonumber \\
&=&\frac{\hbar\beta}{2}\sum_{n=-\infty}^{\infty}\sum_{\vect k}\frac{\tilde{q}_{\alpha}^{\dagger}(\nu_{n})\tilde{q}_{\beta}(\nu_{n})}{m_ {\vect k}(\nu_{n}^2+\omega_{\vect k}^2)}\bigg(-C_{\alpha}^{\dagger}(\vect k)C_{\beta}(\vect k)+(\nu_{n}^2+\omega_{\vect k}^2)\big(\kappa^2\epsilon_{\alpha\lambda}l_{\lambda}(\vect k)\epsilon_{\beta\lambda'}l_{\lambda'}(\vect k)+\omega_{\vect k}^{2} l_{\alpha}^{\dagger}(\vect k)l_{\beta}(\vect k)\big)\bigg)   \nonumber \\
&=&\frac{m\hbar\beta}{2}\sum_{n=-\infty}^{\infty}\tilde{\Sigma}_{\alpha\beta}(\nu_{n})\tilde{q}_{\alpha}^{\dagger}(\nu_{n})\tilde{q}_{\beta}(\nu_{n}), \label{EASIA}
\end{eqnarray}
where we have recognized the imaginary-time Fourier transform of the retarded self-energy $\tilde{\vect \Sigma}(s)$ in the last line. We can go further by elaborating on the latter expression and substituting the system-environment coupling coefficients (\ref{EIT}), that is
\begin{eqnarray}
\tilde{\Sigma}_{\alpha\beta}(\nu_{n})&=&
\frac{1}{m}\sum_{\vect k}\frac{\omega_{\vect k}^2}{m_ {\vect k}(\nu_{n}^2+\omega_{\vect k}^2)}\Big(\nu_{n}^2l_{\alpha}^{\dagger}(\vect k)l_{\beta}(\vect k)+\kappa^2\epsilon_{\alpha\lambda}l_{\lambda}(\vect k)\epsilon_{\beta\lambda'}l_{\lambda'}(\vect k)\big)  \nonumber \\
&+&\nu_{n}\kappa\big(\epsilon_{\alpha\lambda}l_{\lambda}(\vect k)l_{\beta}(\vect k)+\epsilon_{\beta\lambda'}l_{\lambda'}(\vect k) l_{\alpha}^{\dagger}(\vect k)\big)\Big)\nonumber \\
&=&\frac{\delta_{\alpha\beta}}{m}\sum_{\vect k}\frac{\nu_{n}^2}{m_ {\vect k}(\nu_{n}^2+\omega_{\vect k}^2)} \frac{\hbar m_{\vect k} e^2f^2(|\vect k|)}{(2\pi L)^2}  \nonumber \\
&-&\frac{1}{m}\sum_{\vect k}\frac{\nu_{n}^2}{m_ {\vect k}(\nu_{n}^2+\omega_{\vect k}^2)}\Bigg((\nu_{n}^2-\kappa^2)\frac{\hbar m_{\vect k} e^2 f^2(|\vect k|) k_{\alpha}k_{\beta}}{(2\pi L)^2|\vect k|^2} \nonumber \\
&+&\frac{\hbar m_{\vect k}\kappa\nu_{n}  e^2 f^2(|\vect k|) }{(2\pi L)^2|\vect k|^2}\big(\epsilon_{\beta\lambda}k_{\lambda}k_{\alpha}-\epsilon_{\alpha\lambda'}k_{\lambda'}k_{\beta}\big)\Bigg) \nonumber \\
&=&\delta_{\alpha\beta} \frac{\hbar e^2}{m (2\pi L)^2} \sum_{\vect k}\frac{f^2(|\vect k|) }{\nu_{n}^2+\omega_{\vect k}^2}\left(\nu_{n}^2+(\kappa^2-\nu_{n}^2)\frac{k_{\alpha}^2}{|\vect k|^2}\right)  \nonumber\\
&+&\epsilon_{\alpha\beta}\frac{\hbar e^2 \kappa}{m (2\pi L)^2} \sum_{\vect k}\frac{\nu_{n}}{\nu_{n}^2+\omega_{\vect k}^2}f^2(|\vect k|), \label{SEEA}
\end{eqnarray}
where, in the last equality, we have made use of the fact that the crossing terms $f^2(|\vect k|) k_{\alpha}k_{\beta}$ (for $\alpha\neq \beta$) within the discrete sum are odd in the integration variable, and thus, they must vanish (this can be seen more clearly by doing a change of variables to polar coordinates). Furthermore, paying attention to the symmetrical properties of the discrete sum, one may realize that
\begin{equation}
\sum_{\vect k}\frac{\nu_{n}^2f^2(|\vect k|)}{\nu_{n}^2+\omega_{\vect k}^2}\left(\nu_{n}^2+(\kappa^2-\nu_{n}^2)\frac{k_{\alpha}^2}{|\vect k|^2}\right)=\frac{1}{2}\sum_{\vect k}\frac{\nu_{n}^2+\kappa^2}{\nu_{n}^2+\omega_{\vect k}^2}f^2(|\vect k|) \ \ \text{for} \ \alpha=1,2.
\nonumber
\end{equation}
Now by looking at the first and second terms of the right-hand side of (\ref{SEEA}), one may recognize the dissipative kernels, i.e.
\begin{eqnarray}
\tilde{\Delta}(\nu_{n})&=&\frac{\hbar e^2 }{2m(2\pi L)^2} \sum_{\vect k}\frac{\nu_{n}^2}{\nu_{n}^2+\omega_{\vect k}^2}f^2(|\vect k|),\label{SECI} \\
\tilde{\Lambda}_{||}(\nu_{n})&=&\frac{\hbar e^2 \kappa^2}{2m (2\pi L)^2} \sum_{\vect k}\frac{1}{(\nu_{n}^2+\omega_{\vect k}^2)}f^2(|\vect k|)\label{SECII}, \\
\tilde{\Lambda}_{\perp}(\nu_{n})&=&\frac{\hbar e^2 \kappa}{m (2\pi L)^2} \sum_{\vect k}\frac{\nu_{n}}{\nu_{n}^2+\omega_{\vect k}^2}f^2(|\vect k|)\label{SECIII}.
\end{eqnarray}
Introducing the definition of the spectral density $J(\omega)$ (given by (\ref{SPDI})) in the Eqs. (\ref{SECI}), (\ref{SECII}), and (\ref{SECIII}), we are led immediately to the expressions (\ref{SDCI}), (\ref{SDCII}) and (\ref{SDCIII}), respectively. 

Having determined the imaginary-time Fourier transform of the retarded self energy, the reduced partition function (\ref{PFEI}) of the flux-carrying particle is directly obtained from gathering together the Euclidean action of the system $S_{S}^{(E)}[\tilde{\vect q}]$ (see Eq.(\ref{EAS})) and $S_{I}^{(E)}[\tilde{\vect q}]$ given by (\ref{EASIA}), i.e.
\begin{eqnarray}
S_{eff}^{(E)}[\tilde{\vect q}]&=&S_{S}^{(E)}[\tilde{\vect q}]+S_{I}^{(E)}[\tilde{\vect q}] \label{SEASI}\\
&=&\frac{m\beta}{2}\sum_{n=-\infty}^{\infty}\Big(  \delta_{\alpha\beta}\big(\nu_{n}^2+\omega_{0}^2 \big)\tilde{q}_{\alpha}^{\dagger}(\nu_{n})\tilde{q}_{\beta}(\nu_{n})+\tilde{\Sigma}_{\alpha\beta}(\nu_{n}) \tilde{q}_{\alpha}^{\dagger}(\nu_{n})\tilde{q}_{\beta}(\nu_{n}) \Big),\nonumber 
\end{eqnarray}
which yields,
\begin{eqnarray}
&&Z_{flux}(\beta)=\oint \mathcal{D}\vect q(\cdot) \ \text{exp}\Big\{-S_{eff}^{(E)}[\vect q(\cdot)]/\hbar\Big\} \nonumber \\
&=&\frac{m}{2\pi \hbar^2\beta}\int_{-\infty}^{\infty} d \tilde{\vect q}(0) \ \text{exp}\bigg( -\frac{m\beta(\delta_{\alpha\beta}\omega_{0}^{2}+\tilde{\Sigma}_{\alpha\beta}(0))}{2} \tilde{q}_{\alpha}^{\dagger}(0)\tilde{q}_{\beta}(0) \bigg) \nonumber \\
&\times &\prod_{n= 1}^{\infty} \left[ \int_{-\infty}^{\infty} \int_{-\infty}^{\infty} \frac{d \tilde{\vect q}(\nu_{n})d \tilde{\vect q}(\nu_{-n})}{(2\pi/( m\beta \nu_{n}))^2} \ \text{exp}\bigg( -\frac{m\beta}{2}\Big(  \delta_{\alpha\beta}(\nu_{n}^2+\omega_{0}^{2}) +\tilde{\Sigma}_{\alpha\beta}(\nu_{n})\Big)\tilde{q}_{\alpha}^{\dagger}(\nu_{n})\tilde{q}_{\beta}(\nu_{n}) \bigg)\right]   \nonumber\\
&=&\frac{1}{\hbar^2\beta^2\sqrt{\det(\omega_{0}^2\vect I+\tilde{\vect \Sigma}(0))}}\prod_{n=1}^{\infty}\frac{\nu_{n}^{4}}{\sqrt{\det\left(\Big((\nu_{n}^2+\omega_{0}^2)\vect I_{2}+\tilde{\vect \Sigma}(\nu_{n})\Big)\Big((\nu_{n}^2+\omega_{0}^2)\vect I_{2}+\tilde{\vect \Sigma}^{\dagger}(\nu_{n})\Big)  \right)} }\nonumber \\
&=&\frac{1}{\hbar^2\beta^{2}\sqrt{\det(\omega_{0}^2\vect I+\tilde{\vect \Sigma}(0))}}\prod_{n=1}^{\infty}\Bigg[\frac{\nu_{n}^4}{\left(\nu_{n}^2+\omega_{0}^2+\tilde{\Delta}(\nu_{n}) \right)^2 }\Bigg] \nonumber \\
&\times & \prod_{n=1}^{\infty}\Bigg[\frac{\left(\nu_{n}^2+\omega_{0}^2+\tilde{\Delta}(\nu_{n}) \right)^2 }{\left( (\nu_{n}^2+\omega_{0}^2+\tilde{\Delta}(\nu_{n})+\tilde{\Lambda}_{||}(\nu_{n}))^{2}-|\tilde{\Lambda}_{\perp}(\nu_{n})|^2\right) }\Bigg] \nonumber \\
&=&Z_{0}^2(\beta)\left(1+ \frac{\tilde{\Lambda}_{\parallel}(0)}{\omega_{0}^2}\right)^{-1} \prod_{n=1}^{\infty} \Bigg[\frac{\left(\nu_{n}^2+\omega_{0}^2+\tilde{\Delta}(\nu_{n}) \right)^2 }{ (\nu_{n}^2+\omega_{0}^2+\tilde{\Delta}(\nu_{n})+\tilde{\Lambda}_{||}(\nu_{n}))^{2}-|\tilde{\Lambda}_{\perp}(\nu_{n})|^2 }\Bigg],\label{RPFA}
\end{eqnarray}
with $\vect I_{2}$ denoting the $2\times2$ identity matrix, and where we have introduced the well-known Fourier functional measure from the quantum statistical path-integral framework \cite{weiss20121}. As mentioned in Sec.\ref{SRPF}, the Eq.(\ref{RPFA}) clearly consists of a Gaussian path integral \cite{weiss20121} characterized by the (infinite-dimensional) covariance matrix $\bigoplus_{n=1}^{\infty}\tilde{\vect G}^{-1}_{R}(\nu_{n})$, with
\begin{equation}
\tilde{\vect G}^{-1}_{R}(s)=(s^2+\omega_{0}^2)\vect I_{2}+\tilde{\vect \Sigma}(s) , \label{ZSCM}
\end{equation}
where $\tilde{\vect G}_{R}(s)$ essentially being the Laplace transform of the retarded Green's function associated to the quantum open-system dynamics.

Accordingly, the imaginary-time path integral in (\ref{RPFA}) could diverge if the matrix (\ref{ZSCM}) eventually vanishes. Conversely, the $Z_{flux}(\beta)$ provided by (\ref{RPFA}) will represent a partition function physically consistent with a dissipative scenario (where the interested system evolves towards a thermal equilibrium state), when (\ref{ZSCM}) is strictly positive-definite \cite{horn19901}. It is now readily to see that the latter is equivalent to demand that the subsidiary condition (\ref{SCDI}) always holds, in agreement with the discussion in Ref.\cite{valido20191}.

\section{Reduced partition function for the strict Ohmic and Lorentzian-type spectral densities}\label{app2}
In this section, we show the derivation of the reduced partition function (\ref{PFSO}) and (\ref{PFSL}) presented in Secs. \ref{OHSDS} and \ref{LTSDS} respectively, once we have fixed the spectral density. Let us consider first the strict Ohmic example. Replacing the damping kernels (\ref{SDCO}) in the expression (\ref{RPFA}) obtained in the previous section, we find 
\begin{eqnarray}
Z_{flux}(\beta)&=&Z_{0}^2(\beta)\left(1+ \frac{\phi}{\omega_{0}^2}\right)^{-1}\prod_{n=1}^{\infty}\frac{\nu_{n}^2(\nu_{n}^2+\omega_{0}^2+2\gamma_{0}\nu_{n})^2}{(\nu_{n}^3+\Omega_{0}^2\nu_{n}+2\gamma_{0}\nu_{n}^2+2\gamma_{0}\kappa^2)^{2}-(\gamma_{0}\kappa )^2\nu_{n}^2 } \nonumber \\
&=&Z_{0}^2(\beta)\left(1+ \frac{\phi}{\omega_{0}^2}\right)^{-1}\prod_{n=1}^{\infty}\frac{\prod_{i=1,2}\left( 1+\frac{r_{i}}{\nu_{n}}\right)^2}{\prod_{i=1}^{3}\left( 1+\frac{R_{i}}{\nu_{n}}\right) \left(1+\frac{R'_{i}}{\nu_{n}}\right) } \nonumber \\
&=&Z_{0}^2(\beta)\left(1+ \frac{\phi}{\omega_{0}^2}\right)^{-1}\prod_{n=1}^{\infty}\frac{\frac{\prod_{i=1}^3\left( 1+\frac{1}{n}\right)^{R_{i}\nu^{-1}} \left(1+\frac{1}{n}\right)^{R_{i}'\nu^{-1}}}{\prod_{i=1}^{3}\left( 1+\frac{R_{i}\nu^{-1}}{n}\right) \left(1+\frac{R'_{i} \nu^{-1}}{n}\right) }}{\frac{\prod_{i=1,2}\left( 1+\frac{1}{n}\right)^{2r_{i}\nu^{-1}}}{\prod_{i=1,2}\left( 1+\frac{r_{i}\nu^{-1}}{n}\right)^2}} \nonumber \\
&=&Z_{0}^2(\beta) \left(1+ \frac{\phi}{\omega_{0}^2}\right)^{-1}\frac{\prod_{i=1}^{3}\Gamma\left(1+\nu^{-1}R_{i} \right) \Gamma\left(1+\nu^{-1}R_{i}' \right)}{\prod_{i=1,2}\Gamma\left(1+\nu^{-1}r_{i} \right)^2}, \label{PFSOA}
\end{eqnarray}
where $\nu^{-1}=\hbar\beta/2\pi$, and we have made use of the Vieta's relations for the roots $r_{i}$, $R_{i}$, and $R_{i}'$ as well (i.e. $\sum_{i=1}^{3}(R_{i}+R_{i}')=-2\sum_{i=1,2}r_{i}=-4\gamma_{0}$). It is immediate to see that the reduced partition function (\ref{PFSO}) is obtained from (\ref{PFSOA}) after minor manipulation. We can follow an identical procedure to obtain the reduced partition function (\ref{PFSL}) for the choice of the Lorentzian-type spectral density (\ref{SPDII}). That is, 
\begin{eqnarray}
&&Z_{flux}(\beta)=Z_{0}^2(\beta)\left(1+\frac{\phi}{\omega_{0}^2} \right)^{-1}\nonumber \\
&\times&\prod_{n=1}^{\infty}\frac{\nu_{n}^2\left( (\nu_{n}^2+\omega_{0}^2)(\nu_{n}^2+\chi \nu_{n}+\Omega^2)+\frac{2\gamma_{0}\Omega^2}{\chi}\nu_{n}(\nu_{n}+\chi)\right)^2}{\left( \nu_{n}(\nu_{n}^2+\omega_{0}^2)(\nu_{n}^2+\chi \nu_{n}+\Omega^2)+\frac{2\gamma_{0}\Omega^2}{\chi}\nu_{n}^2(\nu_{n}+\chi)+\frac{2\kappa^2\gamma_{0}\Omega^2}{\chi}(\nu_{n}+\chi)\right) ^{2}-\frac{\kappa^2 \gamma_{0}^2\Omega^4}{\chi^2}\nu_{n}^2(\nu_{n}+\chi)^2} \nonumber \\
&=&Z_{0}^2(\beta)\left(1+\frac{\phi}{\omega_{0}^2} \right)^{-1}\prod_{n=1}^{\infty}\frac{\prod_{i=1}^{2}\left(1 +\frac{r_{i}}{\nu_{n}}\right)^2\left(1 +\frac{r'_{i}}{\nu_{n}}\right)^2}{\prod_{i=1}^{5}\left(1 +\frac{R_{i}}{\nu_{n}}\right)\left(1 +\frac{R'_{i}}{\nu_{n}}\right) } \nonumber \\
&=&Z_{0}^2(\beta)\left(1+\frac{\phi}{\omega_{0}^2} \right)^{-1}\frac{\prod_{i=1}^{5}\Gamma\left(1+\nu^{-1}R_{i} \right) \Gamma\left(1+\nu^{-1}R_{i}' \right)}{\prod_{i=1}^{2}\Big[\Gamma\left(1+\nu^{-1}r_{i} \right) \Gamma\left(1+\nu^{-1}r_{i}' \right)  \Big]^2}, \nonumber
\end{eqnarray}
where once again we have employed the Vieta's relations for the roots (i.e. $\sum_{i=1}^{5}(R_{i}+R_{i}')=2\sum_{i=1}^{2}(r_{i}+r_{i}')=-2\chi$).

\section{Fluctuation-dissipation relation}\label{app3}
Here we illustrate the computation of the fluctuation-dissipation relation (\ref{FDRPCC}) appearing in Sec.\ref{EIMP}. We start from the previously computed Euclidean action (\ref{SEASI}) with the additional source term in the imaginary-time Fourier representation, i.e.
\begin{eqnarray}
S_{eff}^{(E)}[\tilde{\vect q},\tilde{\vect F}]&=&\frac{m\beta}{2}\sum_{n=-\infty}^{\infty}\Big(  \delta_{\alpha\beta}(\nu_{n}^2 +\omega_{0}^2) \tilde{q}_{\alpha}^{\dagger}(\nu_{n})\tilde{q}_{\beta}(\nu_{n})+\tilde{\Sigma}_{\alpha\beta}(\nu_{n}) \tilde{q}_{\alpha}^{\dagger}(\nu_{n})\tilde{q}_{\beta}(\nu_{n}) \nonumber \\
& +&\tilde{\vect F}_{\alpha}(\nu_{n})\tilde{q}_{\alpha}^{\dagger}(\nu_{n})+\tilde{\vect F}_{\alpha}^{\dagger}(\nu_{n})\tilde{q}_{\alpha}(\nu_{n})\Big) \nonumber \\
&=&\frac{m\beta}{2}\sum_{n=-\infty}^{\infty} \Big(\tilde{\vect G}_{R}^{-1}\Big)_{\alpha\beta}(\nu_{n}) \tilde{q}_{\alpha}^{\dagger}(\nu_{n})\tilde{q}_{\beta}(\nu_{n})+\tilde{\vect F}_{\alpha}(\nu_{n})\tilde{q}_{\alpha}^{\dagger}(\nu_{n}) +\tilde{\vect F}_{\alpha}^{\dagger}(\nu_{n})\tilde{q}_{\alpha}(\nu_{n}),
\label{SEASIF}
\end{eqnarray}
where $\tilde{\vect F}(\nu_{n})$ are the Fourier coefficients of the external force as similarly defined in the Eq.(\ref{FourRep}). It is readily to see that the solution of the classical equation of motion deduced from the variation of the action (\ref{SEASIF}) is given by $\tilde{q}_{\alpha}(\nu_{n})=\Big(\tilde{\vect G}_{R}^{-1}\Big)_{\alpha\beta}(\nu_{n})\tilde{\vect F}_{\beta}(\nu_{n})$, and similarly $\tilde{q}_{\alpha}^{\dagger}(\nu_{n})=\Big(\tilde{\vect G}_{R}^{-1}\Big)_{\alpha\beta}^{\dagger}(\nu_{n})\tilde{\vect F}_{\beta}^{\dagger}(\nu_{n})$. Plugging the latter result into (\ref{SEASIF}) and after some straightforward manipulation, we arrive to the classical Euclidean action
\begin{equation}
S_{cl}^{(E)}[\vect F ]=-\frac{1}{2m\hbar\beta}\sum_{n=-\infty}^{\infty}\Big(\tilde{\vect G}_{R}^{-1}\Big)_{\alpha\beta}(\nu_{n}) \int_{0}^{\hbar\beta}d\tau\int_{0}^{\hbar\beta}d\sigma \ F_{\alpha}(\tau)F_{\beta}(\sigma)e^{i \nu_{n}(\tau-\sigma)}.
\label{CEAF}
\end{equation}
Following a similar procedure as to compute the decoupled effective action Eq.(\ref{EASI}) in appendix\ref{app1}, we may decompose the expression (\ref{SEASIF}) into the classical Euclidean action (\ref{CEAF}) and the unperturbed Euclidean action (\ref{SEASI}) , i.e. $S_{eff}^{(E)}[\vect q,\vect F]=S_{cl}^{(E)}[\vect F]+S_{eff}^{(E)}[\vect q]$ \cite{weiss20121,atland20101}. This immediately retrieves that the generating functional expresses as the unperturbed partition function weighted by the classical Euclidean action, i.e. $Z_{flux}(\beta,J)=Z_{flux}(\beta)\text{exp}\Big( -S_{cl}^{(E)}[\vect F]/\hbar\Big)$. Having determined the latter, now we can use the identity (\ref{CCFZ}) to obtain the imaginary-time correlation function in position, which yields
\begin{equation}
\left\langle \hat{q}_{\alpha}(\tau)\hat{q}_{\delta}(0) \right\rangle_{\beta} =\frac{1}{m\beta}\sum_{n=-\infty}^{\infty} \Big(\tilde{\vect G}_{R}\Big)_{\alpha\delta}(\nu_{n}) e^{i \nu_{n} \tau}.
\nonumber
\end{equation}
Following the Ref.\cite{ingold20021}, from the above equation we may obtain the thermal correlation function in real time via analytic continuation by rephrasing the infinite sum as a contour integral in the complex frequency domain. This is formally achieved by combining the Fourier representation of the retarded Green's function and its conjugate transpose with the function $(1-e^{-\hbar \omega\beta })^{-1}$ that has poles at frequencies $\omega=i\nu_{n}$ with $n\in \mathbb{Z}$. As the poles of $\tilde{\vect G}_{R}(\omega)$ and $\tilde{\vect G}_{R}^{\dagger}(\omega)$ are respectively in the lower and upper -half complex plane in agreement with Kramers-Kronig relations (notice that $\det(\tilde{\vect G}_{R}^{\dagger}(\omega))=(\det(\tilde{\vect G}_{R}^{\dagger}(\omega)))^{H}$), we may conveniently devise two semicircular contour integrals $C_{+}$ and $C_{-}$ for which solely contribute the poles $i\nu_{n}$, i.e.,
\begin{eqnarray}
\sum_{n=-\infty}^{\infty} \Big(\tilde{\vect G}_{R}\Big)_{\alpha\beta}(\nu_{n}) e^{i \nu_{n} \tau}&=&\frac{-i}{2\pi}\int_{C_{+}} \frac{d\omega \ \hbar \beta \ e^{-\omega\tau}}{1-e^{-\hbar \omega\beta }}\Big(\tilde{\vect G}_{R}\Big)_{\alpha\beta}(\omega) \nonumber \\
&+&\frac{i}{2\pi}\int_{C_{-}} \frac{d\omega \ \hbar \beta \ e^{-\omega\tau}}{1-e^{-\hbar \omega\beta }}\Big(\tilde{\vect G}_{R}\Big)_{\alpha\beta}^{\dagger}(\omega).
\nonumber
\end{eqnarray}
Since the contribution of the semicircular arc eventually cancels owning to $\tilde{\vect G}_{R}(\omega)$ is expected to decay in frequency, after taking the Wick rotation the thermal correlation function can be finally express in the following form
\begin{equation}
\left\langle \hat{q}_{\alpha}(t)\hat{q}_{\delta}(0) \right\rangle_{\beta} =\frac{\hbar}{m \pi}\int_{-\infty}^{\infty} \frac{d\omega \  e^{-i\omega t}}{1-e^{-\hbar \omega\beta }}\text{Im}\Big(\check{\vect G}_{R}\Big)_{\alpha\delta}(\omega),
\nonumber 
\end{equation}
where $\check{\vect G}_{R}(\omega)$ denotes the real-time Fourier transform of the retarded Green's function, and $\text{Im}\vect A$ represents the imaginary part of the complex matrix $\vect A$ \cite{horn19901}. Hence, the symmetrical position correlation function reads
\begin{eqnarray}
\Delta_{\alpha\delta} (t)&=&\frac{\hbar}{2\pi m }\int_{-\infty}^{\infty} \frac{  d\omega  \ e^{-i\omega t}}{1-e^{-\hbar \omega\beta }}\text{Im}\Big(\check{\vect G}_{R}\Big)_{\alpha\delta}(\omega) \nonumber \\
&+&\frac{\hbar}{2\pi m }\int_{-\infty}^{\infty} \frac{d\omega \ e^{i\omega t}}{1-e^{-\hbar \omega\beta }}\text{Im}\Big(\check{\vect G}_{R}\Big)_{\delta\alpha}(\omega)\nonumber \\
&=&\frac{\hbar}{4 m \pi}\int_{-\infty}^{\infty} d\omega e^{-i\omega t}\Bigg(\text{Im}\Big(\check{\vect G}_{R}\Big)_{\alpha\delta}(\omega)+\text{Im}\Big(\check{\vect G}_{R}\Big)_{\delta\alpha}(-\omega)\Bigg)\label{DFES} \\
&+&\frac{\hbar}{4\pi m }\int_{-\infty}^{\infty} d\omega e^{-i\omega t}\coth\left( \frac{\omega\hbar\beta}{2}\right) \Bigg(\text{Im}\Big(\check{\vect G}_{R}\Big)_{\alpha\delta}(\omega)-\text{Im}\Big(\check{\vect G}_{R}\Big)_{\delta\alpha}(-\omega)\Bigg),
 \nonumber
\end{eqnarray}
In deriving the above result, we have made use of the identity $(1-e^{-\hbar \omega\beta })^{-1}=1/2+1/2\coth(\hbar\omega\beta)$. Now we note that the first integral in the last line of Eq.(\ref{DFES}) cancels owning to $\text{Im}\Big(\check{\vect G}_{R}\Big)_{\beta\alpha}(-\omega)=-\text{Im}\Big(\check{\vect G}_{R}\Big)_{\alpha\beta}(\omega)$, which is a direct consequence of the Green's function property $\Big(\check{\vect G}_{R}\Big)_{\alpha\beta}(-\omega)=\Big(\check{\vect G}_{R}\Big)_{\alpha\beta}^{H}(\omega)$ \cite{valido20191}, with $\Big(\check{\vect G}_{R}\Big)_{\alpha\beta}^{H}(\omega)$ being the Hermitian conjugate. Thus, one may see that the fluctuation-dissipation relation (\ref{FDRPCC}) directly follows from Eq.(\ref{DFES}), as we wanted to show.


%


\begin{thebibliography}{00}%
\makeatletter
\providecommand \@ifxundefined [1]{%
 \@ifx{#1\undefined}
}%
\providecommand \@ifnum [1]{%
 \ifnum #1\expandafter \@firstoftwo
 \else \expandafter \@secondoftwo
 \fi
}%
\providecommand \@ifx [1]{%
 \ifx #1\expandafter \@firstoftwo
 \else \expandafter \@secondoftwo
 \fi
}%
\providecommand \natexlab [1]{#1}%
\providecommand \enquote  [1]{``#1''}%
\providecommand \bibnamefont  [1]{#1}%
\providecommand \bibfnamefont [1]{#1}%
\providecommand \citenamefont [1]{#1}%
\providecommand \href@noop [0]{\@secondoftwo}%
\providecommand \href [0]{\begingroup \@sanitize@url \@href}%
\providecommand \@href[1]{\@@startlink{#1}\@@href}%
\providecommand \@@href[1]{\endgroup#1\@@endlink}%
\providecommand \@sanitize@url [0]{\catcode `\\12\catcode `\$12\catcode
  `\&12\catcode `\#12\catcode `\^12\catcode `\_12\catcode `\%12\relax}%
\providecommand \@@startlink[1]{}%
\providecommand \@@endlink[0]{}%
\providecommand \url  [0]{\begingroup\@sanitize@url \@url }%
\providecommand \@url [1]{\endgroup\@href {#1}{\urlprefix }}%
\providecommand \urlprefix  [0]{URL }%
\providecommand \Eprint [0]{\href }%
\providecommand \doibase [0]{http://dx.doi.org/}%
\providecommand \selectlanguage [0]{\@gobble}%
\providecommand \bibinfo  [0]{\@secondoftwo}%
\providecommand \bibfield  [0]{\@secondoftwo}%
\providecommand \translation [1]{[#1]}%
\providecommand \BibitemOpen [0]{}%
\providecommand \bibitemStop [0]{}%
\providecommand \bibitemNoStop [0]{.\EOS\space}%
\providecommand \EOS [0]{\spacefactor3000\relax}%
\providecommand \BibitemShut  [1]{\csname bibitem#1\endcsname}%
\let\auto@bib@innerbib\@empty
\bibitem [{\citenamefont {Wen}(2004)}]{wen20041}%
  \BibitemOpen
  \bibfield  {author} {\bibinfo {author} {\bibfnamefont {X.-G.}\ \bibnamefont
  {Wen}},\ }\href@noop {} {\emph {\bibinfo {title} {Quantum Field Theory of
  Many-body Systems: From the Origin of Sound to an Origin of Light and
  Electrons}}}\ (\bibinfo  {publisher} {Oxford University Press},\ \bibinfo
  {year} {2004})\BibitemShut {NoStop}%
\bibitem [{\citenamefont {Atland}\ and\ \citenamefont
  {Simons}(2010)}]{atland20101}%
  \BibitemOpen
  \bibfield  {author} {\bibinfo {author} {\bibfnamefont {A.}~\bibnamefont
  {Atland}}\ and\ \bibinfo {author} {\bibfnamefont {B.}~\bibnamefont
  {Simons}},\ }\href@noop {} {\emph {\bibinfo {title} {Condensed Matter Field
  Theory (Second edition)}}}\ (\bibinfo  {publisher} {Cambridge University
  Press},\ \bibinfo {year} {2010})\BibitemShut {NoStop}%
\bibitem [{\citenamefont {Goldman}\ \emph {et~al.}(2016)\citenamefont
  {Goldman}, \citenamefont {Budich},\ and\ \citenamefont
  {Zoller}}]{goldman20161}%
  \BibitemOpen
  \bibfield  {author} {\bibinfo {author} {\bibfnamefont {N.}~\bibnamefont
  {Goldman}}, \bibinfo {author} {\bibfnamefont {J.~C.}\ \bibnamefont {Budich}},
  \ and\ \bibinfo {author} {\bibfnamefont {P.}~\bibnamefont {Zoller}},\ }\href
  {\doibase 10.1038/nphys3803} {\bibfield  {journal} {\bibinfo  {journal} {Nat.
  Physcs.}\ }\textbf {\bibinfo {volume} {12}},\ \bibinfo {pages} {639}
  (\bibinfo {year} {2016})},\ \Eprint {http://arxiv.org/abs/1607.03902}
  {arXiv:1607.03902} \BibitemShut {NoStop}%
\bibitem [{\citenamefont {Kitaev}(2005)}]{kitaev20051}%
  \BibitemOpen
  \bibfield  {author} {\bibinfo {author} {\bibfnamefont {A.}~\bibnamefont
  {Kitaev}},\ }\href {\doibase 10.1016/S0003-4916(02)00018-0} {\bibfield
  {journal} {\bibinfo  {journal} {Ann. Phys.}\ }\textbf {\bibinfo {volume}
  {303}},\ \bibinfo {pages} {2} (\bibinfo {year} {2005})},\ \Eprint
  {http://arxiv.org/abs/9707021} {arXiv:9707021 [quant-ph]} \BibitemShut
  {NoStop}%
\bibitem [{\citenamefont {Stormer}(1999)}]{stormer19991}%
  \BibitemOpen
  \bibfield  {author} {\bibinfo {author} {\bibfnamefont {H.~L.}\ \bibnamefont
  {Stormer}},\ }\href {\doibase 10.1103/revmodphys.71.875} {\bibfield
  {journal} {\bibinfo  {journal} {Rev. Mod. Phys.}\ }\textbf {\bibinfo {volume}
  {71}},\ \bibinfo {pages} {875} (\bibinfo {year} {1999})}\BibitemShut
  {NoStop}%
\bibitem [{\citenamefont {Hansson}\ \emph {et~al.}(2017)\citenamefont
  {Hansson}, \citenamefont {Hermanns}, \citenamefont {Simon},\ and\
  \citenamefont {Viefers}}]{hansson20171}%
  \BibitemOpen
  \bibfield  {author} {\bibinfo {author} {\bibfnamefont {T.~H.}\ \bibnamefont
  {Hansson}}, \bibinfo {author} {\bibfnamefont {M.}~\bibnamefont {Hermanns}},
  \bibinfo {author} {\bibfnamefont {S.~H.}\ \bibnamefont {Simon}}, \ and\
  \bibinfo {author} {\bibfnamefont {S.~F.}\ \bibnamefont {Viefers}},\ }\href
  {\doibase 10.1103/RevModPhys.89.025005} {\bibfield  {journal} {\bibinfo
  {journal} {Rev. Mod. Phys.}\ }\textbf {\bibinfo {volume} {89}},\ \bibinfo
  {pages} {025005} (\bibinfo {year} {2017})},\ \Eprint
  {http://arxiv.org/abs/1601.01697} {arXiv:1601.01697} \BibitemShut {NoStop}%
\bibitem [{\citenamefont {Heinonen}(1991)}]{heinonen19911}%
  \BibitemOpen
  \bibfield  {author} {\bibinfo {author} {\bibfnamefont {O.~G.}\ \bibnamefont
  {Heinonen}},\ }\href@noop {} {\emph {\bibinfo {title} {Composite fermions: a
  unified of the quantum Hall regime}}}\ (\bibinfo  {publisher} {World
  Scientific},\ \bibinfo {year} {1991})\BibitemShut {NoStop}%
\bibitem [{\citenamefont {Dunne}(1999)}]{dunne19991}%
  \BibitemOpen
  \bibfield  {author} {\bibinfo {author} {\bibfnamefont {G.~V.}\ \bibnamefont
  {Dunne}},\ }in\ \href@noop {} {\emph {\bibinfo {booktitle} {Aspects
  topologiques de la physique en basse dimension. Topological aspects of low
  dimensional systems}}}\ (\bibinfo  {publisher} {Springer},\ \bibinfo {year}
  {1999})\ pp.\ \bibinfo {pages} {177--263}\BibitemShut {NoStop}%
\bibitem [{\citenamefont {Chen}\ and\ \citenamefont
  {Halperin}(1989)}]{chen19891}%
  \BibitemOpen
  \bibfield  {author} {\bibinfo {author} {\bibfnamefont {W.~F. W.~E.}\
  \bibnamefont {Chen}, \bibfnamefont {Y.-H.}}\ and\ \bibinfo {author}
  {\bibfnamefont {B.}~\bibnamefont {Halperin}},\ }\href {\doibase
  10.1142/S0217979289000725} {\bibfield  {journal} {\bibinfo  {journal} {Int.
  J. Mod. Phys. B}\ }\textbf {\bibinfo {volume} {7}},\ \bibinfo {pages} {1001}
  (\bibinfo {year} {1989})}\BibitemShut {NoStop}%
\bibitem [{\citenamefont {Witten}(1989)}]{witten19891}%
  \BibitemOpen
  \bibfield  {author} {\bibinfo {author} {\bibfnamefont {E.}~\bibnamefont
  {Witten}},\ }\href {\doibase 10.1007/BF01217730} {\bibfield  {journal}
  {\bibinfo  {journal} {Commun. Math. Phys.}\ }\textbf {\bibinfo {volume}
  {121}},\ \bibinfo {pages} {351} (\bibinfo {year} {1989})}\BibitemShut
  {NoStop}%
\bibitem [{\citenamefont {Moura-Melo}\ and\ \citenamefont
  {Helay{\"{e}}l-Neto}(2001)}]{moura20011}%
  \BibitemOpen
  \bibfield  {author} {\bibinfo {author} {\bibfnamefont {W.~A.}\ \bibnamefont
  {Moura-Melo}}\ and\ \bibinfo {author} {\bibfnamefont {J.~A.}\ \bibnamefont
  {Helay{\"{e}}l-Neto}},\ }\href {\doibase 10.1103/PhysRevD.63.065013}
  {\bibfield  {journal} {\bibinfo  {journal} {Phys. Rev. D}\ }\textbf {\bibinfo
  {volume} {63}},\ \bibinfo {pages} {065013} (\bibinfo {year} {2001})},\
  \Eprint {http://arxiv.org/abs/0004143} {arXiv:0004143 [hep-th]} \BibitemShut
  {NoStop}%
\bibitem [{\citenamefont {Deser}\ \emph {et~al.}(1982)\citenamefont {Deser},
  \citenamefont {Jackiw},\ and\ \citenamefont {Templeton}}]{deser19821}%
  \BibitemOpen
  \bibfield  {author} {\bibinfo {author} {\bibfnamefont {S.}~\bibnamefont
  {Deser}}, \bibinfo {author} {\bibfnamefont {R.}~\bibnamefont {Jackiw}}, \
  and\ \bibinfo {author} {\bibfnamefont {S.}~\bibnamefont {Templeton}},\ }\href
  {\doibase 10.1016/0003-4916(82)90164-6} {\bibfield  {journal} {\bibinfo
  {journal} {Ann. Phys.}\ }\textbf {\bibinfo {volume} {140}},\ \bibinfo {pages}
  {372} (\bibinfo {year} {1982})}\BibitemShut {NoStop}%
\bibitem [{\citenamefont {Hosotani}(1993)}]{hosotani19931}%
  \BibitemOpen
  \bibfield  {author} {\bibinfo {author} {\bibfnamefont {Y.}~\bibnamefont
  {Hosotani}},\ }\href {\doibase 10.1016/0370-2693(93)90822-Y} {\bibfield
  {journal} {\bibinfo  {journal} {Phys. Lett. B}\ }\textbf {\bibinfo {volume}
  {319}},\ \bibinfo {pages} {332} (\bibinfo {year} {1993})}\BibitemShut
  {NoStop}%
\bibitem [{\citenamefont {Itoh}\ and\ \citenamefont {Kato}(1998)}]{itoh19981}%
  \BibitemOpen
  \bibfield  {author} {\bibinfo {author} {\bibfnamefont {T.}~\bibnamefont
  {Itoh}}\ and\ \bibinfo {author} {\bibfnamefont {H.}~\bibnamefont {Kato}},\
  }\href {\doibase 10.1103/PhysRevLett.81.30} {\bibfield  {journal} {\bibinfo
  {journal} {Phys. Rev. Lett.}\ }\textbf {\bibinfo {volume} {81}},\ \bibinfo
  {pages} {30} (\bibinfo {year} {1998})}\BibitemShut {NoStop}%
\bibitem [{\citenamefont {Dillenschneider}\ and\ \citenamefont
  {Richert}(2006)}]{dillenschneider20061}%
  \BibitemOpen
  \bibfield  {author} {\bibinfo {author} {\bibfnamefont {R.}~\bibnamefont
  {Dillenschneider}}\ and\ \bibinfo {author} {\bibfnamefont {J.}~\bibnamefont
  {Richert}},\ }\href {\doibase 10.1103/PhysRevB.74.144404} {\bibfield
  {journal} {\bibinfo  {journal} {Phys. Rev. B}\ }\textbf {\bibinfo {volume}
  {74}},\ \bibinfo {pages} {144404} (\bibinfo {year} {2006})}\BibitemShut
  {NoStop}%
\bibitem [{\citenamefont {Matsuyama}(1990)}]{matsuyama19901}%
  \BibitemOpen
  \bibfield  {author} {\bibinfo {author} {\bibfnamefont {T.}~\bibnamefont
  {Matsuyama}},\ }\href {\doibase 10.1103/PhysRevD.42.3469} {\bibfield
  {journal} {\bibinfo  {journal} {Phys. Rev. D}\ }\textbf {\bibinfo {volume}
  {42}},\ \bibinfo {pages} {3469} (\bibinfo {year} {1990})}\BibitemShut
  {NoStop}%
\bibitem [{\citenamefont {Ferreiros}\ and\ \citenamefont
  {Fradkin}(2018{\natexlab{b}})}]{ferreiros20181}%
  \BibitemOpen
  \bibfield  {author} {\bibinfo {author} {\bibfnamefont {Y.}~\bibnamefont
  {Ferreiros}}\ and\ \bibinfo {author} {\bibfnamefont {E.}~\bibnamefont
  {Fradkin}},\ }\href {\doibase 10.1016/j.aop.2018.10.001} {\bibfield
  {journal} {\bibinfo  {journal} {Ann. Phys.}\
  }\textbf {\bibinfo {volume} {399}},\ \bibinfo {pages} {1} (\bibinfo {year}
  {2018}{\natexlab{b}})}\BibitemShut {NoStop}%
\bibitem [{\citenamefont {Pachos}(2012)}]{pachos20121}%
  \BibitemOpen
  \bibfield  {author} {\bibinfo {author} {\bibfnamefont {J.~K.}\ \bibnamefont
  {Pachos}},\ }\href@noop {} {\emph {\bibinfo {title} {Introduction to
  topological quantum computation}}}\ (\bibinfo  {publisher} {Cambridge
  University Press},\ \bibinfo {year} {2012})\BibitemShut {NoStop}%
\bibitem [{\citenamefont {Qi}\ and\ \citenamefont {Zhang}(2011)}]{qi20111}%
  \BibitemOpen
  \bibfield  {author} {\bibinfo {author} {\bibfnamefont {X.~L.}\ \bibnamefont
  {Qi}}\ and\ \bibinfo {author} {\bibfnamefont {S.~C.}\ \bibnamefont {Zhang}},\
  }\href {\doibase 10.1103/RevModPhys.83.1057} {\bibfield  {journal} {\bibinfo
  {journal} {Rev. Mod. Phys.}\ }\textbf {\bibinfo {volume} {83}},\ \bibinfo
  {pages} {1057} (\bibinfo {year} {2011})}\BibitemShut {NoStop}%
\bibitem [{\citenamefont {Callan}\ and\ \citenamefont
  {Freed}(1992)}]{callan19921}%
  \BibitemOpen
  \bibfield  {author} {\bibinfo {author} {\bibfnamefont {C.~G.}\ \bibnamefont
  {Callan}}\ and\ \bibinfo {author} {\bibfnamefont {D.}~\bibnamefont {Freed}},\
  }\href {\doibase 10.1016/0550-3213(92)90400-6} {\bibfield  {journal}
  {\bibinfo  {journal} {Nucl. Phys. B}\ }\textbf {\bibinfo {volume} {374}},\
  \bibinfo {pages} {543} (\bibinfo {year} {1992})}\BibitemShut {NoStop}%
\bibitem [{\citenamefont {Cobanera}\ \emph {et~al.}(2016)\citenamefont
  {Cobanera}, \citenamefont {Kristel},\ and\ \citenamefont
  {Smith}}]{cobanera20161}%
  \BibitemOpen
  \bibfield  {author} {\bibinfo {author} {\bibfnamefont {E.}~\bibnamefont
  {Cobanera}}, \bibinfo {author} {\bibfnamefont {P.}~\bibnamefont {Kristel}}, \
  and\ \bibinfo {author} {\bibfnamefont {C.~M.}\ \bibnamefont {Smith}},\ }\href
  {\doibase 10.1103/PhysRevB.93.245422} {\bibfield  {journal} {\bibinfo
  {journal} {Phys. Rev. B}\ }\textbf {\bibinfo {volume} {93}},\ \bibinfo
  {pages} {245422} (\bibinfo {year} {2016})}\BibitemShut {NoStop}%
\bibitem [{\citenamefont {Viyuela}\ \emph {et~al.}(2015)\citenamefont
  {Viyuela}, \citenamefont {Rivas},\ and\ \citenamefont
  {Martin-Delgado}}]{viyuela20151}%
  \BibitemOpen
  \bibfield  {author} {\bibinfo {author} {\bibfnamefont {O.}~\bibnamefont
  {Viyuela}}, \bibinfo {author} {\bibfnamefont {A.}~\bibnamefont {Rivas}}, \
  and\ \bibinfo {author} {\bibfnamefont {M.}~\bibnamefont {Martin-Delgado}},\
  }\href@noop {} {\bibfield  {journal} {\bibinfo  {journal} {2D Materials}\
  }\textbf {\bibinfo {volume} {2}},\ \bibinfo {pages} {034006} (\bibinfo {year}
  {2015})}\BibitemShut {NoStop}%
\bibitem [{\citenamefont {Sieberer}\ \emph {et~al.}(2016)\citenamefont
  {Sieberer}, \citenamefont {Buchhold},\ and\ \citenamefont
  {Diehl}}]{sieberer20161}%
  \BibitemOpen
  \bibfield  {author} {\bibinfo {author} {\bibfnamefont {L.~M.}\ \bibnamefont
  {Sieberer}}, \bibinfo {author} {\bibfnamefont {M.}~\bibnamefont {Buchhold}},
  \ and\ \bibinfo {author} {\bibfnamefont {S.}~\bibnamefont {Diehl}},\ }\href
  {\doibase 10.1088/0034-4885/79/9/096001} {\bibfield  {journal} {\bibinfo
  {journal} {Rep. Prog. Phys.}\ }\textbf {\bibinfo {volume} {79}},\ \bibinfo
  {pages} {096001} (\bibinfo {year} {2016})}\BibitemShut {NoStop}%
\bibitem [{\citenamefont {Viyuela}\ \emph {et~al.}(2014)\citenamefont
  {Viyuela}, \citenamefont {Rivas},\ and\ \citenamefont
  {Martin-Delgado}}]{viyuela20141}%
  \BibitemOpen
  \bibfield  {author} {\bibinfo {author} {\bibfnamefont {O.}~\bibnamefont
  {Viyuela}}, \bibinfo {author} {\bibfnamefont {A.}~\bibnamefont {Rivas}}, \
  and\ \bibinfo {author} {\bibfnamefont {M.~A.}\ \bibnamefont
  {Martin-Delgado}},\ }\href {\doibase 10.1103/PhysRevLett.113.076408}
  {\bibfield  {journal} {\bibinfo  {journal} {Phys. Rev. Lett.}\ }\textbf
  {\bibinfo {volume} {113}},\ \bibinfo {pages} {076408} (\bibinfo {year}
  {2014})}\BibitemShut {NoStop}%
\bibitem [{\citenamefont {Zhang}\ \emph {et~al.}(2017)\citenamefont {Zhang},
  \citenamefont {Shen},\ and\ \citenamefont {Yi}}]{zhang20171}%
  \BibitemOpen
  \bibfield  {author} {\bibinfo {author} {\bibfnamefont {W.~Q.}\ \bibnamefont
  {Zhang}}, \bibinfo {author} {\bibfnamefont {H.~Z.}\ \bibnamefont {Shen}}, \
  and\ \bibinfo {author} {\bibfnamefont {X.~X.}\ \bibnamefont {Yi}},\ }\href
  {\doibase 10.1038/s41598-017-16061-6} {\bibfield  {journal} {\bibinfo
  {journal} {Scientific Reports}\ }\textbf {\bibinfo {volume} {7}},\ \bibinfo
  {pages} {1} (\bibinfo {year} {2017})}\BibitemShut {NoStop}%
\bibitem [{\citenamefont {Shen}\ \emph {et~al.}(2015)\citenamefont {Shen},
  \citenamefont {Qin}, \citenamefont {Shao},\ and\ \citenamefont
  {Yi}}]{shen20151}%
  \BibitemOpen
  \bibfield  {author} {\bibinfo {author} {\bibfnamefont {H.~Z.}\ \bibnamefont
  {Shen}}, \bibinfo {author} {\bibfnamefont {M.}~\bibnamefont {Qin}}, \bibinfo
  {author} {\bibfnamefont {X.~Q.}\ \bibnamefont {Shao}}, \ and\ \bibinfo
  {author} {\bibfnamefont {X.~X.}\ \bibnamefont {Yi}},\ }\href {\doibase
  10.1103/PhysRevE.92.052122} {\bibfield  {journal} {\bibinfo  {journal}
  {Physical Review E - Statistical, Nonlinear, and Soft Matter Physics}\
  }\textbf {\bibinfo {volume} {92}},\ \bibinfo {pages} {1} (\bibinfo {year}
  {2015})}\BibitemShut {NoStop}%
\bibitem [{\citenamefont {Viyuela}\ \emph {et~al.}(2018)\citenamefont
  {Viyuela}, \citenamefont {Rivas}, \citenamefont {Gasparinetti}, \citenamefont
  {Wallraff}, \citenamefont {Filipp},\ and\ \citenamefont
  {Martin-Delgado}}]{viyuela20181}%
  \BibitemOpen
  \bibfield  {author} {\bibinfo {author} {\bibfnamefont {O.}~\bibnamefont
  {Viyuela}}, \bibinfo {author} {\bibfnamefont {A.}~\bibnamefont {Rivas}},
  \bibinfo {author} {\bibfnamefont {S.}~\bibnamefont {Gasparinetti}}, \bibinfo
  {author} {\bibfnamefont {A.}~\bibnamefont {Wallraff}}, \bibinfo {author}
  {\bibfnamefont {S.}~\bibnamefont {Filipp}}, \ and\ \bibinfo {author}
  {\bibfnamefont {M.~A.}\ \bibnamefont {Martin-Delgado}},\ }\href {\doibase
  10.1038/s41534-017-0056-9} {\bibfield  {journal} {\bibinfo  {journal} {npj
  Quantum Information}\ }\textbf {\bibinfo {volume} {4}},\ \bibinfo {pages}
  {10} (\bibinfo {year} {2018})}\BibitemShut {NoStop}%
\bibitem [{\citenamefont {Santos}\ \emph {et~al.}(2017)\citenamefont {Santos},
  \citenamefont {Almeida},\ and\ \citenamefont {Souza}}]{santos20171}%
  \BibitemOpen
  \bibfield  {author} {\bibinfo {author} {\bibfnamefont {W.~O.}\ \bibnamefont
  {Santos}}, \bibinfo {author} {\bibfnamefont {G.~M.}\ \bibnamefont {Almeida}},
  \ and\ \bibinfo {author} {\bibfnamefont {A.~M.}\ \bibnamefont {Souza}},\
  }\href {\doibase 10.1142/S0217751X17501469} {\bibfield  {journal} {\bibinfo
  {journal} {Int. J. Mod. Phys. A}\ }\textbf {\bibinfo {volume} {32}},\
  \bibinfo {pages} {1750146} (\bibinfo {year} {2017})}\BibitemShut {NoStop}%
\bibitem [{\citenamefont {Misaki}\ \emph {et~al.}(2014)\citenamefont
  {Misaki}, \citenamefont {Miyashita},\ and\ \citenamefont
  {Nagaosa}}]{misaki20181}%
  \BibitemOpen
  \bibfield  {author} {\bibinfo {author} {\bibfnamefont {K.}~\bibnamefont
  {Misaki}}, \bibinfo {author} {\bibfnamefont {S.}~\bibnamefont {Miyshita}}, \
  and\ \bibinfo {author} {\bibfnamefont {N.}\ \bibnamefont
  {Nagaosa}},\ }\href {\doibase 10.1103/PhysRevB.97.075122}
  {\bibfield  {journal} {\bibinfo  {journal} {Phys. Rev. B}\ }\textbf
  {\bibinfo {volume} {97}},\ \bibinfo {pages} {075122} (\bibinfo {year}
  {2018})}\BibitemShut {NoStop}%
\bibitem [{\citenamefont {Cho}\ and\ \citenamefont
  {Moore}(2011{\natexlab{b}})}]{cho20111}%
  \BibitemOpen
  \bibfield  {author} {\bibinfo {author} {\bibfnamefont {G.Y.}~\bibnamefont
  {Cho}}\ and\ \bibinfo {author} {\bibfnamefont {J.E.}~\bibnamefont
  {Moore}},\ }\href {\doibase 10.1016/j.aop.2010.12.011} {\bibfield
  {journal} {\bibinfo  {journal} {Ann. Phys.}\
  }\textbf {\bibinfo {volume} {326}},\ \bibinfo {pages} {1515--1535} (\bibinfo {year}
  {2011}{\natexlab{b}})}\BibitemShut {NoStop}%
\bibitem [{\citenamefont {Weiss}(2012)}]{weiss20121}%
  \BibitemOpen
  \bibfield  {author} {\bibinfo {author} {\bibfnamefont {U.}~\bibnamefont
  {Weiss}},\ }\href@noop {} {\emph {\bibinfo {title} {Quantum dissipative
  systems}}},\ Vol.~\bibinfo {volume} {13}\ (\bibinfo  {publisher} {World
  scientific},\ \bibinfo {year} {2012})\BibitemShut {NoStop}%
\bibitem [{\citenamefont {Caldeira}(2014)}]{caldeira20141}%
  \BibitemOpen
  \bibfield  {author} {\bibinfo {author} {\bibfnamefont {A.~O.}\ \bibnamefont
  {Caldeira}},\ }\href@noop {} {\emph {\bibinfo {title} {An Introduction to
  Macroscopic Quantum Phenomena and Quantum Dissipation}}}\ (\bibinfo
  {publisher} {Cambridge University Press},\ \bibinfo {year}
  {2014})\BibitemShut {NoStop}%
\bibitem [{\citenamefont {Breuer}\ and\ \citenamefont
  {Petruccione}()}]{breuer20021}%
  \BibitemOpen
  \bibfield  {author} {\bibinfo {author} {\bibfnamefont {H.-P.}\ \bibnamefont
  {Breuer}}\ and\ \bibinfo {author} {\bibfnamefont {y.~p.}\ \bibnamefont
  {Petruccione}, \bibfnamefont {Francesco}},\ }\href@noop {} {\emph {\bibinfo
  {title} {The theory of open quantum systems}}}\BibitemShut {NoStop}%
\bibitem [{\citenamefont {Yao}\ \emph {et~al.}(2017)\citenamefont {Yao},
  \citenamefont {Tang},\ and\ \citenamefont {Ao}}]{yao20171}%
  \BibitemOpen
  \bibfield  {author} {\bibinfo {author} {\bibfnamefont {Y.}~\bibnamefont
  {Yao}}, \bibinfo {author} {\bibfnamefont {Y.}~\bibnamefont {Tang}}, \ and\
  \bibinfo {author} {\bibfnamefont {P.}~\bibnamefont {Ao}},\ }\href {\doibase
  10.1103/PhysRevB.96.134414} {\bibfield  {journal} {\bibinfo  {journal}
  {Physical Review B}\ }\textbf {\bibinfo {volume} {96}} (\bibinfo {year}
  {2017}),\ 10.1103/PhysRevB.96.134414}\BibitemShut {NoStop}%
\bibitem [{\citenamefont {Campisi}\ \emph {et~al.}(2012)\citenamefont
  {Campisi}, \citenamefont {Denisov},\ and\ \citenamefont
  {H\"anggi}}]{campisi20121}%
  \BibitemOpen
  \bibfield  {author} {\bibinfo {author} {\bibfnamefont {M.}~\bibnamefont
  {Campisi}}, \bibinfo {author} {\bibfnamefont {S.}~\bibnamefont {Denisov}}, \
  and\ \bibinfo {author} {\bibfnamefont {P.}~\bibnamefont {H\"anggi}},\ }\href
  {\doibase 10.1103/PhysRevA.86.032114} {\bibfield  {journal} {\bibinfo
  {journal} {Phys. Rev. A}\ }\textbf {\bibinfo {volume} {86}},\ \bibinfo
  {pages} {032114} (\bibinfo {year} {2012})}\BibitemShut {NoStop}%
\bibitem [{\citenamefont {Valido}(2019)}]{valido20191}%
  \BibitemOpen
  \bibfield  {author} {\bibinfo {author} {\bibfnamefont {A.~A.}\ \bibnamefont
  {Valido}},\ }\href {\doibase 10.1103/PhysRevD.99.016003} {\bibfield
  {journal} {\bibinfo  {journal} {Phys. Rev. D}\ }\textbf {\bibinfo {volume}
  {99}},\ \bibinfo {pages} {016003} (\bibinfo {year} {2019})}\BibitemShut
  {NoStop}%
  \bibitem [{\citenamefont {Grabert}\ \emph {et~al.}(1984)\citenamefont
  {Grabert}, \citenamefont {Weiss},\ and\ \citenamefont
  {Talkner}}]{grabert19841}%
  \BibitemOpen
  \bibfield  {author} {\bibinfo {author} {\bibfnamefont {H.}~\bibnamefont
  {Grabert}}, \bibinfo {author} {\bibfnamefont {U.}~\bibnamefont {Weiss}}, \
  and\ \bibinfo {author} {\bibfnamefont {P.}~\bibnamefont {Talkner}},\ }\href
  {\doibase 10.1007/BF01307505} {\bibfield  {journal} {\bibinfo  {journal} {Z.
  Phys. B}\ }\textbf {\bibinfo {volume} {55}},\ \bibinfo {pages} {87} (\bibinfo
  {year} {1984})}\BibitemShut {NoStop}%
\bibitem [{\citenamefont {Haake}\ and\ \citenamefont
  {Reibold}(1985)}]{haake19851}%
  \BibitemOpen
  \bibfield  {author} {\bibinfo {author} {\bibfnamefont {F.}~\bibnamefont
  {Haake}}\ and\ \bibinfo {author} {\bibfnamefont {R.}~\bibnamefont
  {Reibold}},\ }\href {\doibase 10.1103/PhysRevA.32.2462} {\bibfield  {journal}
  {\bibinfo  {journal} {Phys. Rev. A}\ }\textbf {\bibinfo {volume} {32}},\
  \bibinfo {pages} {2462} (\bibinfo {year} {1985})}\BibitemShut {NoStop}%
\bibitem [{\citenamefont {H{\"{a}}nggi}\ and\ \citenamefont
  {Ingold}(2005)}]{hanggi20051}%
  \BibitemOpen
  \bibfield  {author} {\bibinfo {author} {\bibfnamefont {P.}~\bibnamefont
  {H{\"{a}}nggi}}\ and\ \bibinfo {author} {\bibfnamefont {G.-l.}\ \bibnamefont
  {Ingold}},\ }\href {\doibase 10.1063/1.1853631} {\bibfield  {journal}
  {\bibinfo  {journal} {Chaos}\ }\textbf {\bibinfo {volume} {15}},\ \bibinfo
  {pages} {026105} (\bibinfo {year} {2005})}\BibitemShut {NoStop}%
\bibitem [{\citenamefont {Feynman}\ and\ \citenamefont
  {Vernon~Jr}(1963)}]{feynman19631}%
  \BibitemOpen
  \bibfield  {author} {\bibinfo {author} {\bibfnamefont {R.~P.}\ \bibnamefont
  {Feynman}}\ and\ \bibinfo {author} {\bibfnamefont {F.}~\bibnamefont
  {Vernon~Jr}},\ }\href@noop {} {\bibfield  {journal} {\bibinfo  {journal}
  {Ann. Phys.}\ }\textbf {\bibinfo {volume} {24}},\ \bibinfo {pages} {118}
  (\bibinfo {year} {1963})}\BibitemShut {NoStop}%
\bibitem [{\citenamefont {Ford}\ \emph {et~al.}(1988)\citenamefont {Ford},
  \citenamefont {Lewis},\ and\ \citenamefont {O'Connell}}]{ford19881}%
  \BibitemOpen
  \bibfield  {author} {\bibinfo {author} {\bibfnamefont {W.}~\bibnamefont
  {Ford}}, \bibinfo {author} {\bibfnamefont {J.}~\bibnamefont {Lewis}}, \ and\
  \bibinfo {author} {\bibfnamefont {R.}~\bibnamefont {O'Connell}},\ }\href
  {\doibase 10.1103/PhysRevA.37.4419} {\bibfield  {journal} {\bibinfo
  {journal} {Phys. Rev. A}\ }\textbf {\bibinfo {volume} {37}},\ \bibinfo
  {pages} {4419} (\bibinfo {year} {1988})}\BibitemShut {NoStop}%
\bibitem [{\citenamefont {Valido}\ \emph
  {et~al.}(2013{\natexlab{a}})\citenamefont {Valido}, \citenamefont {Alonso},\
  and\ \citenamefont {Kohler}}]{valido20131}%
  \BibitemOpen
  \bibfield  {author} {\bibinfo {author} {\bibfnamefont {A.~A.}\ \bibnamefont
  {Valido}}, \bibinfo {author} {\bibfnamefont {D.}~\bibnamefont {Alonso}}, \
  and\ \bibinfo {author} {\bibfnamefont {S.}~\bibnamefont {Kohler}},\ }\href
  {\doibase 10.1103/PhysRevA.88.042303} {\bibfield  {journal} {\bibinfo
  {journal} {Phys. Rev. A}\ }\textbf {\bibinfo {volume} {88}} (\bibinfo {year}
  {2013}{\natexlab{a}}),\ 10.1103/PhysRevA.88.042303}\BibitemShut {NoStop}%
\bibitem [{\citenamefont {Caldeira}\ and\ \citenamefont
  {Leggett}(1983)}]{caldeira19831}%
  \BibitemOpen
  \bibfield  {author} {\bibinfo {author} {\bibfnamefont {A.}~\bibnamefont
  {Caldeira}}\ and\ \bibinfo {author} {\bibfnamefont {A.~J.}\ \bibnamefont
  {Leggett}},\ }\href@noop {} {\bibfield  {journal} {\bibinfo  {journal} {Ann.
  Phys.}\ }\textbf {\bibinfo {volume} {149}},\ \bibinfo {pages} {374} (\bibinfo
  {year} {1983})}\BibitemShut {NoStop}%
\bibitem [{\citenamefont {Grabert}\ \emph {et~al.}(1988)\citenamefont
  {Grabert}, \citenamefont {Schramm},\ and\ \citenamefont
  {Ingold}}]{grabert19881}%
  \BibitemOpen
  \bibfield  {author} {\bibinfo {author} {\bibfnamefont {H.}~\bibnamefont
  {Grabert}}, \bibinfo {author} {\bibfnamefont {P.}~\bibnamefont {Schramm}}, \
  and\ \bibinfo {author} {\bibfnamefont {G.~L.}\ \bibnamefont {Ingold}},\
  }\href {\doibase 10.1016/0370-1573(88)90023-3} {\bibfield  {journal}
  {\bibinfo  {journal} {Physics Reports}\ }\textbf {\bibinfo {volume} {168}},\
  \bibinfo {pages} {115} (\bibinfo {year} {1988})}\BibitemShut {NoStop}%
\bibitem [{\citenamefont {de~Vega}\ and\ \citenamefont
  {Alonso}(2017)}]{devega20171}%
  \BibitemOpen
  \bibfield  {author} {\bibinfo {author} {\bibfnamefont {I.}~\bibnamefont
  {de~Vega}}\ and\ \bibinfo {author} {\bibfnamefont {D.}~\bibnamefont
  {Alonso}},\ }\href {\doibase 10.1103/RevModPhys.89.015001} {\bibfield
  {journal} {\bibinfo  {journal} {Rev. Mod. Phys.}\ }\textbf {\bibinfo {volume}
  {89}},\ \bibinfo {pages} {015001} (\bibinfo {year} {2017})}\BibitemShut
  {NoStop}%
\bibitem [{\citenamefont {Valido}\ \emph
  {et~al.}(2013{\natexlab{b}})\citenamefont {Valido}, \citenamefont {Correa},\
  and\ \citenamefont {Alonso}}]{valido20132}%
  \BibitemOpen
  \bibfield  {author} {\bibinfo {author} {\bibfnamefont {A.~A.}\ \bibnamefont
  {Valido}}, \bibinfo {author} {\bibfnamefont {L.~A.}\ \bibnamefont {Correa}},
  \ and\ \bibinfo {author} {\bibfnamefont {D.}~\bibnamefont {Alonso}},\ }\href
  {\doibase 10.1103/PhysRevA.88.012309} {\bibfield  {journal} {\bibinfo
  {journal} {Phys. Rev. A}\ }\textbf {\bibinfo {volume} {88}},\ \bibinfo
  {pages} {012309} (\bibinfo {year} {2013}{\natexlab{b}})}\BibitemShut
  {NoStop}%
\bibitem [{\citenamefont {Roy}\ and\ \citenamefont {Kumar}(2008)}]{roy20081}%
  \BibitemOpen
  \bibfield  {author} {\bibinfo {author} {\bibfnamefont {D.}~\bibnamefont
  {Roy}}\ and\ \bibinfo {author} {\bibfnamefont {N.}~\bibnamefont {Kumar}},\
  }\href {\doibase 10.1103/PhysRevE.78.052102} {\bibfield  {journal} {\bibinfo
  {journal} {Phys. Rev. E}\ }\textbf {\bibinfo {volume} {78}},\ \bibinfo
  {pages} {052102} (\bibinfo {year} {2008})}\BibitemShut {NoStop}%
\bibitem [{\citenamefont {St\"ohr}(2006)}]{stohr20061}%
  \BibitemOpen
  \bibfield  {author} {\bibinfo {author} {\bibfnamefont {S.~H.}\ \bibnamefont
  {St\"ohr}, \bibfnamefont {J.}},\ }\href@noop {} {\emph {\bibinfo {title}
  {Magnetism: from fundamentals to nanoscale dynamics}}}\ (\bibinfo
  {publisher} {Springer-Verlag},\ \bibinfo {year} {2006})\BibitemShut {NoStop}%
\bibitem [{\citenamefont {Jayannavar}\ and\ \citenamefont
  {Kumar}(1981)}]{jayannavar19811}%
  \BibitemOpen
  \bibfield  {author} {\bibinfo {author} {\bibfnamefont {A.~M.}\ \bibnamefont
  {Jayannavar}}\ and\ \bibinfo {author} {\bibfnamefont {N.}~\bibnamefont
  {Kumar}},\ }\href {\doibase 10.1088/0305-4470/14/6/016} {\bibfield  {journal}
  {\bibinfo  {journal} {J. Phys.: A: Gen. Phys.}\ }\textbf {\bibinfo {volume}
  {14}},\ \bibinfo {pages} {1399} (\bibinfo {year} {1981})}\BibitemShut
  {NoStop}%
\bibitem [{\citenamefont {Dattagupta}\ and\ \citenamefont
  {Singh}(1997)}]{dattagupta19971}%
  \BibitemOpen
  \bibfield  {author} {\bibinfo {author} {\bibfnamefont {S.}~\bibnamefont
  {Dattagupta}}\ and\ \bibinfo {author} {\bibfnamefont {J.}~\bibnamefont
  {Singh}},\ }\href {\doibase 10.1103/PhysRevLett.79.961} {\bibfield  {journal}
  {\bibinfo  {journal} {Phys. Rev. Lett.}\ }\textbf {\bibinfo {volume} {79}},\
  \bibinfo {pages} {961} (\bibinfo {year} {1997})},\ \Eprint
  {http://arxiv.org/abs/9604005} {arXiv:9604005 [quant-ph]} \BibitemShut
  {NoStop}%
\bibitem [{\citenamefont {Kumar}\ \emph {et~al.}(2009)\citenamefont {Kumar},
  \citenamefont {Sreeram},\ and\ \citenamefont {Dattagupta}}]{kumar20091}%
  \BibitemOpen
  \bibfield  {author} {\bibinfo {author} {\bibfnamefont {J.}~\bibnamefont
  {Kumar}}, \bibinfo {author} {\bibfnamefont {P.~A.}\ \bibnamefont {Sreeram}},
  \ and\ \bibinfo {author} {\bibfnamefont {S.}~\bibnamefont {Dattagupta}},\
  }\href {\doibase 10.1103/PhysRevE.79.021130} {\bibfield  {journal} {\bibinfo
  {journal} {Phys. Rev. E}\ }\textbf {\bibinfo {volume} {79}},\ \bibinfo
  {pages} {021130} (\bibinfo {year} {2009})},\ \Eprint
  {http://arxiv.org/abs/arXiv:0901.1502v2} {arXiv:arXiv:0901.1502v2}
  \BibitemShut {NoStop}%
\bibitem [{\citenamefont {Bandyopadhyay}\ and\ \citenamefont
  {Dattagupta}(2006{\natexlab{a}})}]{bandyopadhyay20061}%
  \BibitemOpen
  \bibfield  {author} {\bibinfo {author} {\bibfnamefont {M.}~\bibnamefont
  {Bandyopadhyay}}\ and\ \bibinfo {author} {\bibfnamefont {S.}~\bibnamefont
  {Dattagupta}},\ }\href {\doibase 10.1007/s10955-006-9114-y} {\bibfield
  {journal} {\bibinfo  {journal} {J. Stat. Phys.}\ }\textbf {\bibinfo {volume}
  {123}},\ \bibinfo {pages} {1273} (\bibinfo {year}
  {2006}{\natexlab{a}})}\BibitemShut {NoStop}%
\bibitem [{\citenamefont {Bandyopadhyay}(2009)}]{bandyopadhyay20091}%
  \BibitemOpen
  \bibfield  {author} {\bibinfo {author} {\bibfnamefont {M.}~\bibnamefont
  {Bandyopadhyay}},\ }\href {\doibase 10.1088/1742-5468/2009/05/P05002}
  {\bibfield  {journal} {\bibinfo  {journal} {J. Stats. Mech.}\ }\textbf
  {\bibinfo {volume} {5}},\ \bibinfo {pages} {P05002} (\bibinfo {year}
  {2009})},\ \Eprint {http://arxiv.org/abs/0906.1332} {arXiv:0906.1332}
  \BibitemShut {NoStop}%
\bibitem [{\citenamefont {Bandyopadhyay}\ and\ \citenamefont
  {Dattagupta}(2006{\natexlab{b}})}]{bandyopadhyay20062}%
  \BibitemOpen
  \bibfield  {author} {\bibinfo {author} {\bibfnamefont {M.}~\bibnamefont
  {Bandyopadhyay}}\ and\ \bibinfo {author} {\bibfnamefont {S.}~\bibnamefont
  {Dattagupta}},\ }\href {\doibase 10.1088/0953-8984/18/44/004} {\bibfield
  {journal} {\bibinfo  {journal} {Journal of Physics Condensed Matter}\
  }\textbf {\bibinfo {volume} {18}},\ \bibinfo {pages} {10029} (\bibinfo {year}
  {2006}{\natexlab{b}})}\BibitemShut {NoStop}%
\bibitem [{\citenamefont {Hong}\ and\ \citenamefont
  {Wheatley}(1991)}]{hong19911}%
  \BibitemOpen
  \bibfield  {author} {\bibinfo {author} {\bibfnamefont {T.}~\bibnamefont
  {Hong}}\ and\ \bibinfo {author} {\bibfnamefont {J.}~\bibnamefont
  {Wheatley}},\ }\href {\doibase 10.1103/PhysRevB.43.5702} {\bibfield
  {journal} {\bibinfo  {journal} {Phys. Rev. B}\ }\textbf {\bibinfo {volume}
  {43}},\ \bibinfo {pages} {5702} (\bibinfo {year} {1991})}\BibitemShut
  {NoStop}%
\bibitem [{\citenamefont {Li}\ \emph {et~al.}(1990)\citenamefont {Li},
  \citenamefont {Ford},\ and\ \citenamefont {O'Connell}}]{li19901}%
  \BibitemOpen
  \bibfield  {author} {\bibinfo {author} {\bibfnamefont {X.~L.}\ \bibnamefont
  {Li}}, \bibinfo {author} {\bibfnamefont {G.}~\bibnamefont {Ford}}, \ and\
  \bibinfo {author} {\bibfnamefont {R.~F.}\ \bibnamefont {O'Connell}},\ }\href
  {\doibase 10.1103/PhysRevA.42.4519} {\bibfield  {journal} {\bibinfo
  {journal} {Phys. Rev. A}\ }\textbf {\bibinfo {volume} {42}},\ \bibinfo
  {pages} {4519} (\bibinfo {year} {1990})}\BibitemShut {NoStop}%
\bibitem [{\citenamefont {Kumar}(2014)}]{kumar20141}%
  \BibitemOpen
  \bibfield  {author} {\bibinfo {author} {\bibfnamefont {J.}~\bibnamefont
  {Kumar}},\ }\href {\doibase 10.1002/andp.201400061} {\bibfield  {journal}
  {\bibinfo  {journal} {Annalen der Physik}\ }\textbf {\bibinfo {volume}
  {513}},\ \bibinfo {pages} {499} (\bibinfo {year} {2014})}\BibitemShut
  {NoStop}%
\bibitem [{\citenamefont {Wan}\ and\ \citenamefont {Saglam}(2006)}]{wan20061}%
  \BibitemOpen
  \bibfield  {author} {\bibinfo {author} {\bibfnamefont {K.~K.}\ \bibnamefont
  {Wan}}\ and\ \bibinfo {author} {\bibfnamefont {M.}~\bibnamefont {Saglam}},\
  }\href {\doibase 10.1007/s10773-006-9118-z} {\bibfield  {journal} {\bibinfo
  {journal} {Int. J. Theor. Phycs.}\ }\textbf {\bibinfo {volume} {45}},\
  \bibinfo {pages} {1171} (\bibinfo {year} {2006})}\BibitemShut {NoStop}%
\bibitem [{\citenamefont {Savoie}(2015)}]{savoie20151}%
  \BibitemOpen
  \bibfield  {author} {\bibinfo {author} {\bibfnamefont {B.}~\bibnamefont
  {Savoie}},\ }\href {\doibase 10.1142/S0129055X15500191} {\bibfield  {journal}
  {\bibinfo  {journal} {Rev. Math. Phys.}\ }\textbf {\bibinfo {volume} {27}},\
  \bibinfo {pages} {1550019} (\bibinfo {year} {2015})},\ \Eprint
  {http://arxiv.org/abs/arXiv:1403.2834v3} {arXiv:arXiv:1403.2834v3}
  \BibitemShut {NoStop}%
\bibitem [{\citenamefont {Pradhan}\ and\ \citenamefont
  {Seifert}(2010)}]{pradhan20102}%
  \BibitemOpen
  \bibfield  {author} {\bibinfo {author} {\bibfnamefont {P.}~\bibnamefont
  {Pradhan}}\ and\ \bibinfo {author} {\bibfnamefont {U.}~\bibnamefont
  {Seifert}},\ }\href {\doibase 10.1209/0295-5075/89/37001} {\bibfield
  {journal} {\bibinfo  {journal} {Europhys. Lett.}\ }\textbf {\bibinfo {volume}
  {89}},\ \bibinfo {pages} {37001} (\bibinfo {year} {2010})},\ \Eprint
  {http://arxiv.org/abs/0912.4697} {arXiv:0912.4697} \BibitemShut {NoStop}%
\bibitem [{\citenamefont {Kaplan}\ and\ \citenamefont
  {Mahanti}(2009)}]{kaplan20091}%
  \BibitemOpen
  \bibfield  {author} {\bibinfo {author} {\bibfnamefont {T.~A.}\ \bibnamefont
  {Kaplan}}\ and\ \bibinfo {author} {\bibfnamefont {S.~D.}\ \bibnamefont
  {Mahanti}},\ }\href {\doibase 10.1209/0295-5075/87/17002} {\bibfield
  {journal} {\bibinfo  {journal} {Europhys. Lett.}\ }\textbf {\bibinfo {volume}
  {87}},\ \bibinfo {pages} {17002} (\bibinfo {year} {2009})}\BibitemShut
  {NoStop}%
\bibitem [{\citenamefont {Kohler}\ and\ \citenamefont
  {Sols}(2013)}]{kohler20131}%
  \BibitemOpen
  \bibfield  {author} {\bibinfo {author} {\bibfnamefont {H.}~\bibnamefont
  {Kohler}}\ and\ \bibinfo {author} {\bibfnamefont {F.}~\bibnamefont {Sols}},\
  }\href {\doibase 10.1016/j.physa.2013.01.019} {\bibfield  {journal} {\bibinfo
   {journal} {Physica A}\ }\textbf {\bibinfo {volume} {392}},\ \bibinfo {pages}
  {1089} (\bibinfo {year} {2013})}\BibitemShut {NoStop}%
\bibitem [{\citenamefont {Landau}\ and\ \citenamefont
  {Lifshitz}(1971)}]{landau19711}%
  \BibitemOpen
  \bibfield  {author} {\bibinfo {author} {\bibfnamefont {L.~D.}\ \bibnamefont
  {Landau}}\ and\ \bibinfo {author} {\bibfnamefont {E.~M.}\ \bibnamefont
  {Lifshitz}},\ }\href@noop {} {\emph {\bibinfo {title} {The classical theory
  of fields}}}\ (\bibinfo  {publisher} {Pergamon},\ \bibinfo {year}
  {1971})\BibitemShut {NoStop}%
\bibitem [{\citenamefont {Kobe}(1983)}]{kobe19831}%
  \BibitemOpen
  \bibfield  {author} {\bibinfo {author} {\bibfnamefont {D.~H.}\ \bibnamefont
  {Kobe}},\ }\href {\doibase 10.1088/0305-4470/16/4/012} {\bibfield  {journal}
  {\bibinfo  {journal} {J. Phys. A: Gen. Phys.}\ }\textbf {\bibinfo {volume}
  {16}},\ \bibinfo {pages} {737} (\bibinfo {year} {1983})}\BibitemShut
  {NoStop}%
\bibitem [{\citenamefont {Saha}\ and\ \citenamefont
  {Jayannavar}(2008)}]{saha20081}%
  \BibitemOpen
  \bibfield  {author} {\bibinfo {author} {\bibfnamefont {A.}~\bibnamefont
  {Saha}}\ and\ \bibinfo {author} {\bibfnamefont {A.~M.}\ \bibnamefont
  {Jayannavar}},\ }\href {\doibase 10.1103/PhysRevE.77.022105} {\bibfield
  {journal} {\bibinfo  {journal} {Phys. Rev. E}\ }\textbf {\bibinfo {volume}
  {77}},\ \bibinfo {pages} {022105} (\bibinfo {year} {2008})},\ \Eprint
  {http://arxiv.org/abs/arXiv:0707.2131v1} {arXiv:arXiv:0707.2131v1}
  \BibitemShut {NoStop}%
\bibitem [{\citenamefont {Ao}\ and\ \citenamefont {Zuh}(1999)}]{ao19991}%
  \BibitemOpen
  \bibfield  {author} {\bibinfo {author} {\bibfnamefont {P.}~\bibnamefont
  {Ao}}\ and\ \bibinfo {author} {\bibfnamefont {X.-M.}\ \bibnamefont {Zuh}},\
  }\href {\doibase 10.1103/PhysRevB.60.6850} {\bibfield  {journal} {\bibinfo
  {journal} {Phys Rev. B}\ }\textbf {\bibinfo {volume} {60}},\ \bibinfo {pages}
  {6850} (\bibinfo {year} {1999})},\ \Eprint {http://arxiv.org/abs/9902333v2}
  {arXiv:9902333v2 [arXiv:cond-mat]} \BibitemShut {NoStop}%
\bibitem [{\citenamefont {Gumber}\ \emph {et~al.}(2018)\citenamefont {Gumber},
  \citenamefont {Bhattacherjee},\ and\ \citenamefont {Jha}}]{gumber20181}%
  \BibitemOpen
  \bibfield  {author} {\bibinfo {author} {\bibfnamefont {S.}~\bibnamefont
  {Gumber}}, \bibinfo {author} {\bibfnamefont {A.~B.}\ \bibnamefont
  {Bhattacherjee}}, \ and\ \bibinfo {author} {\bibfnamefont {P.~K.}\
  \bibnamefont {Jha}},\ }\href {\doibase 10.1103/PhysRevB.98.205408} {\bibfield
   {journal} {\bibinfo  {journal} {Phys. Rev. B}\ }\textbf {\bibinfo {volume}
  {98}},\ \bibinfo {pages} {205408} (\bibinfo {year} {2018})}\BibitemShut
  {NoStop}%
\bibitem [{\citenamefont {Kobe}\ and\ \citenamefont {Wen}(1982)}]{kobe19821}%
  \BibitemOpen
  \bibfield  {author} {\bibinfo {author} {\bibfnamefont {D.~H.}\ \bibnamefont
  {Kobe}}\ and\ \bibinfo {author} {\bibfnamefont {E.~C.~T.}\ \bibnamefont
  {Wen}},\ }\href {\doibase 10.1088/0305-4470/15/3/018} {\bibfield  {journal}
  {\bibinfo  {journal} {J. Phys. A: Math. Gen.}\ }\textbf {\bibinfo {volume}
  {15}},\ \bibinfo {pages} {787} (\bibinfo {year} {1982})}\BibitemShut
  {NoStop}%
\bibitem [{\citenamefont {Asorey}\ \emph {et~al.}(1983)\citenamefont {Asorey},
  \citenamefont {Esteve},\ and\ \citenamefont {Pacheco}}]{asorey19831}%
  \BibitemOpen
  \bibfield  {author} {\bibinfo {author} {\bibfnamefont {M.}~\bibnamefont
  {Asorey}}, \bibinfo {author} {\bibfnamefont {J.~G.}\ \bibnamefont {Esteve}},
  \ and\ \bibinfo {author} {\bibfnamefont {A.~F.}\ \bibnamefont {Pacheco}},\
  }\href {\doibase 10.1103/PhysRevD.27.1852} {\bibfield  {journal} {\bibinfo
  {journal} {Phys. Rev. D}\ }\textbf {\bibinfo {volume} {27}},\ \bibinfo
  {pages} {1852} (\bibinfo {year} {1983})}\BibitemShut {NoStop}%
\bibitem [{\citenamefont {Caio}\ \emph {et~al.}(2016)\citenamefont {Caio},
  \citenamefont {Cooper},\ and\ \citenamefont {Bhaseen}}]{caio20161}%
  \BibitemOpen
  \bibfield  {author} {\bibinfo {author} {\bibfnamefont {M.~D.}\ \bibnamefont
  {Caio}}, \bibinfo {author} {\bibfnamefont {N.~R.}\ \bibnamefont {Cooper}}, \
  and\ \bibinfo {author} {\bibfnamefont {M.~J.}\ \bibnamefont {Bhaseen}},\
  }\href {\doibase 10.1103/PhysRevB.94.155104} {\bibfield  {journal} {\bibinfo
  {journal} {Phys. Rev. B}\ }\textbf {\bibinfo {volume} {94}},\ \bibinfo
  {pages} {155104} (\bibinfo {year} {2016})}\BibitemShut {NoStop}%
\bibitem [{\citenamefont {Kardar}(2007)}]{kardar20071}%
  \BibitemOpen
  \bibfield  {author} {\bibinfo {author} {\bibfnamefont {M.}~\bibnamefont
  {Kardar}},\ }\href@noop {} {\emph {\bibinfo {title} {Statistical physics of
  particles}}}\ (\bibinfo  {publisher} {Cambridge University Press},\ \bibinfo
  {year} {2007})\BibitemShut {NoStop}%
\bibitem [{\citenamefont {Callen}(1987)}]{callen19871}%
  \BibitemOpen
  \bibfield  {author} {\bibinfo {author} {\bibfnamefont {H.~B.}\ \bibnamefont
  {Callen}},\ }\href@noop {} {\emph {\bibinfo {title} {Thermodynamics and
  introduction to thermostatistics}}}\ (\bibinfo  {publisher} {John Wiley \&
  Sons},\ \bibinfo {year} {1987})\BibitemShut {NoStop}%
\bibitem [{\citenamefont {Baxter}(1996)}]{baxter19961}%
  \BibitemOpen
  \bibfield  {author} {\bibinfo {author} {\bibfnamefont {C.}~\bibnamefont
  {Baxter}},\ }\href {\doibase 10.1103/PhysRevLett.74.514} {\bibfield
  {journal} {\bibinfo  {journal} {Phys. Rev. Lett.}\ }\textbf {\bibinfo
  {volume} {65}},\ \bibinfo {pages} {256} (\bibinfo {year} {1996})}\BibitemShut
  {NoStop}%
\bibitem [{\citenamefont {Zhang}(1996)}]{zhang19961}%
  \BibitemOpen
  \bibfield  {author} {\bibinfo {author} {\bibfnamefont {J.~Z.}\ \bibnamefont
  {Zhang}},\ }\href {\doibase 10.1103/PhysRevLett.77.44} {\bibfield  {journal}
  {\bibinfo  {journal} {Phys. Rev. Lett.}\ }\textbf {\bibinfo {volume} {77}},\
  \bibinfo {pages} {44} (\bibinfo {year} {1996})}\BibitemShut {NoStop}%
\bibitem [{\citenamefont {Zhang}(2006)}]{zhang20061}%
  \BibitemOpen
  \bibfield  {author} {\bibinfo {author} {\bibfnamefont {J.~Z.}\ \bibnamefont
  {Zhang}},\ }\href {\doibase 10.1103/PhysRevD.74.124005} {\bibfield  {journal}
  {\bibinfo  {journal} {Phys. Rev. D}\ }\textbf {\bibinfo {volume} {74}},\
  \bibinfo {pages} {124005} (\bibinfo {year} {2006})}\BibitemShut {NoStop}%
\bibitem [{\citenamefont {Zhang}(2004)}]{zhang20041}%
  \BibitemOpen
  \bibfield  {author} {\bibinfo {author} {\bibfnamefont {J.~Z.}\ \bibnamefont
  {Zhang}},\ }\href {\doibase 10.1103/PhysRevLett.93.043002} {\bibfield
  {journal} {\bibinfo  {journal} {Phys. Rev. Lett.}\ }\textbf {\bibinfo
  {volume} {93}},\ \bibinfo {pages} {043002} (\bibinfo {year}
  {2004})}\BibitemShut {NoStop}%
\bibitem [{\citenamefont {Saffman}\ \emph {et~al.}(2010)\citenamefont
  {Saffman}, \citenamefont {Walker},\ and\ \citenamefont
  {M{\o}lmer}}]{saffman20101}%
  \BibitemOpen
  \bibfield  {author} {\bibinfo {author} {\bibfnamefont {M.}~\bibnamefont
  {Saffman}}, \bibinfo {author} {\bibfnamefont {T.~G.}\ \bibnamefont {Walker}},
  \ and\ \bibinfo {author} {\bibfnamefont {K.}~\bibnamefont {M{\o}lmer}},\
  }\href {\doibase 10.1103/RevModPhys.82.2313} {\bibfield  {journal} {\bibinfo
  {journal} {Rev. Mod. Phys.}\ }\textbf {\bibinfo {volume} {82}},\ \bibinfo
  {pages} {2313} (\bibinfo {year} {2010})}\BibitemShut {NoStop}%
\bibitem [{\citenamefont {Marino}(1993)}]{marino19931}%
  \BibitemOpen
  \bibfield  {author} {\bibinfo {author} {\bibfnamefont {E.~C.}\ \bibnamefont
  {Marino}},\ }\href {\doibase 10.1016/0550-3213(93)90379-4} {\bibfield
  {journal} {\bibinfo  {journal} {Nuclear Physics, Section B}\ }\textbf
  {\bibinfo {volume} {408}},\ \bibinfo {pages} {551} (\bibinfo {year}
  {1993})}\BibitemShut {NoStop}%
\bibitem [{\citenamefont {Fradkin}\ and\ \citenamefont
  {Schaposnik}(1994)}]{fradldn19941}%
  \BibitemOpen
  \bibfield  {author} {\bibinfo {author} {\bibfnamefont {E.}~\bibnamefont
  {Fradkin}}\ and\ \bibinfo {author} {\bibfnamefont {F.~A.}\ \bibnamefont
  {Schaposnik}},\ }\href {\doibase 10.1016/0370-2693(94)91374-9} {\bibfield
  {journal} {\bibinfo  {journal} {Phys. Lett. B}\ }\textbf {\bibinfo {volume}
  {338}},\ \bibinfo {pages} {253} (\bibinfo {year} {1994})}\BibitemShut
  {NoStop}%
\bibitem [{\citenamefont {H{\"{a}}nggi}\ \emph {et~al.}(2008)\citenamefont
  {H{\"{a}}nggi}, \citenamefont {Ingold},\ and\ \citenamefont
  {Talkner}}]{hanggi20081}%
  \BibitemOpen
  \bibfield  {author} {\bibinfo {author} {\bibfnamefont {P.}~\bibnamefont
  {H{\"{a}}nggi}}, \bibinfo {author} {\bibfnamefont {G.-L.}\ \bibnamefont
  {Ingold}}, \ and\ \bibinfo {author} {\bibfnamefont {P.}~\bibnamefont
  {Talkner}},\ }\href {\doibase 10.1088/1367-2630/10/11/115008} {\bibfield
  {journal} {\bibinfo  {journal} {New J. Phys.}\ }\textbf {\bibinfo {volume}
  {10}},\ \bibinfo {pages} {115008} (\bibinfo {year} {2008})}\BibitemShut
  {NoStop}%
\bibitem [{\citenamefont {Ingold}\ \emph {et~al.}(2009)\citenamefont {Ingold},
  \citenamefont {H{\"{a}}nggi},\ and\ \citenamefont {Talkner}}]{ingold20091}%
  \BibitemOpen
  \bibfield  {author} {\bibinfo {author} {\bibfnamefont {G.-l.}\ \bibnamefont
  {Ingold}}, \bibinfo {author} {\bibfnamefont {P.}~\bibnamefont
  {H{\"{a}}nggi}}, \ and\ \bibinfo {author} {\bibfnamefont {P.}~\bibnamefont
  {Talkner}},\ }\href {\doibase 10.1103/PhysRevE.79.061105} {\bibfield
  {journal} {\bibinfo  {journal} {Phys. Rev. E}\ }\textbf {\bibinfo {volume}
  {79}},\ \bibinfo {pages} {061105} (\bibinfo {year} {2009})}\BibitemShut
  {NoStop}%
\bibitem [{\citenamefont {{G.-L. Ingold}}(2012)}]{ingold20121}%
  \BibitemOpen
  \bibfield  {author} {\bibinfo {author} {\bibnamefont {{G.-L. Ingold}}},\
  }\href {\doibase 10.1140/epjb/e2011-20930-2} {\bibfield  {journal} {\bibinfo
  {journal} {Eur. Phys. J. B}\ }\textbf {\bibinfo {volume} {85}},\ \bibinfo
  {pages} {30} (\bibinfo {year} {2012})}\BibitemShut {NoStop}%
\bibitem [{\citenamefont {{Peter Hanggi}}\ and\ \citenamefont {{Gert-Ludwig
  Ingold}}(2006)}]{hanggi20061}%
  \BibitemOpen
  \bibfield  {author} {\bibinfo {author} {\bibnamefont {{Peter Hanggi}}}\ and\
  \bibinfo {author} {\bibnamefont {{Gert-Ludwig Ingold}}},\ }\href
  {http://www.springerlink.com/index/10.1140/epjb/e2007-00013-y} {\bibfield
  {journal} {\bibinfo  {journal} {Acta Phys. Pol. B}\ }\textbf {\bibinfo
  {volume} {37}},\ \bibinfo {pages} {1537} (\bibinfo {year} {2006})},\ \Eprint
  {http://arxiv.org/abs/0601056} {arXiv:0601056 [quant-ph]} \BibitemShut
  {NoStop}%
\bibitem [{\citenamefont {Ford}\ and\ \citenamefont {R.F.}(2007)}]{ford20071}%
  \BibitemOpen
  \bibfield  {author} {\bibinfo {author} {\bibfnamefont {G.~W.}\ \bibnamefont
  {Ford}}\ and\ \bibinfo {author} {\bibfnamefont {O.}~\bibnamefont {R.F.}},\
  }\href {\doibase 10.1103/PhysRevB.75.134301} {\bibfield  {journal} {\bibinfo
  {journal} {Phys. Rev. B}\ }\textbf {\bibinfo {volume} {75}},\ \bibinfo
  {pages} {134301} (\bibinfo {year} {2007})}\BibitemShut {NoStop}%
\bibitem [{\citenamefont {Bandyopadhyay}\ and\ \citenamefont
  {Dattagupta}(2010)}]{bandyopadhyay20101}%
  \BibitemOpen
  \bibfield  {author} {\bibinfo {author} {\bibfnamefont {M.}~\bibnamefont
  {Bandyopadhyay}}\ and\ \bibinfo {author} {\bibfnamefont {S.}~\bibnamefont
  {Dattagupta}},\ }\href {\doibase 10.1103/PhysRevE.81.042102} {\bibfield
  {journal} {\bibinfo  {journal} {Phys. Rev. E}\ }\textbf {\bibinfo {volume}
  {81}},\ \bibinfo {pages} {042102} (\bibinfo {year} {2010})}\BibitemShut
  {NoStop}%
\bibitem [{\citenamefont {Bandyopadhyay}(2010)}]{bandyopadhyay20102}%
  \BibitemOpen
  \bibfield  {author} {\bibinfo {author} {\bibfnamefont {M.}~\bibnamefont
  {Bandyopadhyay}},\ }\href {\doibase 10.1007/s10955-010-9998-4} {\bibfield
  {journal} {\bibinfo  {journal} {J. Stat Phys.}\ }\textbf {\bibinfo {volume}
  {140}},\ \bibinfo {pages} {603} (\bibinfo {year} {2010})}\BibitemShut
  {NoStop}%
\bibitem [{\citenamefont {Hilt}\ \emph {et~al.}(2011)\citenamefont {Hilt},
  \citenamefont {Thomas},\ and\ \citenamefont {Lutz}}]{hilt20111}%
  \BibitemOpen
  \bibfield  {author} {\bibinfo {author} {\bibfnamefont {S.}~\bibnamefont
  {Hilt}}, \bibinfo {author} {\bibfnamefont {B.}~\bibnamefont {Thomas}}, \ and\
  \bibinfo {author} {\bibfnamefont {E.}~\bibnamefont {Lutz}},\ }\href {\doibase
  10.1103/PhysRevE.84.031110} {\bibfield  {journal} {\bibinfo  {journal} {Phys.
  Rev. E}\ }\textbf {\bibinfo {volume} {84}},\ \bibinfo {pages} {031110}
  (\bibinfo {year} {2011})},\ \Eprint {http://arxiv.org/abs/1106.1775}
  {arXiv:1106.1775} \BibitemShut {NoStop}%
\bibitem [{\citenamefont {Ingold}(2002)}]{ingold20021}%
  \BibitemOpen
  \bibfield  {author} {\bibinfo {author} {\bibfnamefont {G.}~\bibnamefont
  {Ingold}},\ }\href@noop {} {\emph {\bibinfo {title} {Springer Lecture Notes
  in Physics vol 611 pp 1-53}}}\ (\bibinfo  {publisher} {Berlin: Springer},\
  \bibinfo {year} {2002})\BibitemShut {NoStop}%
\bibitem [{\citenamefont {Zinn-Justin}(2010)}]{zinn20101}%
  \BibitemOpen
  \bibfield  {author} {\bibinfo {author} {\bibfnamefont {J.}~\bibnamefont
  {Zinn-Justin}},\ }\href@noop {} {\emph {\bibinfo {title} {Path Integrals in
  Quantum Mechanic}}}\ (\bibinfo  {publisher} {Oxford University Press},\
  \bibinfo {year} {2010})\BibitemShut {NoStop}%
\bibitem [{\citenamefont {Paladino}\ \emph {et~al.}(2014)\citenamefont
  {Paladino}, \citenamefont {Galperin}, \citenamefont {Falci},\ and\
  \citenamefont {Altshuler}}]{paladino20141}%
  \BibitemOpen
  \bibfield  {author} {\bibinfo {author} {\bibfnamefont {E.}~\bibnamefont
  {Paladino}}, \bibinfo {author} {\bibfnamefont {Y.~M.}\ \bibnamefont
  {Galperin}}, \bibinfo {author} {\bibfnamefont {G.}~\bibnamefont {Falci}}, \
  and\ \bibinfo {author} {\bibfnamefont {B.~L.}\ \bibnamefont {Altshuler}},\
  }\href {\doibase 10.1103/RevModPhys.86.361} {\bibfield  {journal} {\bibinfo
  {journal} {Rev. Mod. Phys.}\ }\textbf {\bibinfo {volume} {86}},\ \bibinfo
  {pages} {361} (\bibinfo {year} {2014})},\ \Eprint
  {http://arxiv.org/abs/1304.7925} {arXiv:1304.7925} \BibitemShut {NoStop}%
\bibitem [{\citenamefont {Abramowitz}\ and\ \citenamefont
  {Stegun}(1964)}]{abramowitz19641}%
  \BibitemOpen
  \bibfield  {author} {\bibinfo {author} {\bibfnamefont {M.}~\bibnamefont
  {Abramowitz}}\ and\ \bibinfo {author} {\bibfnamefont {I.~A.}\ \bibnamefont
  {Stegun}},\ }\href@noop {} {\emph {\bibinfo {title} {Handbook of mathematical
  functions: with formulas, graphs, and mathematical tables}}}\ (\bibinfo
  {publisher} {Courier Corporation},\ \bibinfo {year} {1964})\BibitemShut
  {NoStop}%
\bibitem [{\citenamefont {Alamoudi}\ \emph {et~al.}(1999)\citenamefont
  {Alamoudi}, \citenamefont {Boyanovsky},\ and\ \citenamefont
  {Vega}}]{alamoudi19991}%
  \BibitemOpen
  \bibfield  {author} {\bibinfo {author} {\bibfnamefont {S.~M.}\ \bibnamefont
  {Alamoudi}}, \bibinfo {author} {\bibfnamefont {D.}~\bibnamefont
  {Boyanovsky}}, \ and\ \bibinfo {author} {\bibfnamefont {H.~J.~D.}\
  \bibnamefont {Vega}},\ }\href {\doibase 10.1103/PhysRevE.60.94} {\bibfield
  {journal} {\bibinfo  {journal} {Phys. Rev. E}\ }\textbf {\bibinfo {volume}
  {60}},\ \bibinfo {pages} {94} (\bibinfo {year} {1999})}\BibitemShut {NoStop}%
\bibitem [{\citenamefont {Alamoudi}\ \emph {et~al.}(1998)\citenamefont
  {Alamoudi}, \citenamefont {Boyanovsky}, \citenamefont {Vega},\ and\
  \citenamefont {Holman}}]{alamoudi19981}%
  \BibitemOpen
  \bibfield  {author} {\bibinfo {author} {\bibfnamefont {S.~M.}\ \bibnamefont
  {Alamoudi}}, \bibinfo {author} {\bibfnamefont {D.}~\bibnamefont
  {Boyanovsky}}, \bibinfo {author} {\bibfnamefont {H.~J.~D.}\ \bibnamefont
  {Vega}}, \ and\ \bibinfo {author} {\bibfnamefont {R.}~\bibnamefont
  {Holman}},\ }\href {\doibase 10.1103/PhysRevD.59.025003} {\bibfield
  {journal} {\bibinfo  {journal} {Phys. Rev. D}\ }\textbf {\bibinfo {volume}
  {59}},\ \bibinfo {pages} {025003} (\bibinfo {year} {1998})}\BibitemShut
  {NoStop}%
\bibitem [{\citenamefont {Anisimov}\ \emph {et~al.}(2009)\citenamefont
  {Anisimov}, \citenamefont {Buchm{\"{u}}ller}, \citenamefont {Drewes},\ and\
  \citenamefont {Mendizabal}}]{anisimov20091}%
  \BibitemOpen
  \bibfield  {author} {\bibinfo {author} {\bibfnamefont {A.}~\bibnamefont
  {Anisimov}}, \bibinfo {author} {\bibfnamefont {W.}~\bibnamefont
  {Buchm{\"{u}}ller}}, \bibinfo {author} {\bibfnamefont {M.}~\bibnamefont
  {Drewes}}, \ and\ \bibinfo {author} {\bibfnamefont {S.}~\bibnamefont
  {Mendizabal}},\ }\href {\doibase 10.1016/j.aop.2009.01.001} {\bibfield
  {journal} {\bibinfo  {journal} {Ann. Phys.}\ }\textbf {\bibinfo {volume}
  {324}},\ \bibinfo {pages} {1234} (\bibinfo {year} {2009})}\BibitemShut
  {NoStop}%
\bibitem [{\citenamefont {Nakagawa}\ \emph {et~al.}(2016)\citenamefont
  {Nakagawa}, \citenamefont {Misguich},\ and\ \citenamefont
  {Oshikawa}}]{nakagawa20161}%
  \BibitemOpen
  \bibfield  {author} {\bibinfo {author} {\bibfnamefont {Y.~O.}\ \bibnamefont
  {Nakagawa}}, \bibinfo {author} {\bibfnamefont {G.}~\bibnamefont {Misguich}},
  \ and\ \bibinfo {author} {\bibfnamefont {M.}~\bibnamefont {Oshikawa}},\
  }\href {\doibase 10.1103/PhysRevB.93.174310} {\bibfield  {journal} {\bibinfo
  {journal} {Phys. Rev. B}\ }\textbf {\bibinfo {volume} {93}},\ \bibinfo
  {pages} {174310} (\bibinfo {year} {2016})}\BibitemShut {NoStop}%
\bibitem [{\citenamefont {Jarzynski}(2011)}]{jarzynski20111}%
  \BibitemOpen
  \bibfield  {author} {\bibinfo {author} {\bibfnamefont {C.}~\bibnamefont
  {Jarzynski}},\ }\href {\doibase 10.1146/annurev-conmatphys-062910-140506}
  {\bibfield  {journal} {\bibinfo  {journal} {Annu. Rev. Condens. Matter
  Phys.}\ }\textbf {\bibinfo {volume} {2}},\ \bibinfo {pages} {329} (\bibinfo
  {year} {2011})}\BibitemShut {NoStop}%
\bibitem [{\citenamefont {Ford}\ and\ \citenamefont
  {O'Connell}(2006)}]{ford20061}%
  \BibitemOpen
  \bibfield  {author} {\bibinfo {author} {\bibfnamefont {G.~W.}\ \bibnamefont
  {Ford}}\ and\ \bibinfo {author} {\bibfnamefont {R.~F.}\ \bibnamefont
  {O'Connell}},\ }\href {\doibase 10.1103/PhysRevLett.96.020402} {\bibfield
  {journal} {\bibinfo  {journal} {Phys. Rev. Lett.}\ }\textbf {\bibinfo
  {volume} {96}},\ \bibinfo {pages} {020402} (\bibinfo {year} {2006})},\
  \Eprint {http://arxiv.org/abs/0602050} {arXiv:0602050 [quant-ph]}
  \BibitemShut {NoStop}%
\bibitem [{\citenamefont {Gradshteyn}\ and\ \citenamefont
  {Ryzhik}(2014)}]{gradshteyn20141}%
  \BibitemOpen
  \bibfield  {author} {\bibinfo {author} {\bibfnamefont {I.~S.}\ \bibnamefont
  {Gradshteyn}}\ and\ \bibinfo {author} {\bibfnamefont {I.~M.}\ \bibnamefont
  {Ryzhik}},\ }\href@noop {} {\emph {\bibinfo {title} {Table of integrals,
  series, and products}}}\ (\bibinfo  {publisher} {Academic press},\ \bibinfo
  {year} {2014})\BibitemShut {NoStop}%
\bibitem [{\citenamefont {Dykman}\ \emph {et~al.}(1991)\citenamefont {Dykman},
  \citenamefont {McClintock}, \citenamefont {Stein},\ and\ \citenamefont
  {Stocks}}]{dykman19911}%
  \BibitemOpen
  \bibfield  {author} {\bibinfo {author} {\bibfnamefont {M.~I.}\ \bibnamefont
  {Dykman}}, \bibinfo {author} {\bibfnamefont {P.~V.~E.}\ \bibnamefont
  {McClintock}}, \bibinfo {author} {\bibfnamefont {N.~D.}\ \bibnamefont
  {Stein}}, \ and\ \bibinfo {author} {\bibfnamefont {N.~G.}\ \bibnamefont
  {Stocks}},\ }\href {\doibase 10.1103/PhysRevLett.67.933} {\bibfield
  {journal} {\bibinfo  {journal} {Phys. Rev. Lett.}\ }\textbf {\bibinfo
  {volume} {67}},\ \bibinfo {pages} {933} (\bibinfo {year} {1991})}\BibitemShut
  {NoStop}%
\bibitem [{\citenamefont {Nieuwenhuizen}\ and\ \citenamefont
  {Allahverdyan}(2002)}]{nieuwenhuizen20021}%
  \BibitemOpen
  \bibfield  {author} {\bibinfo {author} {\bibfnamefont {T.~M.}\ \bibnamefont
  {Nieuwenhuizen}}\ and\ \bibinfo {author} {\bibfnamefont {A.~E.}\ \bibnamefont
  {Allahverdyan}},\ }\href {\doibase 10.1103/PhysRevE.66.036102} {\bibfield
  {journal} {\bibinfo  {journal} {Phys. Rev. E}\ }\textbf {\bibinfo {volume}
  {66}},\ \bibinfo {pages} {036102} (\bibinfo {year} {2002})},\ \Eprint
  {http://arxiv.org/abs/0011389} {arXiv:0011389 [cond-mat]} \BibitemShut
  {NoStop}%
\bibitem [{\citenamefont {Horn}(2006)}]{horn19901}%
  \BibitemOpen
  \bibfield  {author} {\bibinfo {author} {\bibfnamefont {J.~C.}\ \bibnamefont
  {Horn}, \bibfnamefont {R.A.}},\ }\href@noop {} {\emph {\bibinfo {title}
  {Matrix analysis}}}\ (\bibinfo  {publisher} {Springer-Verlag},\ \bibinfo
  {year} {2006})\BibitemShut {NoStop}%
\bibitem [{\citenamefont {Spiegel}(1993)}]{spiegel19931}%
  \BibitemOpen
  \bibfield  {author} {\bibinfo {author} {\bibfnamefont {M.~R.}\ \bibnamefont
  {Spiegel}},\ }\href@noop {} {\emph {\bibinfo {title} {Theory and Problems of
  Complex Variables (Schaum's Outline)}}}\ (\bibinfo  {publisher}
  {McGraw-Hill},\ \bibinfo {year} {1993})\BibitemShut {NoStop}%
\bibitem [{\citenamefont {H{\"{a}}nggi}\ and\ \citenamefont
  {Marchesoni}(2009)}]{hanggi20091}%
  \BibitemOpen
  \bibfield  {author} {\bibinfo {author} {\bibfnamefont {P.}~\bibnamefont
  {H{\"{a}}nggi}}\ and\ \bibinfo {author} {\bibfnamefont {F.}~\bibnamefont
  {Marchesoni}},\ }\href {\doibase 10.1103/RevModPhys.81.387} {\bibfield
  {journal} {\bibinfo  {journal} {Rev. Mod. Phys.}\ }\textbf {\bibinfo {volume}
  {81}},\ \bibinfo {pages} {387} (\bibinfo {year} {2009})}\BibitemShut
  {NoStop}%
\bibitem [{\citenamefont {Li}\ \emph {et~al.}(2010)\citenamefont {Li},
  \citenamefont {Kheifets}, \citenamefont {Medellin},\ and\ \citenamefont
  {Raizen}}]{li20101}%
  \BibitemOpen
  \bibfield  {author} {\bibinfo {author} {\bibfnamefont {T.}~\bibnamefont
  {Li}}, \bibinfo {author} {\bibfnamefont {S.}~\bibnamefont {Kheifets}},
  \bibinfo {author} {\bibfnamefont {D.}~\bibnamefont {Medellin}}, \ and\
  \bibinfo {author} {\bibfnamefont {M.~G.}\ \bibnamefont {Raizen}},\
  }\href@noop {} {\bibfield  {journal} {\bibinfo  {journal} {Science}\ }\textbf
  {\bibinfo {volume} {328}},\ \bibinfo {pages} {1673} (\bibinfo {year}
  {2010})}\BibitemShut {NoStop}%
\bibitem [{\citenamefont {Millen}\ \emph {et~al.}(2014)\citenamefont {Millen},
  \citenamefont {Deesuwan}, \citenamefont {Barker},\ and\ \citenamefont
  {Anders}}]{millen20141}%
  \BibitemOpen
  \bibfield  {author} {\bibinfo {author} {\bibfnamefont {J.}~\bibnamefont
  {Millen}}, \bibinfo {author} {\bibfnamefont {T.}~\bibnamefont {Deesuwan}},
  \bibinfo {author} {\bibfnamefont {P.}~\bibnamefont {Barker}}, \ and\ \bibinfo
  {author} {\bibfnamefont {J.}~\bibnamefont {Anders}},\ }\href {\doibase
  10.1038/nnano.2014.82} {\bibfield  {journal} {\bibinfo  {journal} {Nature
  Nanotechnology}\ }\textbf {\bibinfo {volume} {9}},\ \bibinfo {pages} {425}
  (\bibinfo {year} {2014})}\BibitemShut {NoStop}%
\bibitem [{\citenamefont {Otsuka}\ \emph {et~al.}(2009)\citenamefont {Otsuka},
  \citenamefont {Ohtomo}, \citenamefont {Makino}, \citenamefont {Sudo},\ and\
  \citenamefont {Ko}}]{otsuka20091}%
  \BibitemOpen
  \bibfield  {author} {\bibinfo {author} {\bibfnamefont {K.}~\bibnamefont
  {Otsuka}}, \bibinfo {author} {\bibfnamefont {T.}~\bibnamefont {Ohtomo}},
  \bibinfo {author} {\bibfnamefont {H.}~\bibnamefont {Makino}}, \bibinfo
  {author} {\bibfnamefont {S.}~\bibnamefont {Sudo}}, \ and\ \bibinfo {author}
  {\bibfnamefont {J.~Y.}\ \bibnamefont {Ko}},\ }\href {\doibase
  10.1063/1.3156826} {\bibfield  {journal} {\bibinfo  {journal} {Appl. Phys.
  Lett.}\ }\textbf {\bibinfo {volume} {94}},\ \bibinfo {pages} {241117}
  (\bibinfo {year} {2009})}\BibitemShut {NoStop}%
\end{thebibliography}
\end{document}